\documentclass[aps,pra,secnumarabic,longbibliography,lengthcheck,superscriptaddress,nopreprintnumbers,reprint,groupedaddress]{revtex4-1}
\usepackage{natbib,units}
\usepackage{graphicx}
\usepackage{dcolumn}
\usepackage{bm}        
\usepackage{amssymb}   

\usepackage{bm}
\usepackage[dvipsnames]{xcolor}
\usepackage{color}
\usepackage{url}
\usepackage{dsfont}
\usepackage{slashed}
\usepackage{epsfig}
\usepackage{multirow,array}
\usepackage{float}
\usepackage{fancyhdr}
\usepackage{indentfirst}
\usepackage{cancel}
\usepackage{tabularx}
\usepackage{simplewick}
\usepackage{amsmath}
\usepackage[normalem]{ulem}
\usepackage{slashed}
\usepackage{upgreek}
\usepackage{xtab,afterpage,longtable}
\RequirePackage[colorlinks=true
  ,urlcolor=Green
  ,anchorcolor=blue
  ,citecolor=orange
  ,filecolor=blue
  ,linkcolor=blue
  ,menucolor=blue
  ,linktocpage=true
  ,pdfproducer=medialab
  ,pdfa=true
]{hyperref}

\newcommand{\mnras}{Mon. Not. R. Astron. Soc.}

\newcommand{\apjs}{Astrophys. J. Suppl. Ser.}
\newcommand{\araa}{Annu. Rev. Astron. Astrophys.}
\newcommand{\aap}{Astron. Astrophys.}

\newcommand{\na}{Nat. Astron.}

\begin{document}

\title{Path Measures for Stochastic Galaxy Formation on Layered Halo Graphs}

\author{Daneng Yang}
\email{yangdn@pmo.ac.cn}

\affiliation{$^1$Purple Mountain Observatory, Chinese Academy of Sciences, Nanjing 210033, China}
\affiliation{$^2$School of Astronomy and Space Sciences, University of Science and Technology of China, Hefei 230026, China}

\date{\today}

\begin{abstract}
Galaxy formation, viewed as an inference problem from incomplete information, is inherently stochastic. Reducing the full simulation state to a coarse-grained set of variables integrates out unresolved degrees of freedom, while the retained quantities remain coupled by the underlying dynamics, motivating an effective stochastic description of galaxy formation in reduced variables. Existing approaches have achieved substantial predictive success, but generally lack a unified statistical framework for trajectory-level galaxy assembly and history-conditioned fluctuations. We introduce a Graph Path Likelihood Model (GPLM) that formulates galaxy assembly histories as stochastic dynamical trajectories on hierarchical halo merger graphs, where temporal edges encode causal transport and coeval host edges encode environmental conditioning. Within this formulation, galaxy evolution is described by graph-conditioned path measures and effective actions, from which observables, likelihoods, and response diagnostics emerge from a common probabilistic description. As a first realization, we train a graph neural likelihood model for stellar and gas mass assembly histories on layered halo graphs extracted from hydrodynamic simulations. We show that it reproduces the main statistics of these histories while capturing environmentally conditioned correlated fluctuations. The path measure formulation also provides a natural setting for example fixed-graph applications, which we illustrate with the fraction of dark-matter-deficient galaxies, controlled gas-response deformations, and nonequilibrium diagnostics of environmentally dependent evolution. In particular, the present construction also admits extensions in which merger-history statistics and baryonic evolution are treated within a unified probabilistic description, potentially enabling studies of how graph structure, assembly histories, and galaxy observables respond jointly to variations in the underlying theory, including dark matter microphysics. 
\end{abstract}

\maketitle

\tableofcontents

\section{Introduction}

Galaxy formation is a hierarchical process shaped by merger histories, environmental interactions, local baryonic physics, and the substantial dark matter structures. Without access to full simulation-level information, galaxy formation as an inference problem is intrinsically uncertain in its coarse-grained outcomes, necessitating a stochastic effective dynamics for modeling its evolution. 
Existing galaxy formation models successfully reproduce galaxy populations and catalogs by calibrating key physical processes to match observed galaxy abundances and properties~\cite{White:1991mr,Cole:2000ex,2012NewA...17..175B,2015ARA&A..53...51S,Guo:2015xeg,2016MNRAS.462.3854L,2019MNRAS.490.3234N,Nelson:2018uso,2026OJAp....955306R}. 
However, the same observed distributions can arise from different assembly histories and effective prescriptions, so a path-level statistical measure is needed to describe the underlying variability consistently.
In this work, we introduce a stochastic model based on layered halo graphs that make both assembly transport and local environmental dependence explicit~\cite{Yang:2022oqe,Makinen:2022jsc,2022ApJ...941....7J}. Within this construction, temporal edges encode progenitor-to-descendant transport, while coeval host edges encode satellite-to-host conditioning.
The resulting likelihood-based description marginalizes unresolved baryonic and environmental variability into an effective graph-conditioned stochastic dynamics, shifting the emphasis from a deterministic forward map to a path measure over layered merger histories.

Related work in cosmology covers the reconstruction of dark-matter density, velocity, tidal fields, and even initial conditions from survey data and group catalogs. Methods include Wiener-filtering, halo-domain reconstruction, ELUCID, and recent machine-learning pipelines~\cite{Zaroubi:1994mx,Erdogdu:2006nd,Wang:2008wx,2012MNRAS.420.1809W,Wang:2013ep,Wang:2014hia,Wang:2016qbz,2023MNRAS.519.2199E,Qin:2023dew,Dai:2023kip,2024MNRAS.531.4990P,Shi:2025zoz,Krywonos:2025isr,2020PNAS..11730055C,Li:2020vor}. 
CAMELS-style benchmark suites motivate robust cross-model evaluation by exposing subgrid and domain shift in a controlled setting~\cite{CAMELS:2020cof,CAMELS:2022xwh}. The CAMELS-SAM extension supports large SAM-based catalog ensemble comparisons within the same benchmarks~\cite{Perez:2022nlv}.
Probabilistic approaches to the halo–galaxy connection are also being developed at the level of conditional or joint distributions of galaxy properties given halo features~\cite{Yang:2007pg,Stiskalek:2022nsr,Rodrigues:2023scm,2023mla..confE..21L,Zhang:2023oem,Xu:2023fgg,Rodrigues:2024rbp,2025MNRAS.542.2560J}.

Graph-based studies have grown in recent years. This includes merger-tree and catalog graph neural network (GNN) emulators~\cite{2022ApJ...941....7J,Huertas-Company:2022wni,Villanueva-Domingo:2022rvn,Nguyen:2025nip,Lee:2025tql}, graph IMNN compression for simulation-based cosmological inference~\cite{Makinen:2022jsc}, graph models for halo clustering~\cite{Yang:2022oqe}, and recent cross-model or multimodal inference pipelines built around graph-structured cosmological data~\cite{Jo:2025ndh,Nguyen:2024ndo,Rose:2024xcb}. 
Notably, Mangrove applies GNN to halo merger trees to predict galaxy properties from halo assembly histories~\cite{2022ApJ...941....7J}, and FLORAH uses recurrent neural networks together with normalizing flows to learn the probability distribution of dark matter halo assembly histories~\cite{Nguyen:2025nip}.
These methods connect to optimal-compression techniques for likelihood-free inference~\cite{Heavens:1999am,Alsing:2018eau,2018PhRvD..97h3004C}.
They mainly focus on Eulerian fields, catalog predictions, or compressed parameter summaries. In comparison, our goal is different. We seek an explicit trajectory-level likelihood on a fixed layered graph, allowing transport and environment channels to be represented separately. This enables targeted deformations in path space.

To achieve this, we introduce a Graph Path Likelihood Model (GPLM) on layered halo graphs with conditional path weight
\begin{equation}
P(\mathbf{x}\mid G) \propto p_{\rm attach}(\mathbf{x}\mid G) \exp\left[-S(\mathbf{x};G)\right],
\end{equation}
where $S$ is the within-graph effective action for dynamical increments, and $p_{\rm attach}$ is a boundary measure for entry or infall interfaces. It defines a conditional ensemble of baryonic histories on a fixed layered graph with both assembly and environment information encoded. 

More generally, the fixed-graph construction can be embedded into a graph-trajectory measure
\begin{equation}
P(\mathbf{x},G)=P(\mathbf{x}\mid G)P(G),
\end{equation}
where $P(G)$ is the graph-sector probability in a graph ensemble. 
As a minimal closure of the formalism, we introduce a preferential attachment-detachment (PAD) mechanism extending the attachment- and host-edge-only construction of Ref.~\cite{Yang:2022oqe}. More generally, the graph sector itself could also be treated stochastically. Excursion-set theory offers a stochastic basis for halo formation using random walks and first-crossing statistics~\cite{Bond:1990iw}. However, its standard form is too limited to produce the layered halo graphs discussed here, which also represent same-layer host structure and environment-conditioned connectivity. Extending the framework so that the graph prior $P(G)$ is generated by an explicit stochastic graph dynamics is therefore a natural next step.

We train and test on layered graphs reconstructed from IllustrisTNG-50-1~\cite{2019MNRAS.490.3234N,Pillepich:2019bmb,Nelson:2018uso}. For this paper, we use stellar and gas masses as a minimal two-field realization, but the construction itself is not tied to those variables. The likelihood is realized by a discrete-time Gaussian Onsager-Machlup (OM) action~\cite{1953PhRv...91.1505O} on top of deterministic merger transport. Building on the trained model, we then perform several example calculations on fixed graphs.
We use a DMDG (dark-matter-deficient galaxy) operator to demonstrate quenched path-integral evaluation with and without $p_{\rm attach}$. We then introduce a controlled gas-channel deformation by interpolating between the full learned host-conditioned gas drift and an environment-off reference, and use that same family of deformed dynamics to define nonequilibrium diagnostics. 

Because galaxy formation is highly complex, any controlled attempt to study the impact of dark matter microphysics must work with an effective description built on a reliable backbone while also allowing for intrinsic modeling uncertainty. The present formalism is designed for that purpose. It factorizes the problem into a graph prior for merger-history structure, an attachment sector for node entry, and a conditional path measure for graph-conditioned baryonic evolution, while naturally incorporating effective stochasticity from unresolved physics. Within this framework, dark matter microphysics can then be introduced as controlled deformations of the graph sector, the baryonic path dynamics, or both, making the setup suitable for systematic response studies rather than only forward prediction. 
Self-interacting dark matter (SIDM) provides a concrete example of this broader use case. In the parametric SIDM model of Refs.~\cite{Yang:2023jwn,Yang:2024uqb,Yang:2024tba}, the effects of SIDM are
  incorporated as controlled deformations of the halo structural evolution in $V_{\max}$, the maximum circular velocity, and $R_{\max}$, the radius at which that maximum is attained. From these quantities, the full SIDM density profile can then be reconstructed. The present framework provides a natural graph-conditioned stochastic setting for extending that idea to assembly histories
  and environmental dependence. 

The paper is organized as follows. Section~\ref{sec:graphs} defines layered halo graphs and graph-distance environments. Section~\ref{sec:stochastic_model} sets the likelihood formulation and the physical rationale for an effective stochastic description. Section~\ref{sec:gplm} introduces the GPLM construction and presents validation results. Section~\ref{sec:ensemble_path_integral} develops the graph-conditioned path integrals and the graph-ensemble formalism. Sections~\ref{sec:example_dmdg} and~\ref{sec:lambda_response} give the DMDG and response examples. Section~\ref{sec:noneq} presents path-space likelihood diagnostics. Section~\ref{sec:discussion} summarizes with implications and future directions.

\section{Layered halo graphs from simulations}
\label{sec:graphs}

The geometric object central to this work is a \emph{layered halo graph}. We represent it as
\begin{equation}
G = (\mathcal{V},\mathcal{E}^{\rm time}\cup\mathcal{E}^{\rm host}),
\end{equation}
where $\mathcal{V}$, $\mathcal{E}^{\rm time}$, and $\mathcal{E}^{\rm host}$ represent the set of halos, temporal edges, and spatial host edges. 
$\mathcal{V}$ comprises selected halos across snapshots that form the main halo of interest at redshift zero ($z=0$). These are linked by temporal edges in $\mathcal{E}^{\rm time}$ and spatial edges in $\mathcal{E}^{\rm host}$. Temporal edges $\mathcal{E}^{\rm time}$ connect progenitors to descendants across snapshots and encode merger-tree transport. Same-snapshot host edges $\mathcal{E}^{\rm host}$ connect satellites to enclosing hosts at fixed snapshot, following the assignment rule introduced in Ref.~\cite{Yang:2022oqe}. Building on this structure, the layered halo graph also stores halo and galaxy properties as node attributes. The temporal and environment-dependent relations can also be stored as edge attributes. When emphasizing the graph as a latent random variable in the hierarchical model, we use $\mathcal{G}$.

Given halo catalogs and merger trees from a cosmological simulation, we construct $G$ by traversing snapshots backward from a $z=0$ host and including halos above a mass threshold $M_{\rm cut}$. For the results in this paper we construct
layered graphs for the 100 most massive $z=0$ hosts in IllustrisTNG-50-1~\cite{2019MNRAS.490.3234N,Pillepich:2019bmb,Nelson:2018uso}. 
We split the sample by mass rank, assigning even-rank masses to the training set and odd-rank masses to the test set.
The temporal edges are taken from the simulation merger tree catalog SubLink~\cite{2015MNRAS.449...49R} for IllustrisTNG, while host edges are defined within each snapshot from the group catalog.

The construction proceeds as follows:
\begin{itemize}
\item \textbf{1. Initial node assignment.} Initialize $G$ by including the main halo and all its subhalos in the $z=0$ snapshot as nodes.
\item \textbf{2. Progenitor assignments.} Create temporal edges by linking each halo to its progenitors in the nearest earlier snapshot. Include these progenitors as new nodes in $G$.
This temporal edge creation is applied only to an ``active'' set of nodes, which consists of the halos already included in the graph at the frontier snapshot, ensuring that progenitor assignments are restricted to the causal ancestry of the $z=0$ system.
Each halo may have multiple progenitor halos, allowing several temporal links to be created.
The assumed direction, which is taken to point from the earlier snapshot to the later one.
\item \textbf{3. Host assignments.} For each node corresponding to a subhalo, create a spatial edge from this node to the one corresponding to its least-massive host halo within the same snapshot. A direction from the subhalo node to the host node is assumed, but it may or may not be effective in application.
A host halo is by definition heavier than its subhalos. The latters reside within its virial radius.
Unless otherwise specified, we define the virial radius as the radius where the mean enclosed density equals 200 times that of the critical energy density of the universe.
\item \textbf{4. Domain expansion.} Move to the nearest earlier snapshot.
Search for halos residing within the virial radii of the previously included progenitor halo nodes, and include them in $G$.
Go back to 2 and \textbf{iterate} until no nodes corresponding to halos of masses above $M_{\rm cut}$ in the earlier snapshot can be included.
\end{itemize}

\begin{figure*}[htbp]
  \centering
  \includegraphics[width=5.5cm, trim=30 25 30 25, clip]{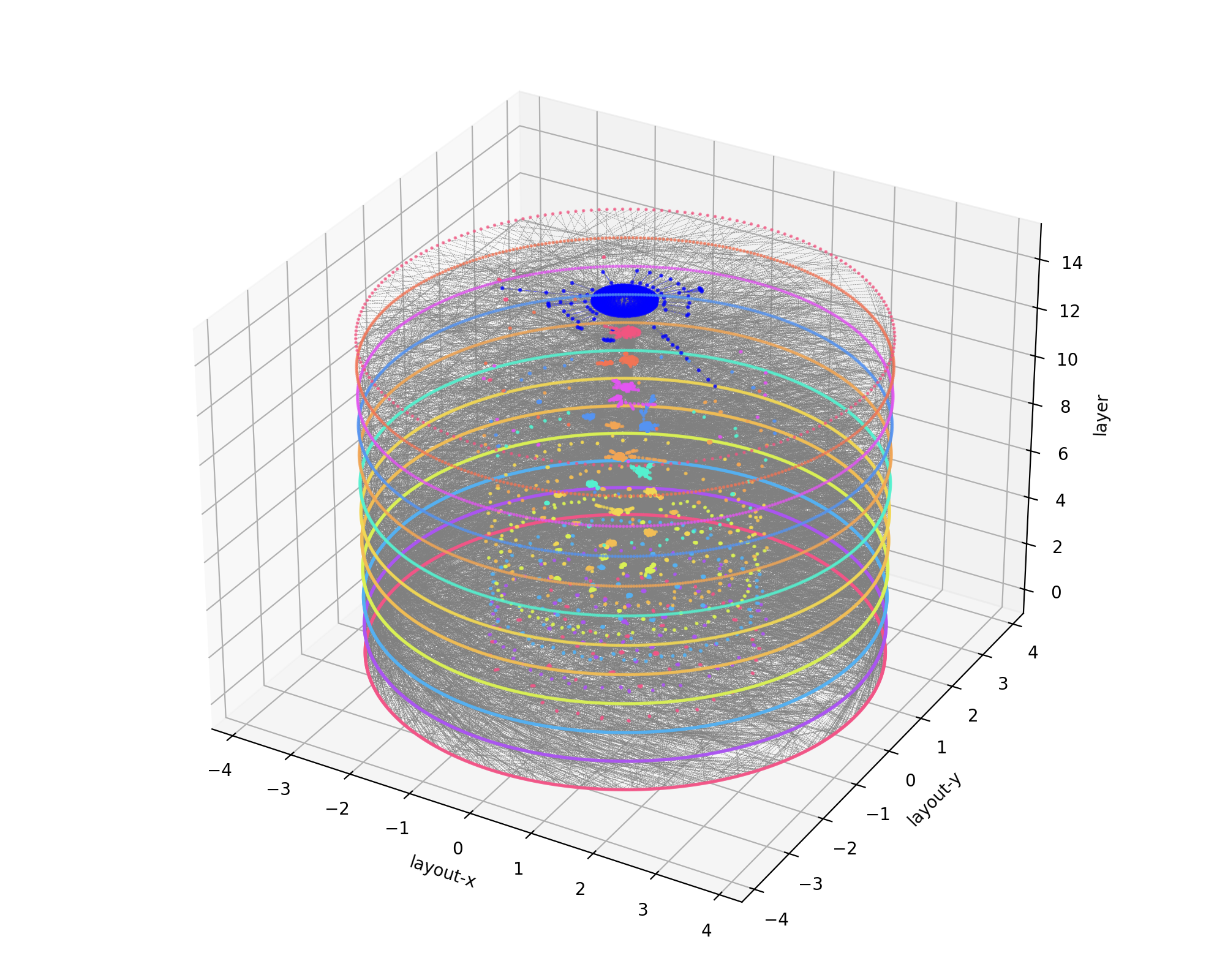}
  \includegraphics[width=5.5cm, trim=30 25 30 25, clip]{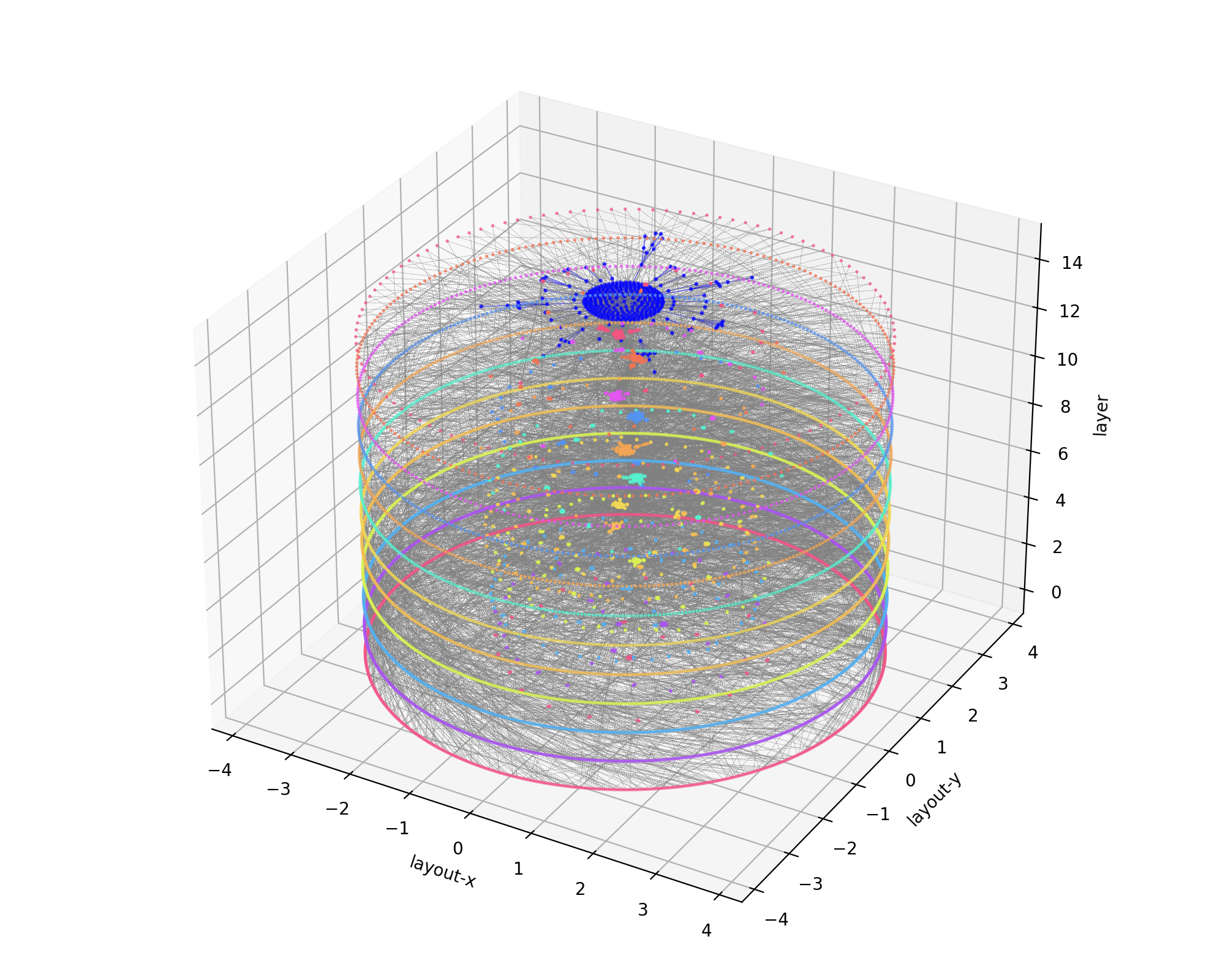}
  \includegraphics[width=5.5cm, trim=30 25 30 25, clip]{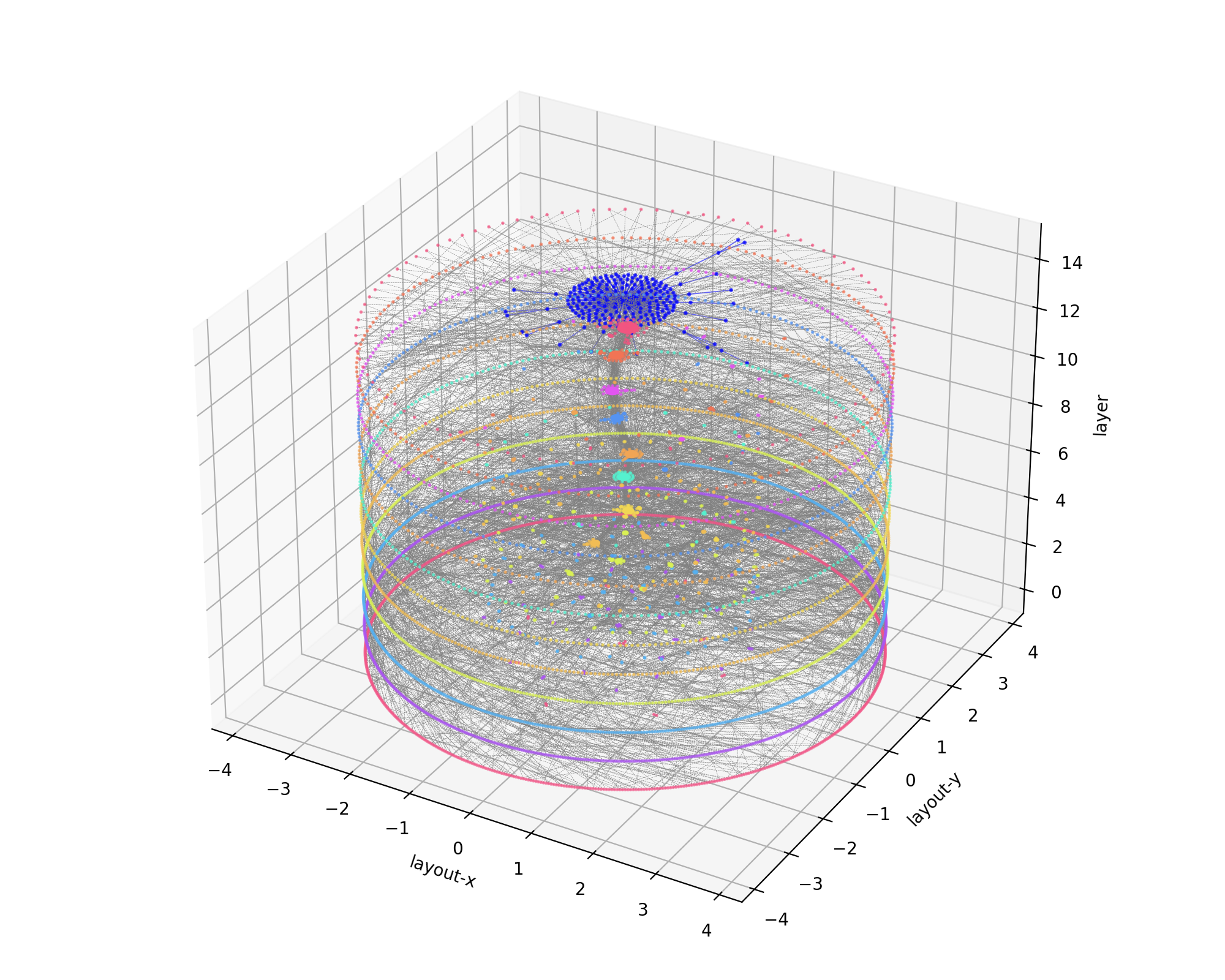}
  \caption{\label{fig:LHG}
Layered halo graphs constructed for the three most massive $z{=}0$ hosts (left to right: rank~1, rank~2, rank~3) in the TNG-50-1 simulation. Each panel shows the layered topology obtained by the backward construction.  Nodes correspond to halos of mass higher than $M_{\rm cut}=10^9~\rm M_{\odot}/h$ and are illustrated with different colors at different layers. Temporal and host edges are both illustrated in dashed gray. They connect all the nodes across the layers into a single object.
The illustrated 13 layers correspond to redshifts $z = 5.0, 4.0, 3.0, 2.0, 1.5, 1.0, 0.7, 0.5, 0.4, 0.3, 0.2, 0.1,$ and $0$, as stacked from the bottom to top. At layers with redshifts greater than zero, rings of nodes surrounding the central graphs correspond to isolated low-mass halos that accrete into their hosts in the next layer. At redshift zero, there is only one halo graph that corresponds to the host halo and its subhalos. These graphs, together with the attributes carried by the nodes and edges, are inputs to the GPLM.
}
\end{figure*}

Following these steps, we obtain a connected graph that is layered in time and acyclic when edge directions are considered.
Notably, multiple host-centered components can appear at early times. These correspond to distinct progenitors that later merge into the final $z=0$ system.
The layered graph structure explicitly distinguishes infall and mergers. Halos that fall into the host are treated as satellites while they orbit. Only if multiple progenitors map to a common descendant at the next snapshot does a merger occur, triggering reservoir summation along temporal edges.

Figure~\ref{fig:LHG} presents the three highest-ranked layered halo graphs reconstructed from the IllustrisTNG TNG-50-1 run. These examples illustrate the layered data backbone on which we evaluate the graph-conditioned trajectory likelihood. We construct the layers from 13 simulation snapshots 17, 21, 25, 33, 40, 50, 59, 67, 72, 78, 84, 91, and 99, corresponding to redshifts $z = 5.0, 4.0, 3.0, 2.0, 1.5, 1.0, 0.7, 0.5, 0.4, 0.3, 0.2, 0.1,$ and $0$. Lower redshift snapshots appear as higher layers in the graph. We selected halos with mass above $M_{\rm cut}=10^9~\rm M_{\odot}/h$ and display their corresponding nodes from different layers in distinct colors. In the high-redshift layers, circular rings of nodes represent isolated halos that later merge into a common host in the next layer. At redshift zero, host edges connect all nodes into a single structure. At earlier times, several disconnected hosts coexist and later merge into one dominant halo. Appendix~\ref{sec:graph1} further illustrates selected layers of the rank-1 halo in both a topological graph representation and a physical view that places each node at its position in the simulation box.

While the construction is designed to include the resolved halos relevant to the $z=0$ host together with their evolution trajectories, the present implementation based on the existing TNG50-1 merger trees and halo catalogs should be regarded as an effective representation of a fully closed merger forest rather than an exact realization of it. 
In practice, some physically continuous branches in the full simulation could still appear with missing resolved merger history in the layered graphs. The GPLM developed in this work is therefore conditioned on the constructed graph, and inherits the corresponding technical and numerical limitations. Integrating over the missing branches would generally broaden the effective scatter of the graph-conditioned likelihood, so improved layered-graph construction should lead to a cleaner boundary sector and potentially better predictive performance in future work.

\begin{figure*}[htbp]
  \centering
  \includegraphics[width=16.5cm]{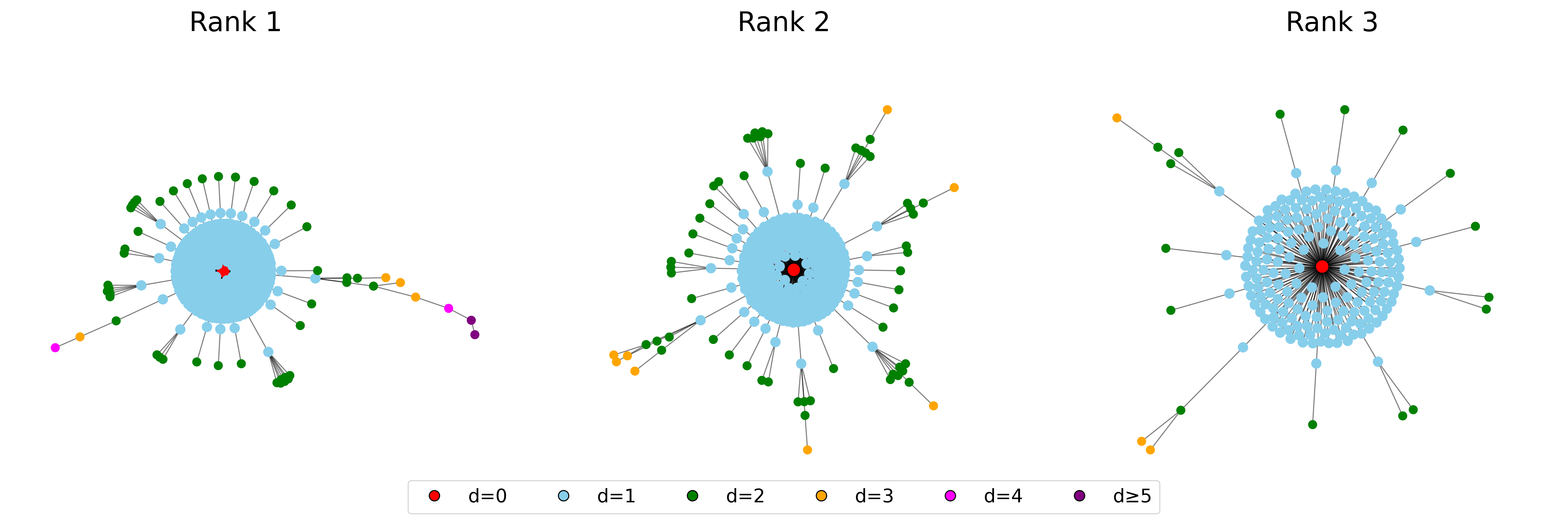}
  \caption{\label{fig:graphs}
  Layered halo graph panels at $z{=}0$ for the three most massive hosts
  (ranks~1-3). Each panel shows the coeval host-subhalo network in a
  spring layout; node colors encode the graph distance $d_G$ defined on the
  connected host-edge components (legend), with $d_G{=}0$ marking the
  component roots, $d_G{=}1$ the immediate satellites, and $d_G\ge 2$
  outer/higher-order satellites. These bins define the environment categories
  used throughout the paper.}
\end{figure*}

To show that layered halo graphs help build likelihood models for environmental effects, we use graph metrics for later comparison. Many topological metrics can be defined on graphs. For example, Ref.~\cite{Yang:2022oqe} studied degree distributions, the $s$-metric, graph distances, and normalized Laplacian or adjacency spectra, along with spectral distances between graphs. 

For simplicity, we use the graph distance $d_G(i)$ to separate the host nodes, immediate satellites, and outer satellites. The $d_G(i)$ is defined on the host-edge subgraph at each snapshot. Specifically, for any layer, we treat each connected component as an undirected host-edge graph.
We define a root set by the node or nodes of maximal degree in that component (in case there are multiple), and then define $d_G(i)$ as the shortest path distance from node $i$ to this root set within its own connected component. 
The environment bins are therefore $d_G=0$ (component roots / host nodes), $d_G=1$ (immediate satellites), and $d_G>1$ (outer/higher-order satellites). Isolated nodes therefore satisfy $d_G=0$. For diagnostics restricted to $z{=}0$, we compute $d_G$ on the $z{=}0$ host graph. For stacked, all-snapshot metrics, we compute $d_G$ separately on each layer. This choice yields a simple, topology-based environmental separation that matches a graph-conditioned likelihood framework.

Figure~\ref{fig:graphs} shows the nodes corresponding to different $d_G$ values, color-coded for the three most massive (highest rank) halos at $z=0$. The connectivity among their constituent nodes already differs at the topological level, indicating that the halos are structurally distinct. Thus, the host-edge graph provides a natural backbone for encoding environmental effects.

\section{Coarse-Grained Stochastic Dynamics and Hierarchical Likelihood}
\label{sec:stochastic_model}

The problem of galaxy formation is often described as largely deterministic. Given a realization of the primordial density field, gravitational motion and baryonic processes are, in principle, governed by deterministic equations. Indeed, most semi-analytic models (SAMs) are formulated as coupled systems of ordinary differential equations that propagate galaxy properties forward in time along merger trees~\cite{Press:1973iz,Somerville:1998bb,1993MNRAS.262..627L,Lacey:1994su,Cole:2000ex,2006MNRAS.370..645B,2012NewA...17..175B,2015MNRAS.451.2663H,2016MNRAS.462.3854L,Ando:2019xlm,2018MNRAS.474..492M,Jiang:2020rdj,Barrera:2022jgo,Hadzhiyska:2021kmt,Alarcon:2022hzg,Alarcon:2025cij}.

However, galaxy formation as an inference problem is intrinsically uncertain. For each observed galaxy, we measure only a late-time snapshot of a high-dimensional evolutionary history. Even in simulations, we record only a finite set of halo and baryonic descriptors at discrete times. Any practical forward model therefore discards part of the available information.

On a layered halo graph, we evolve a finite set of resolved variables along temporal edges while conditioning on host edges. All remaining degrees of freedom, such as feedback frequency, multiphase gas structure, time of infall, and external influences outside the graph domain, are integrated out. 
The resulting coarse-grained dynamics is Markovian in the retained graph-conditioned state, but this state is not memoryless. It contains assembled information such as transported resolved quantities, graph connectivity, and environment-conditioned structure. Unresolved physics can also leave temporally extended imprints through effective time- and state-dependent drifts and scatters. As a result, the unresolved aspects of assembly dependence are captured in an effective stochastic sector, whose influence can persist over multiple steps, even within a Markovian reduced description.

To see how stochasticity emerges, it is convenient to work in increment space. For a node $i$ on layer $k$, we denote the resolved state by $x_{i,k}$ and the one-step increment by $\Delta x_{i,k} \equiv x_{i,k+1} - x_{i,k}$. Suppose that the increment is effectively deterministic once $(s,\zeta)$ is specified, where $s$ is the coarse state and $\zeta$ collects unresolved information. We have
\begin{equation}
\Delta x = f(s,\zeta),
\qquad
P(\Delta x\mid s,\zeta)=\delta\bigl(\Delta x - f(s,\zeta)\bigr).
\end{equation}
Conditioning only on $s$ corresponds to marginalizing over $\zeta$,
\begin{equation}
P(\Delta x\mid s) = \int d\zeta \delta\bigl(\Delta x - f(s,\zeta)\bigr) P(\zeta\mid s),
\end{equation}

which is not definite because
\begin{equation}
\mathrm{Var}(\Delta x\mid s) = \underbrace{\mathbb{E}\left[ \mathrm{Var}(\Delta x\mid s,\zeta)\right]}_{=0} 
+ \underbrace{\mathrm{Var}\left[ \mathbb{E}(\Delta x\mid s,\zeta) \right]}_{\neq 0}.
\end{equation}
Since $\Delta x$ is deterministic at fixed $\zeta$, the first term vanishes and we obtain 
$\mathrm{Var}(\Delta x\mid s) = \mathrm{Var}_{\zeta}\left[f(s,\zeta)\right]$. 
This derivation is generic. Hence, integrating over unresolved degrees of freedom introduces a finite level of stochasticity, even when the underlying microphysics is deterministic~\cite{2019MNRAS.487.3845C}.

To further understand the stochasticity in the continuum limit, we examine how the increment formulation behaves as the layer spacing $\Delta t$ is refined. 
We assume the unresolved contributions between adjacent layers lie in the diffusive regime,
\begin{equation}
\mathbb{E}[\Delta x_k] \sim \Delta t, \qquad \mathrm{Var}(\Delta x_k) \sim \Delta t.
\end{equation}
Although the variance of a single increment vanishes as $\Delta t \to 0$, over a fixed macroscopic interval $T$, the number of increments scales as $N \sim T/\Delta t$, so that
\begin{equation}
\mathrm{Var}\left(\sum_{k=1}^{N} \Delta x_k\right) \sim N \Delta t \sim T,
\end{equation}
which remains finite. 

Under this scaling, the evolution is approximated by a Langevin-type stochastic differential equation (SDE) in the continuum limit, 
\begin{equation}
dx = b(x)dt + \sqrt{2D(x)}dW_t,
\end{equation}
where $b(x)$ is the drift field, $D(x)$ is the diffusion rate, and $W_t$ is the standard Wiener process for idealized Brownian motion. 

For an uncorrelated random walk, the characteristic spread scales as $\sqrt{\Delta t}$. In the present reduced description, however, unresolved dynamics can induce temporal correlations through the evolving retained state, and the effective drift and diffusion coefficients can depend explicitly on time. A Langevin-type effective dynamics is therefore needed not merely to add noise, but to encode how unresolved fluctuations are propagated, regulated, and imprinted on later coarse-grained evolution. 

While the Langevin equation governs the stochastic evolution of individual realizations, it is useful to reformulate the problem in a field-theoretical framework. This can be achieved in the Martin-Siggia-Rose-Janssen-De Dominicis (MSRJD) formalism, which constructs a path integral representation of the stochastic process~\cite{Martin:1973zz,Janssen:1976qag,etde_7140916}.
This formalism introduces an auxiliary response field $\hat x_i$ that enforces the stochastic equations of motion within the path integral.

In this framework, the joint weight can be written as
\begin{equation}
P(x,\hat x \mid G) \propto \exp\left[-S(x,\hat x ; G)\right],
\end{equation}
where the MSRJD action admits the decomposition
\begin{equation}
S[x,\hat x;G] = S_0[x,\hat x;G] + S_{\rm NG}[x,\hat x;G],
\end{equation}
with 
\begin{eqnarray}
S_0[x,\hat x;G] &=& \int dt \left[
\hat x_i \left(\dot x_i-b_i\right)-\frac{1}{2}\hat x_i D_{ij}\hat x_j
\right], \\ \nonumber
S_{\rm NG}[x,\hat x;G] &=& \int dt \left[
\frac{1}{3!}C^{(3)}_{ijk}(G) \hat x_i \hat x_j \hat x_k \right. \\ \nonumber
&& \left. +\frac{1}{4!}C^{(4)}_{ijkl}(G) \hat x_i \hat x_j \hat x_k \hat x_l
+\cdots
\right]. 
\end{eqnarray}
Here and in the following, repeated indices are summed. 
The quadratic contribution $S_0$ reproduces the standard Langevin stochastic differential equation with Gaussian noise, while $S_{\rm NG}$ captures non-Gaussian corrections. The term $S_{\rm NG}$ encodes effects such as skewness, heavy tails, and other signatures of unresolved physics, and is generically generated under coarse-graining when small-scale degrees of freedom are integrated out.

An effective action $S_{\rm eff}(x;G)$ is obtained by integrating out the response field,
\begin{equation}
P(x \mid G) \propto \int \mathcal{D}\hat x e^{-S[x,\hat x;G]}
\equiv \exp\left[-S_{\rm eff}(x;G)\right].
\end{equation}
In general, this procedure induces non-Gaussian contributions inherited from $S_{\rm NG}$. 
In this work, we focus on the Gaussian sector and retain only $S_0$, resulting in a quadratic effective action. The implications will be discussed in Sec.~\ref{sec:gplm}.

Our framework can be formulated through three ingredients: the layered halo graph $\mathcal{G}$, the latent galaxy state $\mathbf{x}(t)$ defined on its nodes, and an endpoint data vector $d_{\rm end}$ that summarizes the observed or externally inferred late-time halo-galaxy state. The corresponding joint distribution can be written schematically as
\begin{equation}
P(d_{\rm end}, \mathbf{x}, \mathcal{G}) = P(d_{\rm end}\mid \mathbf{x}, \mathcal{G}) P(\mathbf{x}\mid \mathcal{G}) P(\mathcal{G}),
\end{equation}
where $P(\mathbf{x}\mid \mathcal{G}) = p_{\rm attach}(\mathbf{x}\mid\mathcal{G})\exp\left[-S(\mathbf{x};\mathcal{G})\right]$ describes graph-conditioned baryonic dynamics. 
Here, $d_{\rm end}$ stands for endpoint information that is available for conditioning, such as observed galaxy properties and inferred halo features. 
The central latent uncertainty then lies in the unobserved assembly history and the corresponding graph-conditioned baryonic trajectory.
The remaining factor $P(d_{\rm end}\mid \mathbf{x}, \mathcal{G})$ denotes a separate endpoint-data model, to be specified according to the observational or inference setting of interest.

From this perspective, the natural inference problem is not just forward prediction of endpoint properties, but the construction of a conditional measure over histories consistent with the endpoint information. In the present work, this history dependence enters through the fixed-graph path measure $P(\mathbf{x}\mid\mathcal{G})$. In a more complete extension, one would also specify a graph-sector probability $P(\mathcal{G})$, or more generally an endpoint-conditioned prior over graph histories. The corresponding inverse problem can then be written schematically as
\begin{equation}
P(\mathcal{G},\mathbf{x}\mid d_{\rm end}) \propto P(d_{\rm end}\mid \mathbf{x},\mathcal{G})
P(\mathbf{x}\mid \mathcal{G}) P(\mathcal{G}).
\end{equation}
Marginalizing over $\mathbf{x}$ then yields a posterior over assembly histories,
\begin{equation}
P(\mathcal{G}\mid d_{\rm end}) \propto P(\mathcal{G}) \int \mathcal{D}\mathbf{x} P(d_{\rm end}\mid \mathbf{x},\mathcal{G}) P(\mathbf{x}\mid \mathcal{G}),
\end{equation}
showing explicitly how an extended GPLM could support inverse inference of assembly history from endpoint information. The associated forward marginal likelihood is
\begin{equation}
\mathcal{L}(d_{\rm end}) = \int \mathcal{D}\mathcal{G} P(\mathcal{G}) \int \mathcal{D}\mathbf{x} P(d_{\rm end}\mid \mathbf{x},\mathcal{G}) P(\mathbf{x}\mid \mathcal{G}).
\end{equation}
This makes clear why a path measure is useful even when endpoint information is incomplete or noisy. It defines the latent conditional distribution needed to ask which aspects of assembly history and environment remain statistically recoverable from the available data. 

\section{Graph Path Likelihood Model}
\label{sec:gplm}

The GPLM is a concrete realization of the conditioned probability function $P(\mathbf{x}\mid G)$. 
We introduce it as a graph-conditioned trajectory-likelihood model based on a shared message-passing graph neural network on layered halo graphs. The local update kernel acts on successive layer pairs, while training is carried out through autoregressive windows so that the learned update is exposed to its own predicted states before full-graph inference. A deterministic backbone first propagates extensive quantities along temporal edges. A residual network then uses the transported state together with same-layer host-edge context to predict state-dependent drift corrections and covariances. The same update kernel is reused across nodes, snapshots, and graphs and is iterated sequentially through the full history at inference time. This section describes the model construction, training objective, and inference procedure.

\begin{figure*}[htbp]
  \centering
  \includegraphics[width=16.5cm]{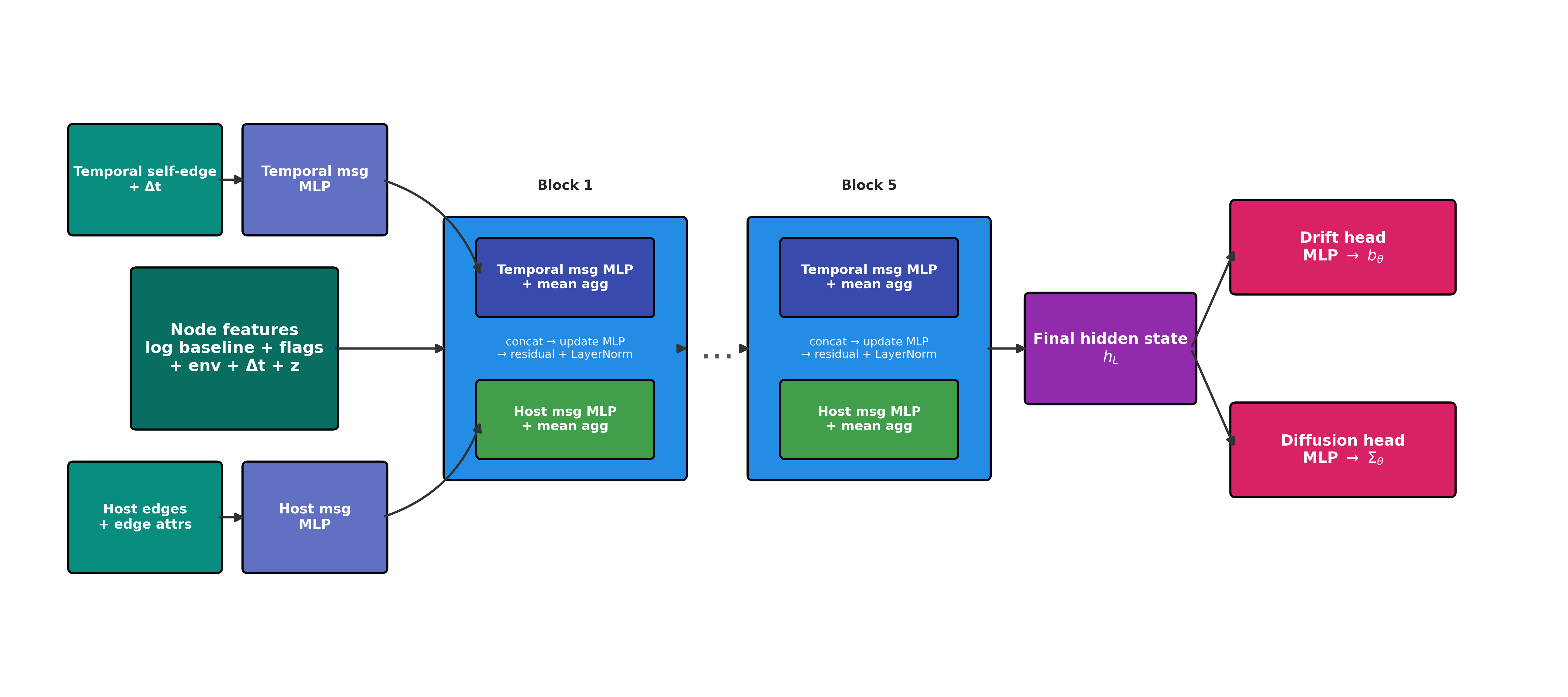}
  \caption{\label{fig:gneft_arch} Overview of the GPLM architecture. Successive layer pairs extracted from a layered halo graph feed one shared message-passing backbone during both training and inference. The deterministic baseline transports extensive quantities along temporal edges. A residual GNN then predicts drift corrections and covariances conditioned on the transported state and the current-layer graph context, with host-edge message passing explicit and temporal history entering through transported node features and predecessor-conditioned temporal inputs. Colors indicate the transport-conditioned temporal channel and the host-edge channel. Temporal and host messages are computed by separate edge-conditioned MLPs, aggregated by destination-node means, concatenated with the incoming hidden state, passed through an update MLP, and then added back through a residual connection followed by LayerNorm.}
\end{figure*}

\subsection{Transport Baseline and Residual Architecture}

For any extensive fields, such as stellar and gas masses, the temporal edges in the layered halo graph define a transport operator $T^{(k)}$ that aggregates progenitor information over the interval $\Delta t_k=t_k-t_{k-1}>0$. For a mass-like field $q$ ($M_{\star}$ and $M_{\rm gas}$ in this work), we write the transported progenitor state as
\begin{equation}
\begin{aligned}
 \widetilde{q}^{(k)}_i    &= \sum_{j} T^{(k)}_{ij} q^{(k-1)}_j   = \sum_{j\in\mathcal{P}(i)} q^{(k-1)}_j,\\
 T^{(k)}_{ij}&\in\{0,1\},
\end{aligned}
\end{equation}
where $\mathcal{P}(i)$ is a set of progenitor nodes with temporal edges into $(i,k)$. If a subhalo disappears between two layers, its reservoir is propagated along an explicit temporal edge to the descendant assigned by the merger tree at the next layer, i.e., the host remnant. If a satellite survives to the next layer, it retains its own node and its own $q$. Nodes without progenitors, either newly formed halos or the first layer of a layered graph, are seeded from the catalog for training and masked from the loss for that transition. This construction ensures that only genuine mergers sum reservoirs into a common descendant, and the training learns information encoded in the tidal evolution of satellites. 

In practice, all modeled mass-like fields are represented in offset-log variables $x_{i,f}=\ln(M_{i,f}+\varepsilon)$ with a fixed $\varepsilon=0.1$, and the training targets are increments relative to the transported state. 
The current variable transform and merger-transport rule work for only extensive fields. Non-extensive quantities may still admit merger transport, but generally through separate prescriptions rather than simple summation, and are not included in the present implementation. 
For this first implementation, we consider only $M_{\rm gas}$ and $M_\star$ in the transport-only backbone.  

The GPLM is constructed as a residual network to learn all additional increments relative to that field-dependent baseline. 
In this setup, any improvement beyond the deterministic backbone must come from learned residual dynamics conditioned on state and environment. 
Notably, host edges do not induce explicit mass exchange. Thus, effects such as tidal stripping enter through graph-conditioned residual drift and covariance rather than through a separate transport channel.

The residual increments are conditioned on the deterministic baseline and the local graph environment. Let $\widetilde{M}_{i,f}^{(k)}$ denote the field-dependent baseline mass defined above for field $f$ at node $i$ and layer $k$. Drift outputs are parameterized in rate units so that nonuniform $\Delta t_k$ can be handled cleanly, but the supervised target remains the residual increment relative to that baseline. For a node $i$ in layer $k$, the dataset stores
\begin{equation}
  \Delta x_{i,f,k} \equiv \log\big(M_{i,f}^{(k)}+\varepsilon\big)  - \log\big(\widetilde{M}_{i,f}^{(k)}+\varepsilon\big).
\end{equation}
Thus $\Delta x_{i,f,k}$ is the residual log-increment beyond the transport-only backbone. 
The residual network outputs a drift-rate vector $\mathbf{b}_{\theta_{\rm GPLM},i,k}$ and a diffusion matrix $\mathbf{D}_{\theta_{\rm GPLM},i,k}$, with $\theta_{\rm GPLM}$ referring to the trainable GPLM parameters. 
The implied mean increment over the step is
\begin{equation}
  \boldsymbol{\delta}_{i,k} \equiv \mathbf{b}_{\theta_{\rm GPLM},i,k}\Delta t_k,
\end{equation}
and the increment covariance is
$\widehat{\mathbf{\Sigma}}_{i,k} \equiv \mathbf{D}_{\theta_{\rm GPLM},i,k}\Delta t_k$. 
The likelihood compares the residual target vector $\boldsymbol{\Delta x}_{i,k}$ against $\boldsymbol{\delta}_{i,k}$ with covariance
$\widehat{\mathbf{\Sigma}}_{i,k}$.
\begin{equation}
\boldsymbol{\Delta x}_{i,k}\mid G  \sim  \mathcal{N}\left(\boldsymbol{\delta}_{i,k},\widehat{\mathbf{\Sigma}}_{i,k}\right). 
\end{equation}

Figure~\ref{fig:gneft_arch} presents an overview of the GPLM architecture. For each adjacent layer pair, deterministic transport along the temporal edges first provides a baseline state on the current layer. The shared message-passing backbone then acts on the current-layer nodes through stacked edge-conditioned message-passing blocks. Within each block, a transport-conditioned temporal channel and a same-layer host-edge channel are processed separately, aggregated by destination-node means, combined with the current hidden state, and passed through an update MLP whose output is added back residually before LayerNorm. Two lightweight MLP heads sit on top of the shared backbone. The drift head predicts a residual rate per field, while the covariance head predicts either diagonal log-variances or a Cholesky factor corresponding to a full covariance.

The feature set denotes the full snapshot-aligned node representation, reused identically at training and inference. For each node in layer $k$, the input vector consists of the transported baseline log-fields, binary presence flags indicating whether each modeled field remains nonzero after transport, the host-graph indegree and outdegree, the node halo mass and assigned host-halo mass together with their logarithmic transforms, a binary satellite flag, the host-relative distance and speed, the current redshift, and the time increment $\Delta t_k$ since the previous retained layer. Among these quantities, only the transported target-field components are evolved by the prediction. The remaining node features serve as conditioning information. 

Temporal and host-edge information are supplied separately through edge attributes. The temporal channel uses self-edges for propagated nodes and carries only $\Delta t_k$. Host edges are represented bidirectionally within layer $k$ and carry the logarithms of the sender and receiver halo masses, $\Delta t_k$, and a direction flag distinguishing the child-to-host and host-to-child directions of the same subhalo-host pair. Environmental information therefore enters GPLM through several complementary channels, namely same-layer host-edge message passing, local node attributes, and the deterministic baseline propagated from the previous layer, which preserves the effects of earlier interactions.

With the drift and diffusion explicitly modeled in the architecture, GPLM realizes the discrete-time Gaussian OM action motivated in Sec.~\ref{sec:stochastic_model}. For the supervised node-layer entries, the action takes the form~\cite{1953PhRv...91.1505O}
\begin{widetext}
\begin{equation}
S_{\rm OM}[\mathbf{x};G]
= \frac{1}{2}\sum_{(i,k)\in\mathcal{V}_{\rm sup}}
   \Big[
     (\boldsymbol{\Delta x}_{i,k}-\boldsymbol{\delta}_{i,k})^{\mathsf T}
     \widehat{\mathbf{\Sigma}}_{i,k}^{-1}
     (\boldsymbol{\Delta x}_{i,k}-\boldsymbol{\delta}_{i,k})
     + \log\det\widehat{\mathbf{\Sigma}}_{i,k}
   \Big],
\label{eq:gneft_action}
\end{equation}
\end{widetext}
up to additive constants independent of $(\mathbf{x},G)$, where $\mathcal{V}_{\rm sup}$ denotes the set of supervised node-layer entries with defined targets. The action is evaluated only for transitions with defined residual targets in $M_\star$ and $M_{\rm gas}$. Nodes without temporal progenitors are excluded, and trivial zero-to-zero mass segments are masked channel by channel.

The same conditional Onsager-Machlup structure also defines the training objective. Parameter estimation minimizes a discrete OM loss proportional to the summed supervised contributions of $S_{\rm OM}$, so the learned drift and diffusion are fitted within the same path-likelihood model that is later used for inference and path-space diagnostics. The learned path measure is therefore not obtained by reinterpreting the predictor afterward, but is the probabilistic model directly encoded by the training procedure itself.

We train on the even-rank subset of the top 100 TNG50-1 hosts and test on the odd-rank subset, yielding 50 layered halo graphs in each split as described in Sec.~\ref{sec:graphs}. 
This separation ensures that all the inferences are computed on unseen layered halo graphs. The finite size of this training sample also guides the choice of training strategy.

The layered graph data are decomposed into successive layer pairs $(k-1,k)$, each containing the temporal transport map between layers and the host edges within layer $k$. Instead of treating these pairs independently, the training pipeline uses overlapping autoregressive windows, advancing the model within each window. In this implementation, each window covers 6 retained layers. Given that each graph has 13 retained layers and the training set includes only 50 layered halo graphs, this choice balances two competing considerations. The window must be long enough for the shared local update kernel to capture multi-step temporal propagation, while remaining short enough to ensure stable and statistically meaningful optimization with a limited number of trajectories. Consequently, the model is trained on its own predicted intermediate states, which reduces the train-inference mismatch associated with truth-seeded inputs.

Designed this way, GPLM can learn nontrivial physical structure in galaxy formation. The same weights are shared across all layered pairs, so the model learns a single conditional update kernel that is reused across nodes, snapshots, and unseen graphs. Redshift, $\Delta t$, and environmental effects then enter through the conditioning variables rather than through separate time- or graph-specific fits.

\begin{figure*}[htbp]
  \centering
  \includegraphics[width=0.78\textwidth]{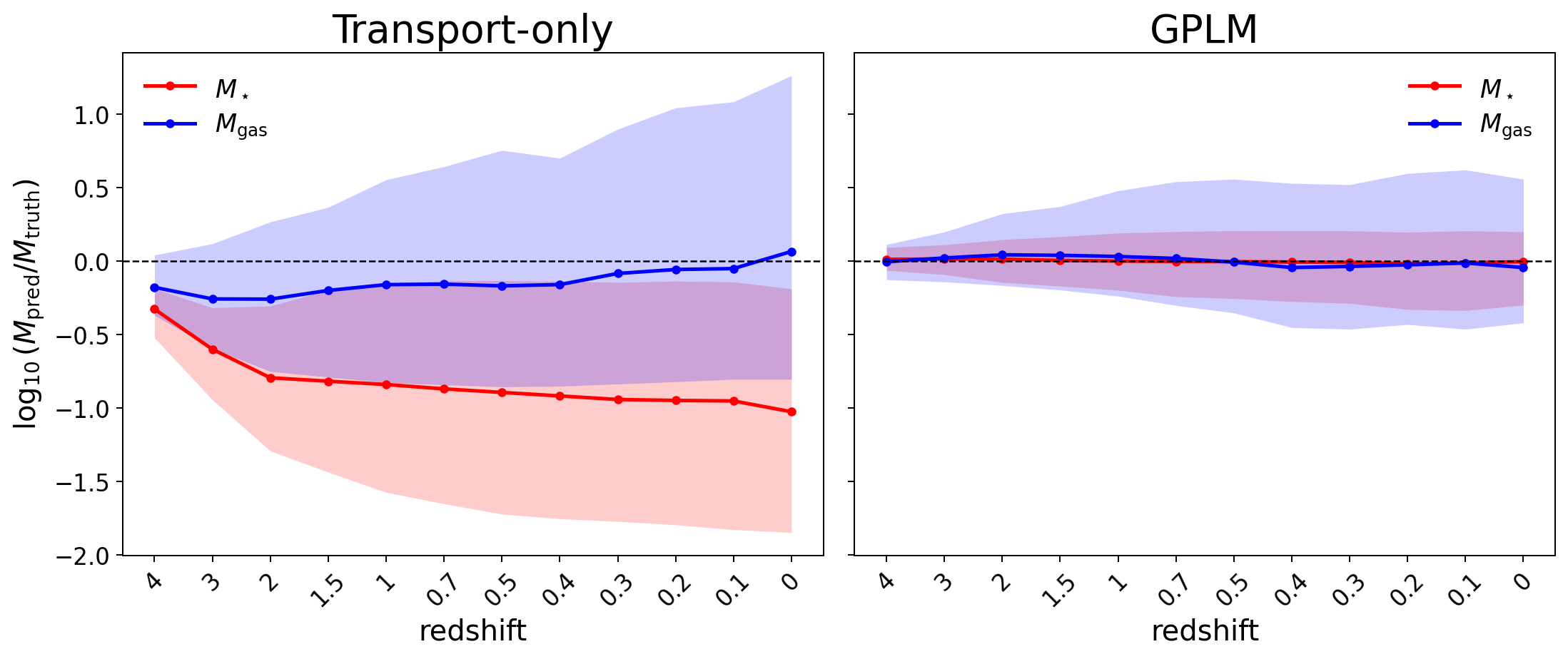}
  \caption{\label{fig:residuals_redshift_main}
  Residual evolution by redshift for the transport-only case (left) and GPLM (right), using only propagated nodes. For each retained snapshot we summarize the residual $\log_{10}(M_{\rm pred}/M_{\rm truth})$ with its median and 16-84 percentile band, shown for $M_{\star}$ (red) and $M_{\rm gas}$ (blue). The transport-only stellar residual becomes increasingly negative toward low redshift because merger transport propagates inherited stellar mass but omits in-situ star formation, so the missing contribution accumulates over time. The gas residuals remain broader and less coherent, reflecting the competing effects of accretion, consumption, feedback, and stripping. GPLM regulates both the bias and the scatter through learned graph-conditioned drift corrections on top of the transport backbone.}
\end{figure*}

\begin{figure*}[htbp]
  \centering
  \includegraphics[width=0.78\textwidth]{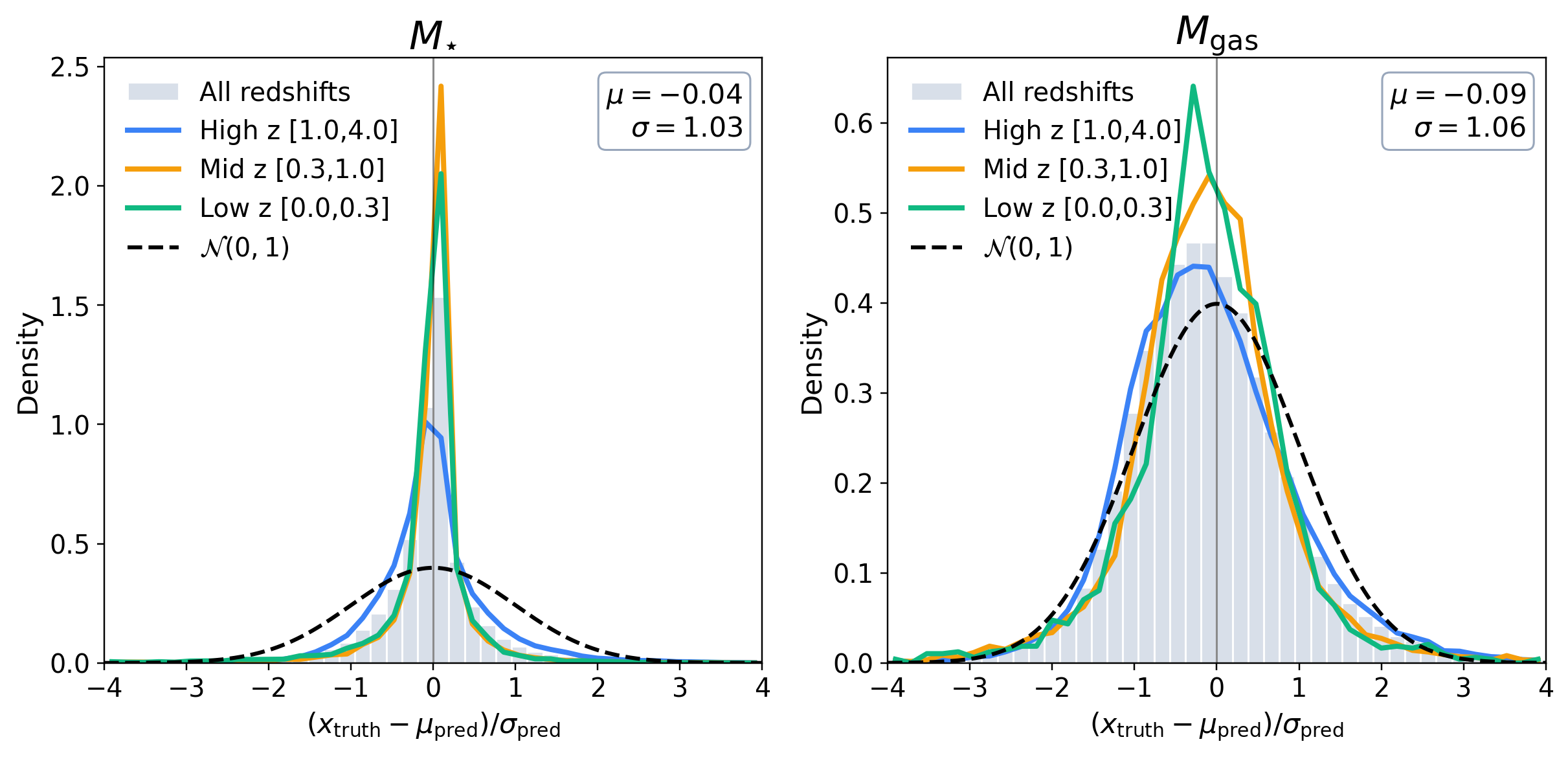}
  \caption{\label{fig:diffusion_scatter_main}
  One-step diffusion calibration for the GPLM residual sector, shown separately for $M_{\star}$ (left) and $M_{\rm gas}$ (right). In each panel the histogram gives the distribution of the normalized residual $(x_{\rm truth}-\mu_{\rm pred})/\sigma_{\rm pred}$, while the colored curves show the same quantity split into the three redshift ranges used in the analysis. The dashed curve shows the unit normal reference distribution $\mathcal{N}(0,1)$. The present Gaussian conditional diffusion provides a broadly reasonable description of the residual distribution, although the $M_{\star}$ channel still deviates visibly from the unit normal reference.}
\end{figure*}

\begin{figure*}[htbp]
  \centering
  \includegraphics[width=5.35cm]{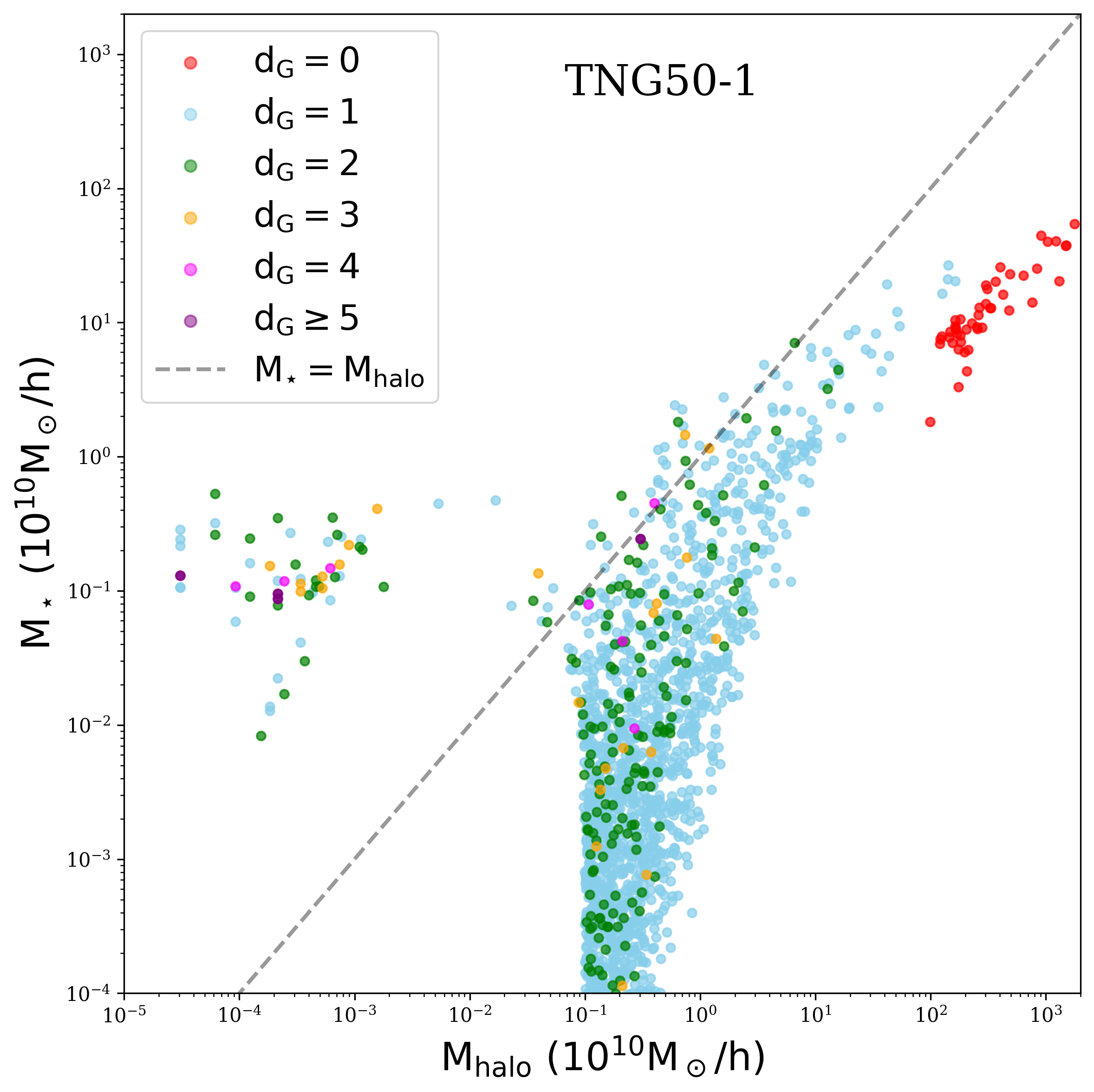}
  \includegraphics[width=5.35cm]{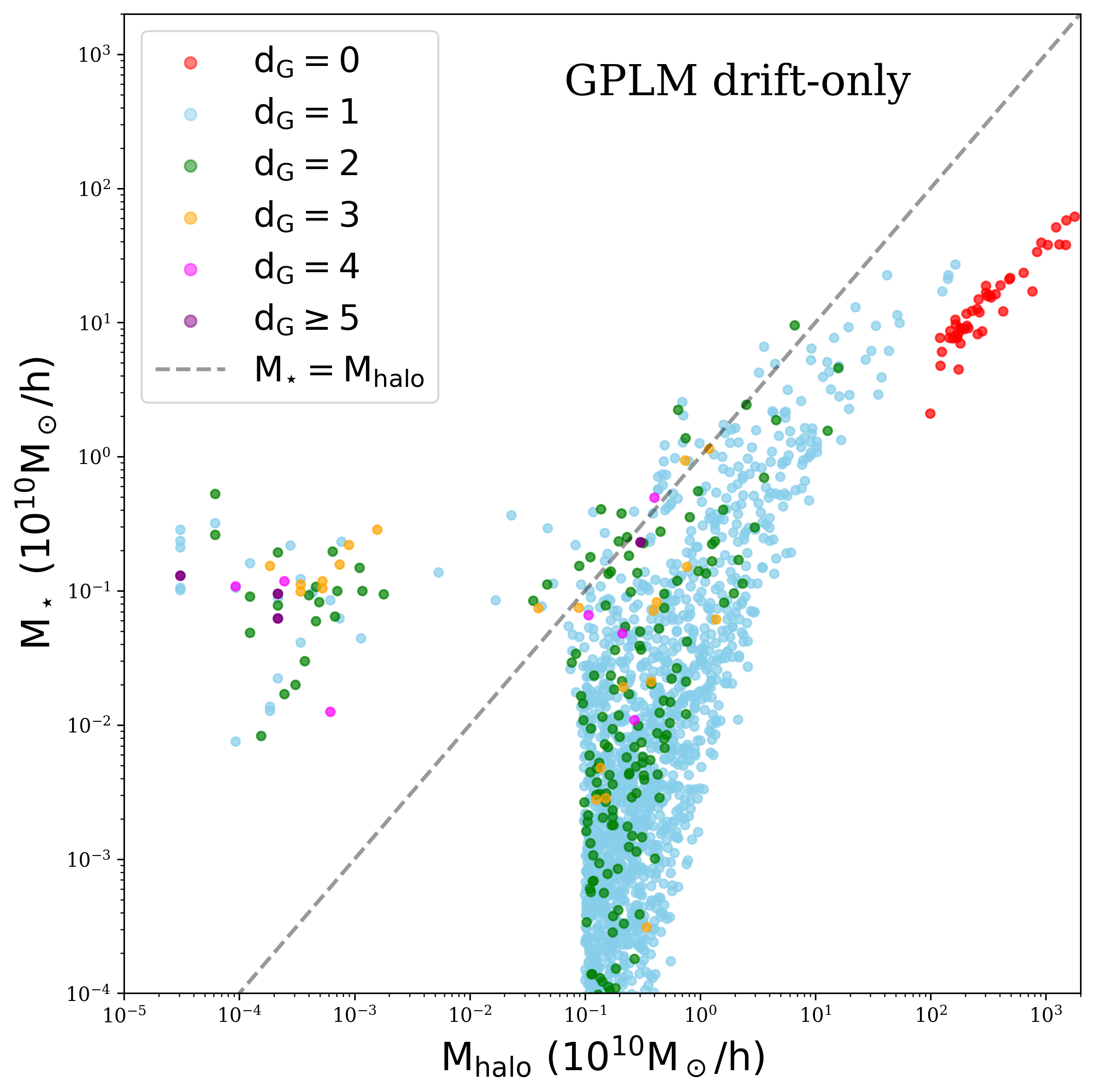}
  \includegraphics[width=5.35cm]{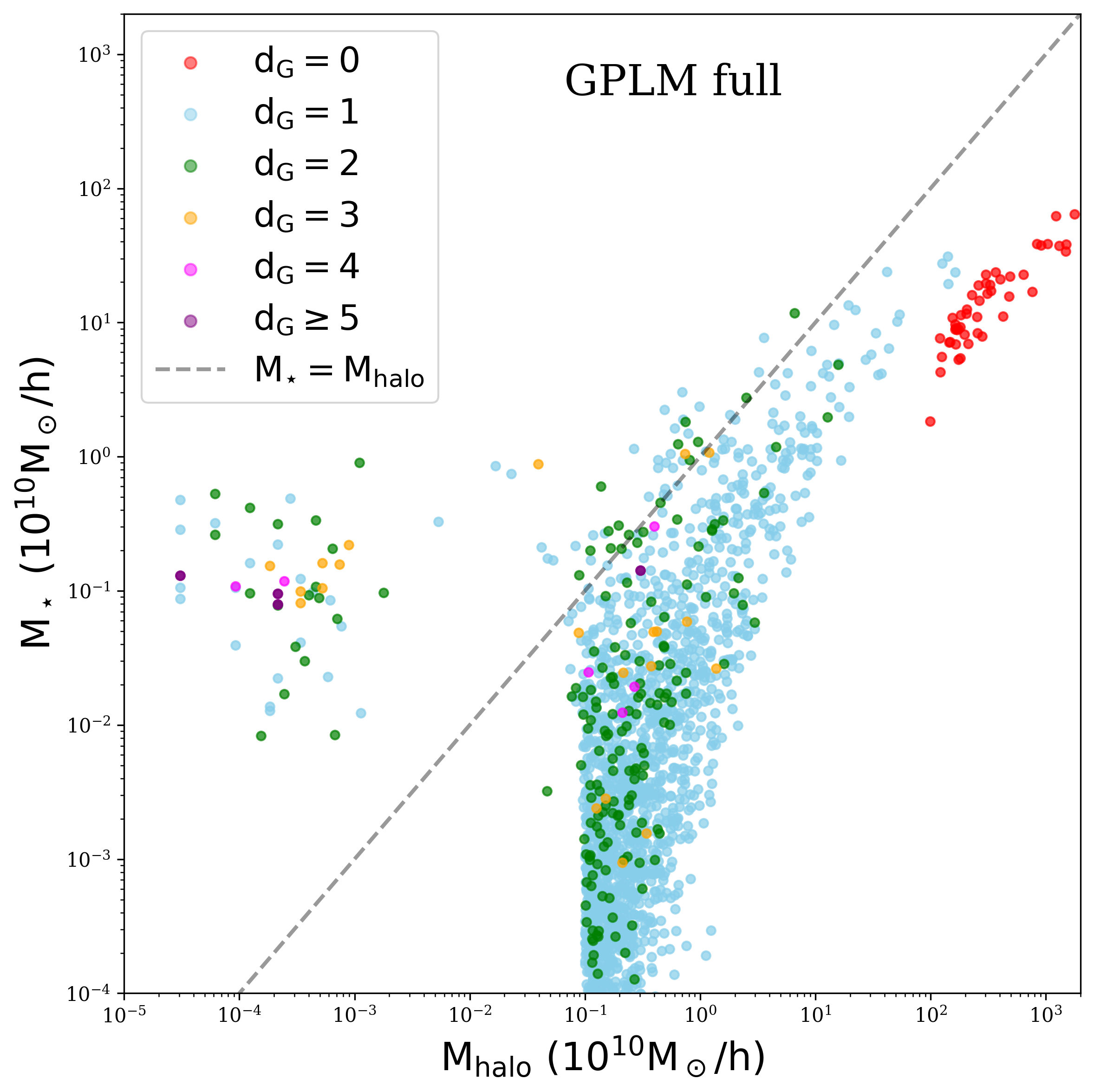} \\
  \includegraphics[width=5.35cm]{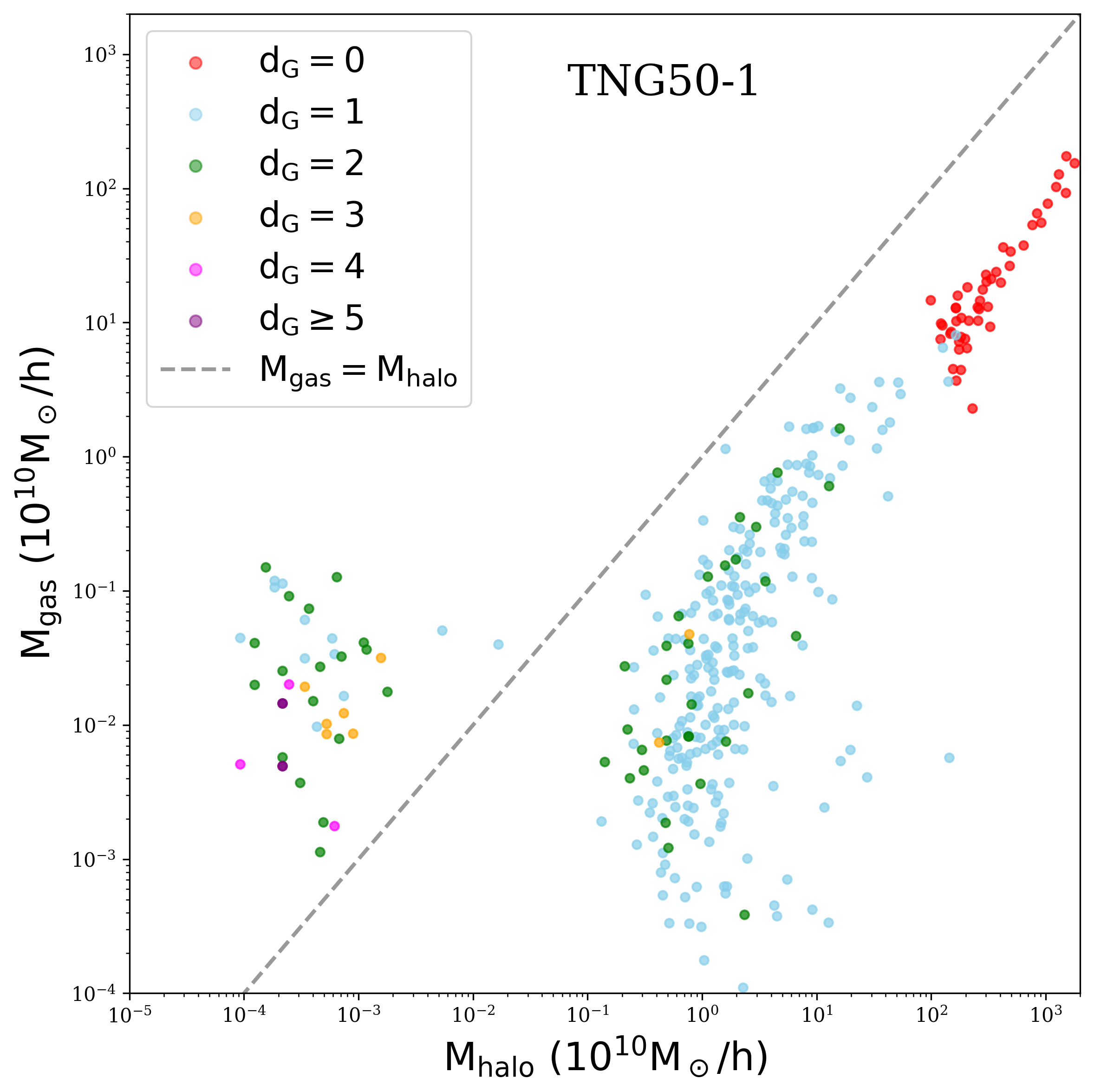}
  \includegraphics[width=5.35cm]{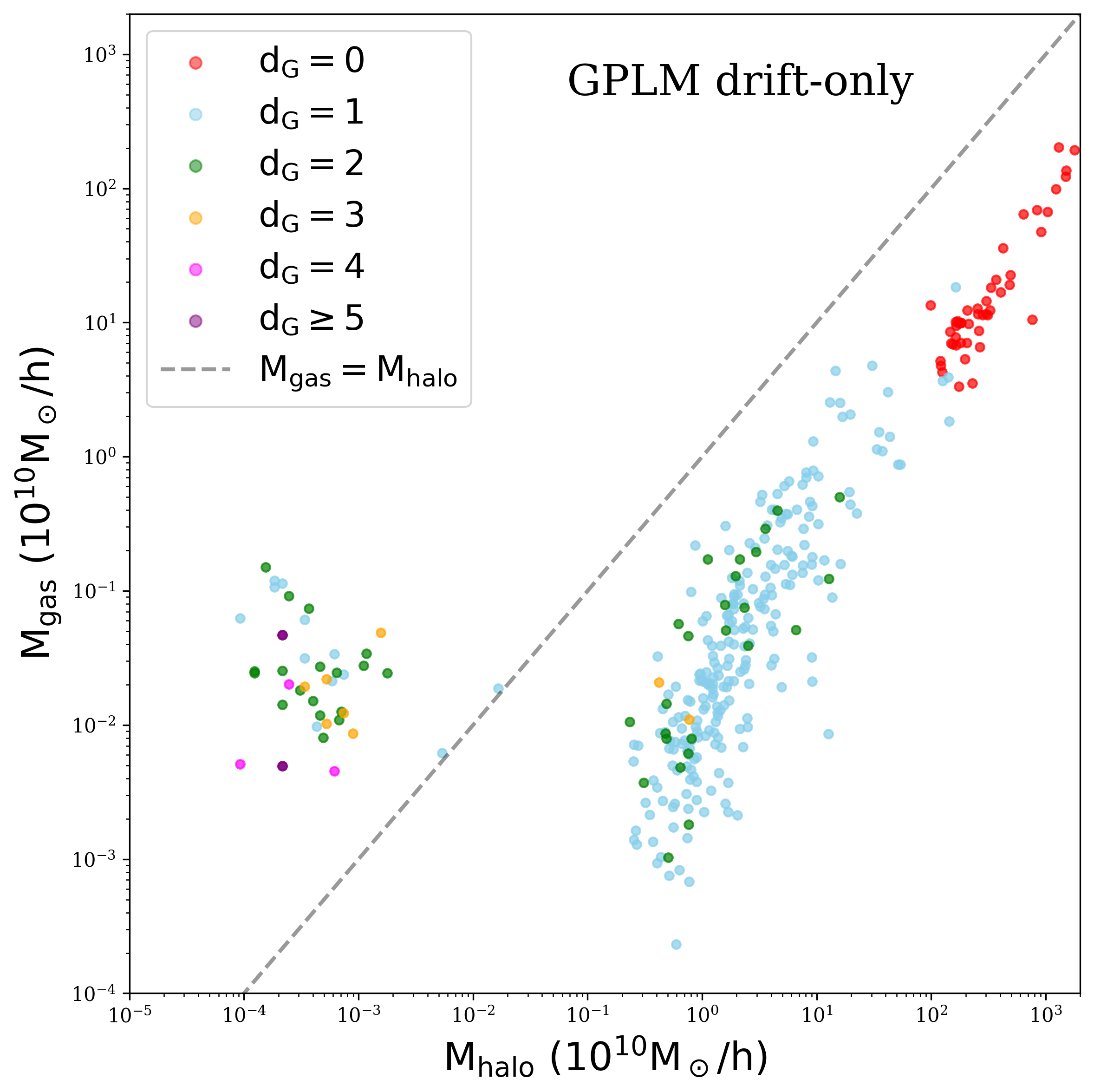}
  \includegraphics[width=5.35cm]{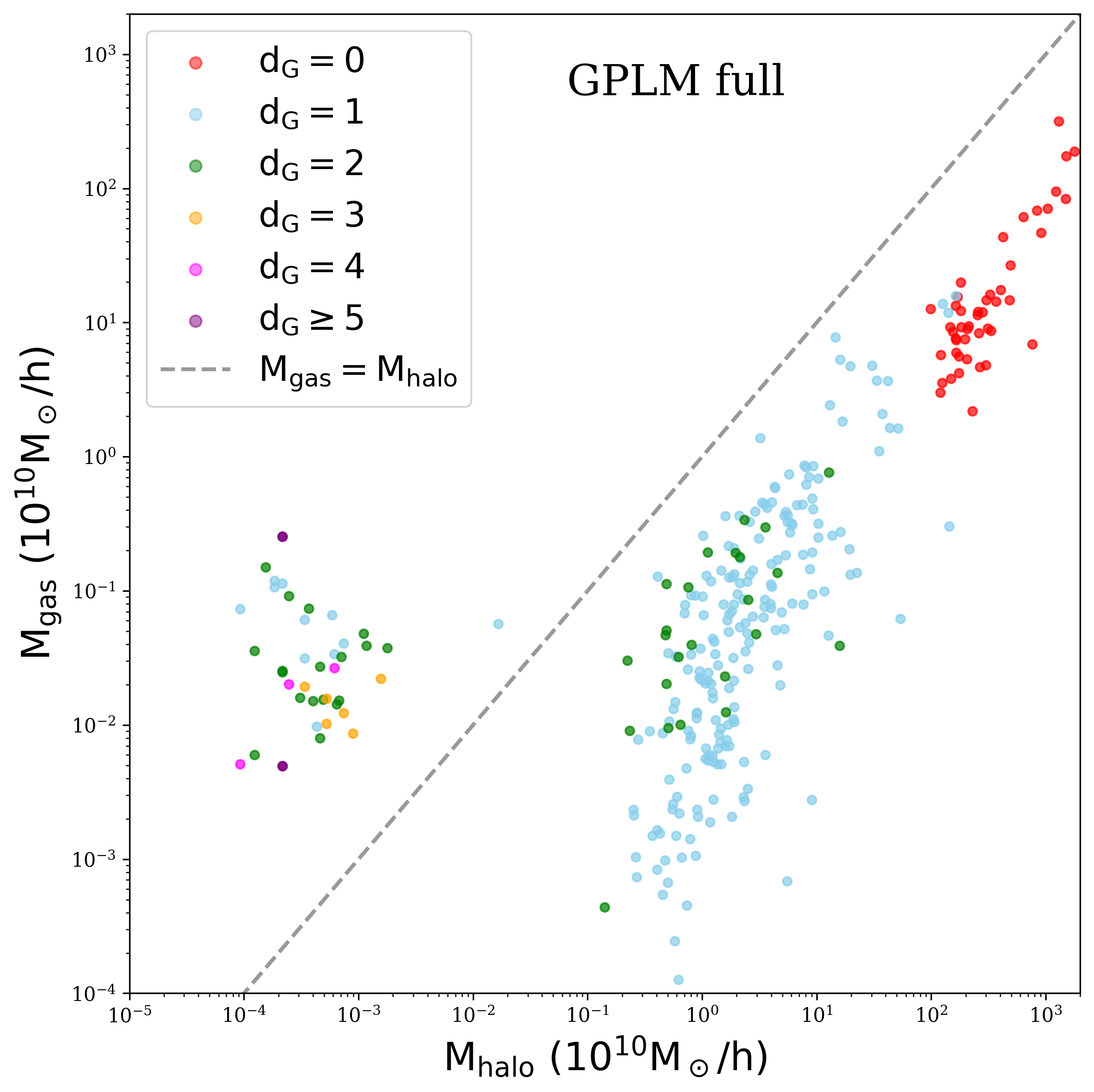}
  \caption{\label{fig:scatter_env}
Environment-colored $z=0$ scatter plots for the TNG-50-1 simulation (left), the deterministic GPLM rollout (middle), and a sampled GPLM rollout (right). The top row shows $M_{\star}$ versus $M_{\rm halo}$ and the bottom row shows $M_{\rm gas}$ versus $M_{\rm halo}$. All panels are colored by the host-edge graph distance $d_G$. The figure demonstrates that GPLM reproduces the environment-split population structure observed in the simulation. The sampled rollout generally preserves a comparable endpoint scatter pattern.}
\end{figure*}

\subsection{Inference and Environmental Diagnostics}

Inference reuses the layer-pair construction from training, with the same features and edges. Starting from the first retained layer, each subsequent layer is propagated autoregressively by transporting the previously predicted state along temporal edges and evaluating the learned residual drift on the current-layer graph in offset-log variables. The resulting predictions are then converted back to nonnegative stellar and gas masses and carried forward to define the next-step state. Repeating this procedure advances the model through the full graph history.

We focus on testing GPLM on its capability to model field trajectories within each graph. Hence, we seed newly appearing nodes, i.e., those with no temporal progenitors, directly from the simulation catalog at their first retained layer. 
In the parity and residual diagnostics discussed below, these entry-seeded nodes are then excluded so that the reported comparisons probe only nodes whose masses are propagated from earlier retained layers rather than copied from the truth at entry. This seeding is replaced by sampling when we incorporate $p_{\rm attach}$ into the likelihood. Details of the $p_{\rm attach}$ calibration are described in Appendix~\ref{sec:attachment_fits}. 

To show how well our trained model predicts stellar and gas masses, we present diagnostics for the learned drift and diffusion in Fig.~\ref{fig:residuals_redshift_main} and Fig.~\ref{fig:diffusion_scatter_main}, respectively. 

Figure~\ref{fig:residuals_redshift_main} tests the learned drift through rollout residuals as a function of redshift. In the transport-only case, the median stellar-mass residual becomes increasingly negative toward low redshift. This is expected because merger-driven transport propagates inherited stellar mass but omits in-situ star formation~\cite{Rodriguez-Gomez:2015ixu,2017MNRAS.470.4698A,Fitts:2018eww}, so the missing contribution accumulates over time and drives $M_{\star}$ below the truth. For $M_{\rm gas}$, the residuals are broader and more weakly organized. This does not indicate better modeling. Instead, gas reflects a balance among accretion, consumption, feedback, and stripping~\cite{2025A&A...695A.121W,Shen:2013wva}, so competing effects can partially cancel in the median even when the overall correlation remains weak. GPLM improves both channels by learning residual drifts on top of the deterministic transport backbone. In Fig.~\ref{fig:residuals_redshift_main}, the scatter of the residuals increases in both the stellar and gas channels for the transport-only reference, whereas it is regulated in GPLM. 

Figure~\ref{fig:diffusion_scatter_main} then tests the learned diffusion on transitions between adjacent layers, with each step evaluated using the true state of the previous layer. The normalized residuals are broadly consistent with the target unit normal distribution, and the redshift-split curves show that the learned diffusion captures nontrivial redshift dependence in the residual scatter. At the same time, the stellar-channel residual distribution deviates visibly from the unit normal reference, making clear that the current Gaussian diffusion is only an effective coarse-grained approximation.  This motivates future extensions of the stochastic description to include non-Gaussian terms beyond the present quadratic OM form. 

When the learned diffusion is included during rollout, the resulting sampled paths should be interpreted as stochastic realizations of the same learned effective dynamics on the same layered halo graphs, instead of point predictions competing with the deterministic drift backbone. 

To illustrate that the endpoint ensemble generated by the sampled rollout remains comparable to the simulation ensemble, we show in Fig.~\ref{fig:scatter_env} a comparison of simulation and sampled rollout in $M_{\star}$ versus $M_{\rm halo}$ (top), and $M_{\rm gas}$ versus $M_{\rm halo}$ (bottom). In each row, the left, middle, and right panels show the TNG50-1 simulation, the deterministic rollout, and a sampled rollout, respectively. The data points are colored by the host-edge graph distance $d_G$ defined in Sec.~\ref{sec:graphs}, as in Fig.~\ref{fig:graphs}. In both rows, the $d_G$ split organizes the population into visibly distinct clouds, and this environment-resolved structure is retained in both the deterministic and sampled GPLM realizations. The diagonal $1:1$ lines denote the thresholds above which galaxies become dark-matter-deficient in the stellar and gas channels. At the level of the $z=0$ scatter structure, the sampled rollout remains broadly comparable to both the deterministic GPLM realization and the TNG50-1 ensemble. This illustrates that the deterministic rollout should be interpreted as one realization drawn from the learned path measure.

For completeness, Appendix~\ref{sec:parity_appendix} collects stacked truth-versus-prediction parity plots across all snapshots for $\log_{10}(M_{\star}/(\rm M_{\odot}/h))$ and $\log_{10}(M_{\rm gas}/(\rm M_{\odot}/h))$, using only nodes propagated from retained layers, together with a rollout residual trajectory diagnostic that visualizes how sampled trajectories spread around the deterministic backbone.

\section{Graph-Conditioned Path Integrals and Graph-Ensemble Extension}
\label{sec:ensemble_path_integral}

For a fixed layered halo graph $G$ and fixed initial layer $x_0$, the trained GPLM defines a normalized discrete-time path measure over baryonic trajectories. 
Combining the dynamical sector with the attachment boundary factor, we can write the conditioned probability as
\begin{equation}
P_{\rm tot}(\mathbf{x}\mid G,x_0) =
\frac{ p_{\rm attach}(\mathbf{x}\mid G) e^{-S_{\rm OM}[\mathbf{x}; G,x_0]} 
}{Z_0[G]},
\label{eq:P_tot}
\end{equation}
with fixed-graph conditional partition function
\begin{equation}
Z_0[G] \equiv \int \mathcal D x
p_{\rm attach}(\mathbf{x}\mid G)
e^{-S_{\rm OM}[\mathbf{x}; G,x_0]} .
\end{equation}

Formally, we can introduce a total weight 
\begin{equation}
S_{\rm tot}=S_{\rm OM}+S_{\rm attach}, \qquad
S_{\rm attach}\equiv -\log p_{\rm attach},
\end{equation}
where $S_{\rm OM}$ governs within-graph dynamical increments, and $S_{\rm attach}$ is a boundary contribution associated with node entry or
infall. 
This total weight is a functional of the full graph-conditioned trajectory $\mathbf{x}$, enabling path integral calculations of operator averages.

For any trajectory functional $\mathcal{O}[\mathbf{x}]$, we can compute the graph-conditioned expectation as
\begin{equation}
\langle \mathcal{O} \rangle_{G} \equiv \int \mathcal{D}x \mathcal{O}[\mathbf{x}]
P_{\rm tot}(\mathbf{x}\mid G,x_0),
\label{eq:O_fixedG_norm}
\end{equation}
with $\mathcal{D}x$ the discrete path measure over node-layer variables. 
This measure also defines a fixed-graph evidence for partially observed
data
\begin{equation}
Z[G;\mathbf{y}] \equiv \int \mathcal{D}x
P(\mathbf{y}\mid \mathbf{x})
P_{\rm tot}(\mathbf{x}\mid G,x_0),
\label{eq:ZG_evidence}
\end{equation}
allowing data-conditioned evidence calculations once an observational map is specified.

Averaging graphs facilitate the gathering of evidence about operators across a population of observed galaxies. This calculation requires an explicit probability measure over merger histories. 
In principle, any calibrated merger history generator can be used in combination with halo catalogs to construct graphs for this purpose. 
However, these constructions are not graph-native and require additional pipelines to obtain layered halo graphs in this work. 
Here, we show that preferential attachment-detachment (PAD) can provide a natural approach to directly obtain layered graphs. 
The probabilistic nature of this approach enables simultaneous assignment of each graph a calculable probability weight and generation of explicit ensembles of merger histories anchored to a fixed $z=0$ host configuration.
A full calibration to simulation data is beyond the scope of this work. Hence, we provide only a minimal explicit example here. 

Our construction starts from graphs generated using the preferential-attachment kernels calibrated in Ref.~\cite{Yang:2022oqe}. 
Graphs generated in this way have been shown to reproduce graph metric distributions that are well matched to those reconstructed in simulations. 
Next, we generate a merger history by running a preferential detachment process that produces a \emph{hierarchical cut tree}. 
We parameterize the kernel weight as $w(u,v) \propto (k_u k_v)^\beta$, where $k_u$ and $k_v$ are the degrees of the nodes $u$ and $v$ connected by the edge $(u,v)$ and $\beta$ is calibrated to be $0.5$. 
At each step of the cut, we select an edge $e_s=(u,v)$ from component $C_s$ to cut based on the probability 
$w_{e_s}/(\sum_{e\in C_s} w_e)$, remove it, and interpret the resulting split into two components as a reverse-time branching event. 
Repeating this procedure recursively builds a binary fragmentation tree whose forward-time interpretation is a merger history.
We use an iterative all-components implementation:
\begin{enumerate}
\item Start with the full graph in a queue.
\item Remove a component from the queue. If it has only one node, skip it. Otherwise, calculate edge weights.
\item Randomly select and remove an edge based on the weight, then add the two resulting components back to the queue.
\item Repeat until only single nodes remain in the queue.
\end{enumerate}
The final split record defines the layered halo graph $G$ at the topological level. 
The product of all the split probabilities defines a total detachment probability 
$P_{\rm prune}(G) = \prod_s \frac{w_{e_s}}{\sum_{e\in C_s}w_e}$.
In this work, we use $P(G)\propto P_{\rm prune}(G)$ as a minimal graph prior. 

\begin{figure*}[t]
  \centering
  \includegraphics[width=0.445\textwidth]{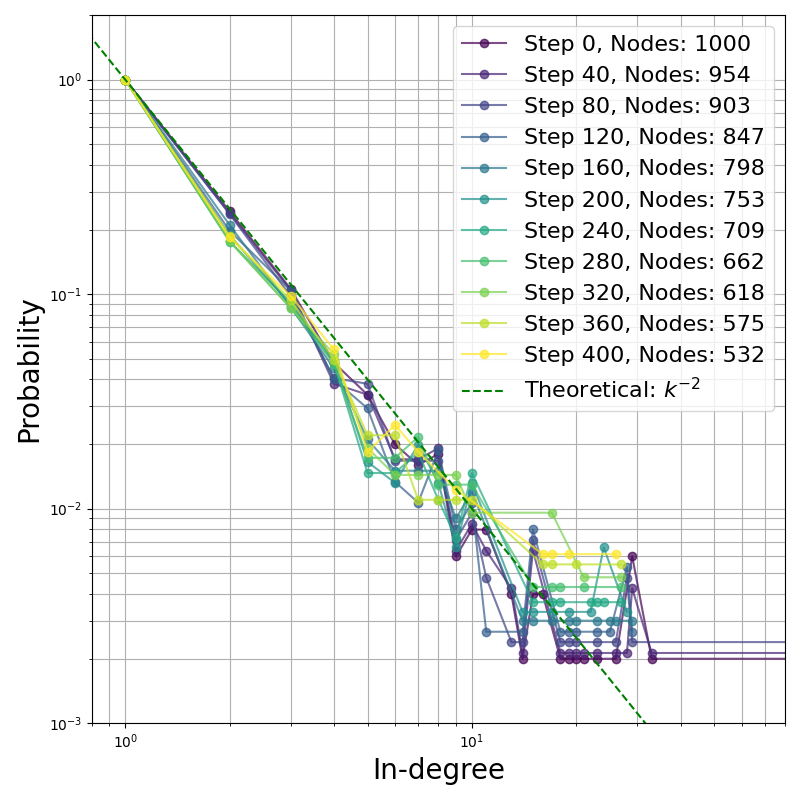}
  \hfill
  \includegraphics[width=0.44\textwidth]{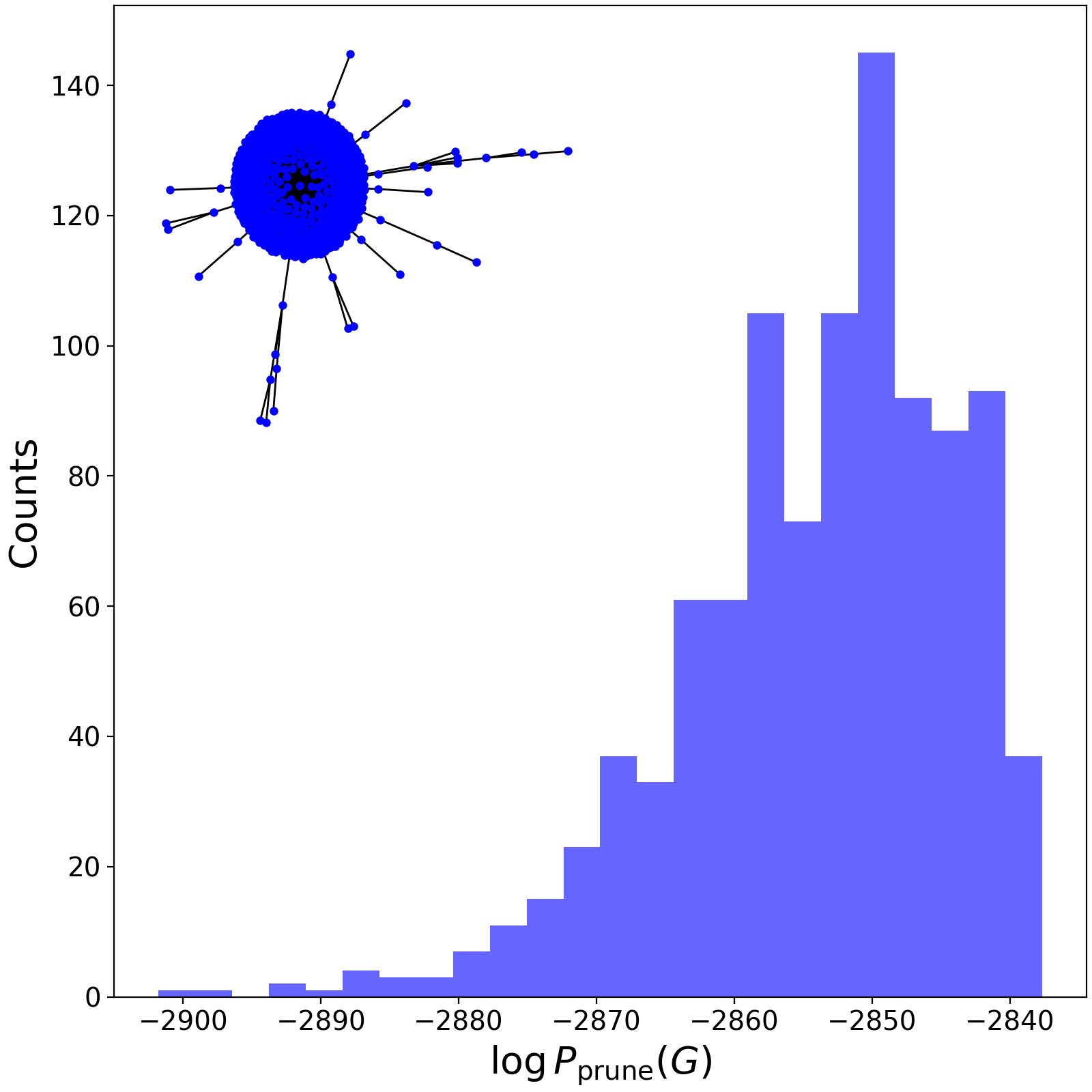}
  \caption{\label{fig:pada}
  Left: in-degree distributions of preferential attachment-detachment graphs. The solid curves illustrate the in-degree distributions at different pruning steps. They are obtained for $\beta=0.5$ by averaging over $20$ endpoint graphs with $N=1000$ nodes at $z=0$. The dashed green line shows the power-law $k^{-2}$ distribution. Right: histogram of $\log P_{\rm prune}(G)$ for one $z=0$ graph using $1000$ independent detachment histories.}
\end{figure*}

The left panel of Fig.~\ref{fig:pada} shows that the detachment process following the proposed kernel with $\beta=0.5$ preserves the $k^{-2}$ in-degree distribution across a wide range of the pruning steps. 
This result implies that the simple detachment kernel leads to host-edge graphs that roughly align with the graphs in Ref.~\cite{Yang:2022oqe}, which agree with simulations. 
The PAD algorithm is simple, enabling efficient generation of a family of merger histories that share the same $z=0$ host-edge graph structure while varying the backward assembly history.
The right panel of Fig.~\ref{fig:pada} illustrates the distribution of $\log P_{\rm prune}(G)$ for one example $z=0$ graph, as displayed in the inset, using $1000$ independent detachment histories. 
Such ensembles are the natural starting point for evaluating $x_0$-conditioned marginal likelihoods of the form introduced in
Sec.~\ref{sec:stochastic_model}. 

For PAD-generated graphs, the missing piece is an explicit prescription for attaching time steps and state variables to the graph nodes and edges.
The completion of these processes will also enable a fully generative model to create layered halo graphs and paint galaxies on top of it, which we leave to future work. 

Once $P(G)$ is specified in the same layout as the GPLM, one can combine the graph and trajectory parts into a joint graph-trajectory partition function
\begin{equation}
\mathcal Z \equiv \int \mathcal D G P(G) Z_0[G],
\end{equation}
based on which one can compute graph ensemble operator averages. 
In the present paper, all reported numerical results correspond to the \emph{fixed-graph} case. The graph sector is introduced here primarily to make the full graph-trajectory structure explicit and to indicate how graph-ensemble extensions can be defined once a calibrated prior $P(G)$ is specified alongside GPLM.

\section{Graph-Conditioned Dark-Matter-Deficient Galaxy Probabilities}
\label{sec:example_dmdg}

As illustrated in Fig.~\ref{fig:scatter_env}, the appearance of dark-matter-deficient galaxies (DMDG) clearly depends on environment, specifically the graph distance $d_G$. Hence, we apply the GPLM to evaluate the DMDG probabilities as a diagnostic of DMDGs' $d_G$ dependence.  
We compare graph-conditioned quenched expectations with and without the attachment factor $p_{\rm attach}$. 
Observations of a population of DMDGs have been reported and debated~\cite{2018Natur.555..629V,2020NatAs...4..246G,2019MNRAS.488.3298J,vanDokkum:2022zdd,2025A&A...695A.124B}.
Proposed formation channels include tidal stripping, dwarf galaxy collisions, potentially under the influence of SIDM~\cite{Ogiya:2018jww,Yang:2020iya,Zhang:2024qem,Wang:2025xre}.

We define the DMDG operator as
\begin{equation}
O_{\rm DMDG}^{(i)}(\mathbf{x}_i) =
\Theta\!\left(\frac{M_\star^{(i)}(z=0)+M_{\rm gas}^{(i)}(z=0)}{M_{\rm halo}^{(i)}(z=0)}-1\right),
\end{equation}
which selects systems of baryonic masses higher than the halo masses at $z=0$.
We evaluate it in two $d_G$ bins: $d_G=1$ for immediate satellites and $d_G>1$ for outer/higher-order satellites. 
In the current test set, no $d_G=0$ node satisfies the DMDG criterion at $z{=}0$.

For each fixed graph $G$, we estimate the graph-conditioned DMDG fraction through Monte Carlo trajectory sampling from the learned $S_{\rm OM}$. Each Monte Carlo sample paints stellar and gas masses on the full layered halo graph, ensuring correlated layer-by-layer evolution. The $O_{\rm DMDG}$ specifies only the $z=0$ DMDGs, and earlier layers enter only through the trajectory weight.
To split the nodes by $d_G$ bin, we introduce the $z=0$ node sets $E_1\equiv\{i:d_G(i)=1\}$ and $E_{>1}\equiv\{i:d_G(i)>1\}$.
The graph-conditioned DMDG fraction is then obtained as 
\begin{equation}
\begin{aligned}
\langle O_{\rm DMDG}\rangle_{G,E} &\equiv
\int \mathcal{D}x 
\left[ \frac{\sum_{i\in E} O_{\rm DMDG}^{(i)}[\mathbf{x}]}{N_E}
\right] \\
&\qquad\times P_{\rm tot}(\mathbf{x}\mid G,x_0).
\end{aligned}
\label{eq:ODMDG_bin}
\end{equation}
In practice, trajectories are sampled from $P_{\rm tot}(\mathbf{x}\mid G,x_0)$ both with and without the $p_{\rm attach}$ factor (Eq.~\eqref{eq:P_tot}). In the case without $p_{\rm attach}$, the newly appearing nodes are inserted with their values from the simulation catalogs at the layer where they first enter the graph. With $p_{\rm attach}$, those entry values are instead sampled from the attachment model.

Specifically, we fit lognormal conditional distributions for $M_\star$ and $M_{\rm gas}$ as functions of $(M_{\rm halo},z)$ using the subset of newly appearing nodes in the TNG graphs, and then sample from those fitted distributions at each layer. This isolates outside-the-graph uncertainty without altering learned transport dynamics on already connected nodes. Details of the fits are given in Appendix~\ref{sec:attachment_fits}.

Applying the evaluator to the 50 test layered graphs yields Monte Carlo estimates of $\langle O_{\rm DMDG}\rangle_{G,E}$ per graph and environment bin. In all cases we apply the same $z=0$ halo mass cut $M_{\rm halo}\ge 10^{5}~\rm M_{\odot}$. The results reveal a higher median DMDG fraction in the $d_G>1$ bin and broad scatter across graphs. Without $p_{\rm attach}$, the graph medians are
\begin{eqnarray}
&&\mathrm{median}~\langle O_{\rm DMDG}\rangle_{G,d_G=1}\simeq 0.008,
\quad\text{and } \\ \nonumber
&&\mathrm{median}~\langle O_{\rm DMDG}\rangle_{G,d_G>1}\simeq 0.071.
\end{eqnarray}
Including $p_{\rm attach}$ shifts these to
\begin{eqnarray}
&&\mathrm{median}~\langle O_{\rm DMDG}\rangle_{G,d_G=1}\simeq 0.006,
\quad\text{and } \\ \nonumber
&&\mathrm{median}~\langle O_{\rm DMDG}\rangle_{G,d_G>1}\simeq 0.040,
\end{eqnarray}
respectively. The attachment model therefore lowers the median DMDG fraction in both satellite bins, while preserving the excess incidence and broader graph-to-graph variation in the $d_G>1$ population.

\begin{figure*}[t]
  \centering
  \includegraphics[width=0.48\textwidth]{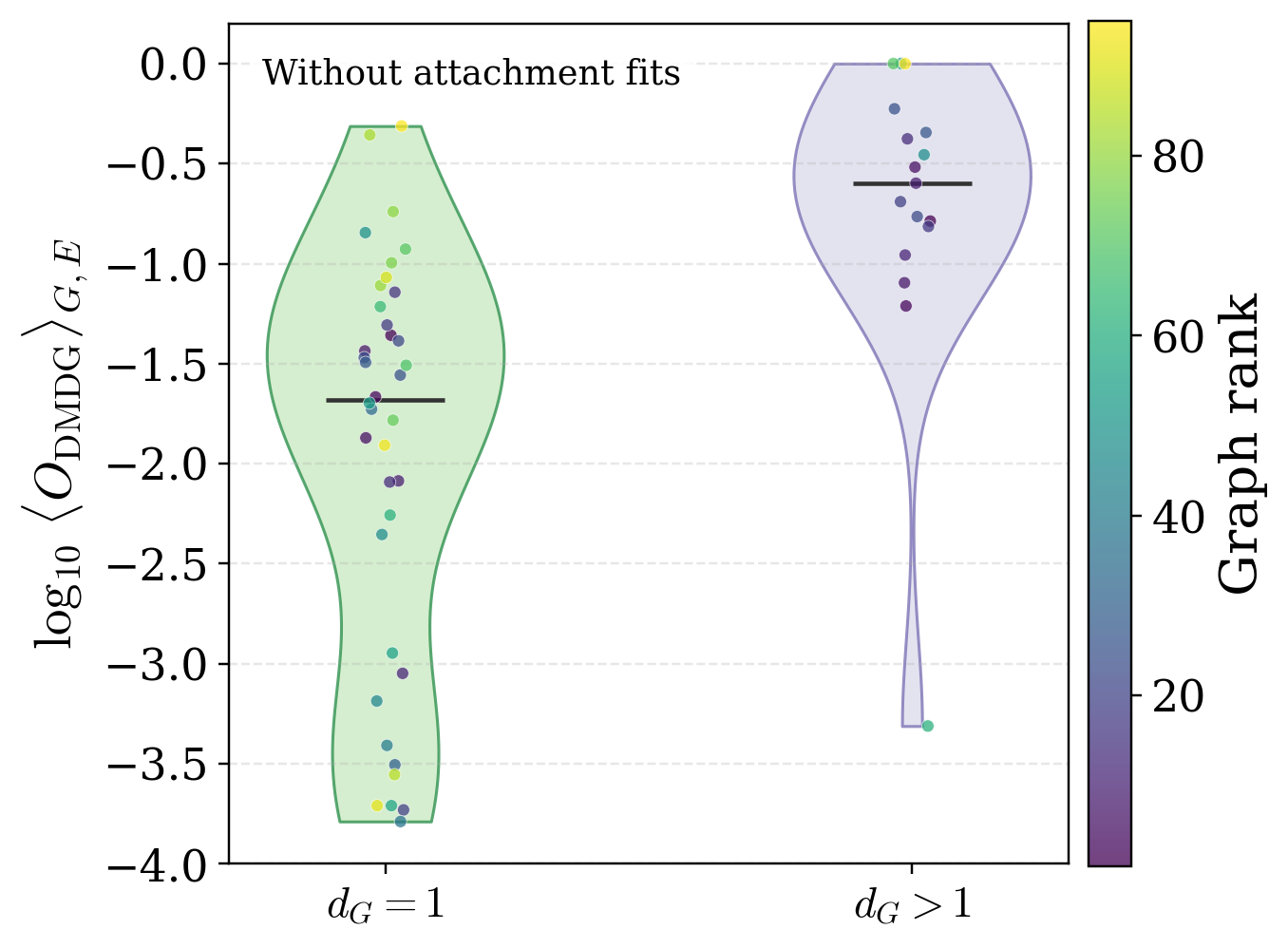}
  \hfill
  \includegraphics[width=0.48\textwidth]{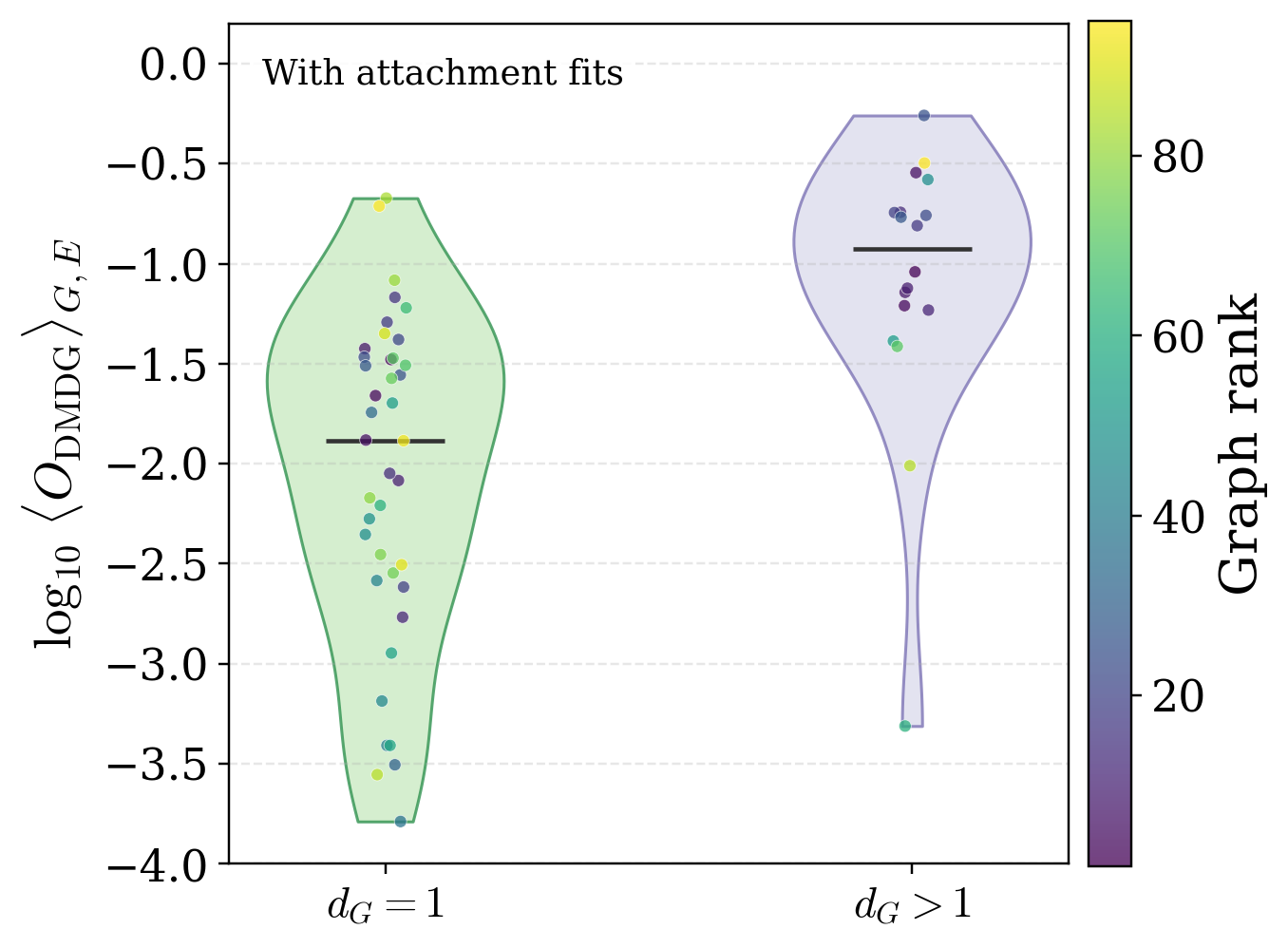}
  \caption{\label{fig:eval_dmdg_violin}
Graph-conditioned $\langle O_{\rm DMDG}\rangle_{G,E}$ shown in logarithmic space. Left: baseline path measure with newly appearing nodes seeded from the simulation catalogs. Right: the same calculation including the attachment likelihood for newly added nodes. Colored points indicate individual graphs and are coded by Graph rank. In both panels the $d_G>1$ violin is shifted upward relative to $d_G=1$, indicating a higher DMDG incidence together with larger graph-to-graph diversity in the outer or higher-order satellite population. }
\end{figure*}

Figure~\ref{fig:eval_dmdg_violin} compares graph-conditioned fractions with and without the attachment likelihood. The median fractions are consistently higher in the $d_G>1$ bins, with broader scatters (in linear scale) than the $d_G=1$ bin. This outcome aligns with our physical intuition, where $d_G>1$ halos receive a more complicated tidal environment, in particular, stronger tidal stripping. Since the distribution of dark matter is more extended relative to the galaxy, stronger tidal stripping causes larger dark matter mass loss, resulting in more DMDGs. The larger scatter quantifies the complexity of the trajectories, which could be explored through trajectory-to-trajectory comparisons. We consider 50 graph contributions in the $d_G=1$ bins and 30 in the $d_G>1$ ones. In the plotted violins, the visible point counts are 38 and 17 without $p_{\rm attach}$, and 40 and 18 with $p_{\rm attach}$, because the omitted entries have no DMDGs in the corresponding bins. Our test graphs span more than two orders of magnitude in halo mass. To indicate this variation, we color-code the graph rank in the displayed points, using darker colors for larger, lower-rank graphs.

Intriguingly, including $p_{\rm attach}$ reduces the broadest tails and the overall graph-to-graph dispersion in both environment bins. This indicates that much of the broad dispersion in the baseline result is already carried by the diversity of the test graphs themselves, while the present attachment model acts as a regularized boundary prescription. In its current form, $p_{\rm attach}$ is calibrated with a factorized lognormal model that depends only on the infall halo mass $M_{\rm halo}$ and redshift $z$. It therefore captures the bulk trend of node entry, but not the full graph-specific dependence of boundary fluctuations.
This observation motivates further stochastic dynamical development of the $p_{\rm attach}$ sector to more faithfully encode the physics of node attachment. For the present paper, however, the calibration in Appendix~\ref{sec:attachment_fits} provides a reasonable formal closure and is sufficient for the graph-conditioned applications reported below.

\section{Gas-Rich Response as a Controlled Deformation}
\label{sec:lambda_response}

The gas to stellar mass ratio $M_{\rm gas}/M_\star$ offers another sensitive probe of environmental effect because gas receives ram-pressure stripping aside from tidal stripping. In the massive host halos considered here, the host-side gas content is significant, and the host-edge environment channel therefore has a natural opportunity to regulate satellite gas depletion. To isolate that effect, we introduce the gas-rich operator
\begin{equation}
O_{\rm GR}(i) \equiv \Theta\left[M_{{\rm gas},i}(z{=}0)-M_{\star,i}(z{=}0)\right],
\end{equation}
and evaluate its response under a controlled gas-channel deformation. All calculations are performed on fixed layered halo graphs constructed from the simulation, and we include the effect of the $p_{\rm attach}$ factor for newly entering nodes.

We introduce the deformation using two forward predictions derived from the same trained checkpoint: a full prediction that preserves all temporal and host‑edge conditioning, and an environment‑off reference generated by ablating host edges and zeroing the static features that encode host-side environment information, including host in- and out-degrees, host mass, satellite status, host distance, and host-relative speed.
Let $b_{{\rm gas},{\rm full}}(x_k,G)$ denote the gas-channel drift rate of the full model and $b_{{\rm gas},{\rm off}}(x_k,G)$ the corresponding gas-channel drift of the environment-off reference. The gas-channel deformation is defined through
\begin{equation}
\begin{aligned}
b_{{\rm gas},\lambda}(x_k,G) &= b_{{\rm gas},{\rm off}}(x_k,G) \\
&\quad +\lambda\left[b_{{\rm gas},{\rm full}}(x_k,G)-b_{{\rm gas},{\rm off}}(x_k,G)\right].
\end{aligned}
\label{eq:lambda_drift}
\end{equation}
Thus $\lambda=0$ removes the learned environmental gas contribution, $\lambda=1$ recovers the fiducial learned dynamics, and $\lambda>1$ amplifies the environmental gas response. 

For a fixed graph and endpoint environment bin $E$, we estimate the endpoint gas-rich fraction as 
\begin{equation}
\langle O_{\rm GR}\rangle_{G,E,\lambda} = \left\langle
\frac{1}{N_{G,E}}\sum_{i\in E} O_{\rm GR}^{(i)}
\right\rangle_{P_\lambda},
\end{equation}
and the corresponding time-averaged occupancy as
\begin{eqnarray}
&&\langle \tau_{\rm GR}\rangle_{G,E,\lambda} = \\ \nonumber
&& \left\langle \frac{1}{T}\sum_{k=1}^{T}\frac{1}{N_{G,E,k}}\sum_{i\in E_k}
\Theta\left[M_{{\rm gas},i}^{(k)}-M_{\star,i}^{(k)}\right] \right\rangle_{P_\lambda}. 
\end{eqnarray}
Here $P_\lambda$ denotes the path measure generated by the deformed gas drift of Eq.~\eqref{eq:lambda_drift} with the same learned diffusion model. The normalization factors $N_{G,E}$ and $N_{G,E,k}$ denote the numbers of nodes in the chosen environment bin at the endpoint and at retained layer $k$, respectively. The application evaluates these observables graph by graph on the 50 test graphs and then plots the graph-level distributions in the three environment bins $d_G=0$, $d_G=1$, and $d_G>1$.

\begin{figure*}[t]
  \centering
  \includegraphics[width=0.48\textwidth]{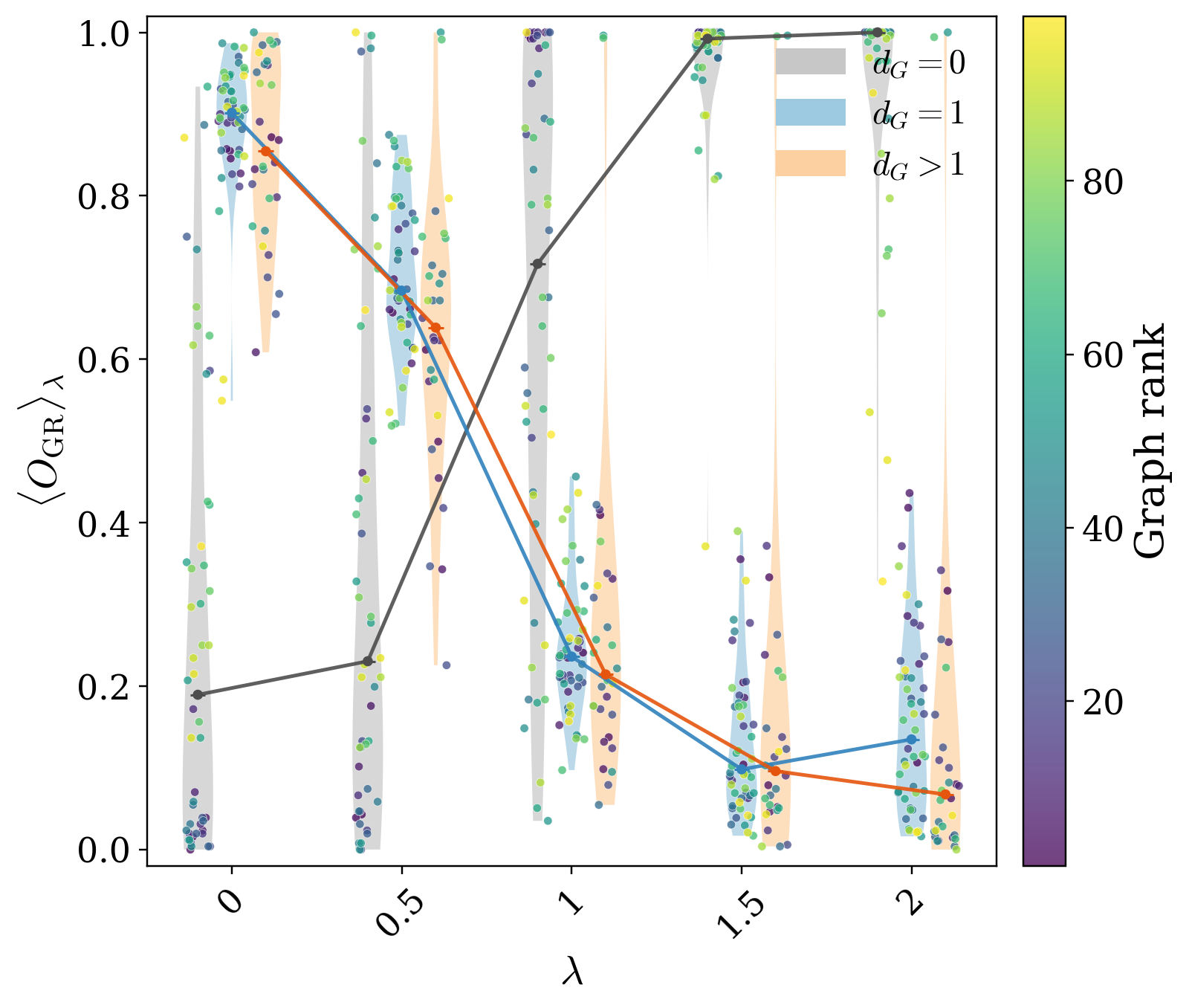}
  \hfill
  \includegraphics[width=0.48\textwidth]{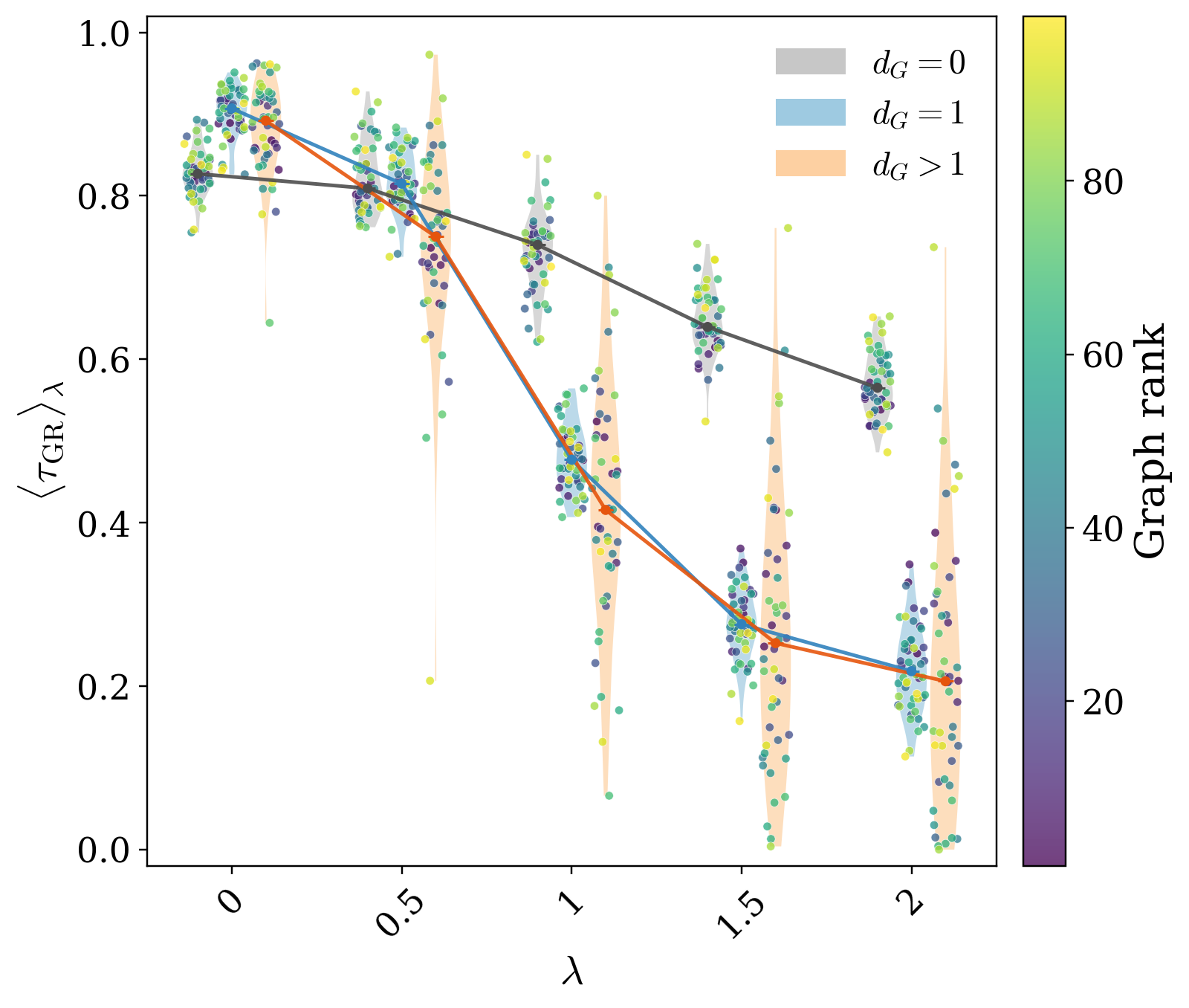}
  \caption{\label{fig:pd_violin_lambda}
Operator response for the endpoint gas-rich fraction $\langle O_{\rm GR}\rangle_{G,E,\lambda}$ (left) and the time-averaged gas-rich occupancy $\langle \tau_{\rm GR}\rangle_{G,E,\lambda}$ (right). 
The deformation controls the gas-channel environmental effects. The reference $\lambda=0$ removes such effects, $\lambda=1$ restores the fiducial learned dynamics, and larger $\lambda$ amplifies the same host-conditioned gas response. Results are shown for the three environment bins $d_G=0$, $d_G=1$, and $d_G>1$. These runs include $p_{\rm attach}$ by default.
}
\end{figure*}

Figure~\ref{fig:pd_violin_lambda} shows how $\langle O_{\rm GR}\rangle$ (left) and $\langle \tau_{\rm GR}\rangle_{G,E,\lambda}$ (right) vary under $\lambda$, considering $\lambda\in[0,2]$ with an increment of $0.5$ and in the $d_G=0$, $d_G=1$ and $d_G>1$ bins. In the host bin $d_G=0$, $\langle O_{\rm GR}\rangle_{G,d_G=0}$ increases monotonically with $\lambda$, showing a moderate gas richness at $\lambda=1$ with a huge scatter, and quickly saturates at higher $\lambda$. A natural interpretation of this outcome is ram-pressure stripping, which causes the gas content of satellites to feed into the host-side reservoir, so the central gas-rich fraction is enhanced by the host-edge channel. Accordingly, gas removal from satellites leads to decreasing $\langle O_{\rm GR}\rangle$ in increasing $\lambda$, as illustrated for the $d_G=1$ and $d_G>1$ bins. The $d_G>1$ violins remain broader than the $d_G=1$ violins, indicating a more diverse environmental response history in the higher-order satellites.

In the right panel, the $\langle \tau_{\rm GR}\rangle_{G,E,\lambda}$ decline in all three bins, particularly for the satellites. This means that stronger environment coupling reduces the fraction of retained history spent with $M_{\rm gas}>M_\star$, for example, by causing satellites to leave the gas-rich state earlier, enter it later, or return to it less often as $\lambda$ increases. Unlike the endpoint fraction $\langle O_{\rm GR}\rangle$, the occupancy $\langle \tau_{\rm GR}\rangle$ receives contributions from a broader population, including galaxies that are gas rich during part of their evolution history even if they are not gas rich at $z=0$.

The obtained results agree with our qualitative understanding of tidal stripping and ram-pressure stripping on gas depletion for satellites. They imply that environmental information carried by the host edges is successfully encoded in the learned residual dynamics of the GPLM.

\section{Path-Space Likelihood Diagnostics}
\label{sec:noneq}

The DMDG and gas-rich examples examine endpoint or time-averaged observables, but do not reveal how the host-edge gas channel alters the full set of satellite histories. GPLM allows us to ask this directly because it assigns an explicit likelihood to each trajectory on a fixed layered graph. Here, we analyze two diagnostics under the same gas-channel deformation: the first quantifies the deviation of the deformed trajectory ensemble from the environment-off reference, and the second assesses changes in local forward-reverse asymmetry. As in the gas-rich response example, all trajectories are sampled directly from the deformed forward dynamics.

The first diagnostic is the (per-trajectory) path log-ratio
\begin{equation}
W_\lambda(\mathbf{x}\mid G) \equiv \log\frac{P_\lambda(\mathbf{x}\mid G)}{P_0(\mathbf{x}\mid G)},
\end{equation}
where $P_0$ is the environment-off reference obtained by removing the learned host-edge gas contribution while keeping the same graph, transport backbone, and diffusion model. Intuitively, this is a path-level KL-like distance from the $\lambda=0$ reference. The code accumulates this quantity over propagated dynamical terms and reports the intensive graph-level diagnostic
\begin{equation}
\bar w_{\rm dyn}^{(G)}(\lambda) = \frac{\sum_{m=1}^{M_G} W_\lambda^{(m,G)}}{\sum_{m=1}^{M_G} N_{{\rm dyn}}^{(m,G)}},
\end{equation}
where $N_{{\rm dyn}}^{(m,G)}$ counts the propagated one-step dynamical likelihood terms that contribute to sampled path $m$ on graph $G$.
By construction, $\bar w_{\rm dyn}$ vanishes at $\lambda=0$. As $\lambda$ increases, it measures how far the sampled trajectory ensemble moves away from the environment-off reference.

The second diagnostic is a local asymmetry proxy built from the learned forward increment law and a reverse proxy using the sign-flipped increment:
\begin{equation}
\begin{split}
\Delta s_{{\rm dyn},i,k}^{\rm proxy} &= \log \frac{p_{i,k}^{\rm fwd}}{p_{i,k}^{\rm rev,proxy}}, \\
p_{i,k}^{\rm rev,proxy} &\equiv P_{\rm OM,\lambda}(-\Delta x_{i,k}\mid x_{i,k}^{+}),
\end{split}
\end{equation}
where the reverse proxy is evaluated by applying the same deformed forward kernel at the post-increment state and flipping the realized increment. This is a local measure of forward-reverse asymmetry, or irreversibility, for the learned increment law. At graph level we report the average density
\begin{equation}
a_{{\rm dyn},E}^{{\rm proxy},(G)}(\lambda) =
\left\langle
\frac{1}{N_{{\rm dyn},E}}\sum_{(i,k)\in E}\Delta s_{{\rm dyn},i,k}^{\rm proxy}
\right\rangle_{P_\lambda},
\end{equation}
for the environment bins $E\in\{d_G=0,d_G=1,d_G>1\}$.
It should be understood as a local reverse-step asymmetry proxy evaluated on the forward rollout, not as a full-time reversed path action difference.

\begin{figure*}[t]
\centering
\includegraphics[height=0.45\textwidth]{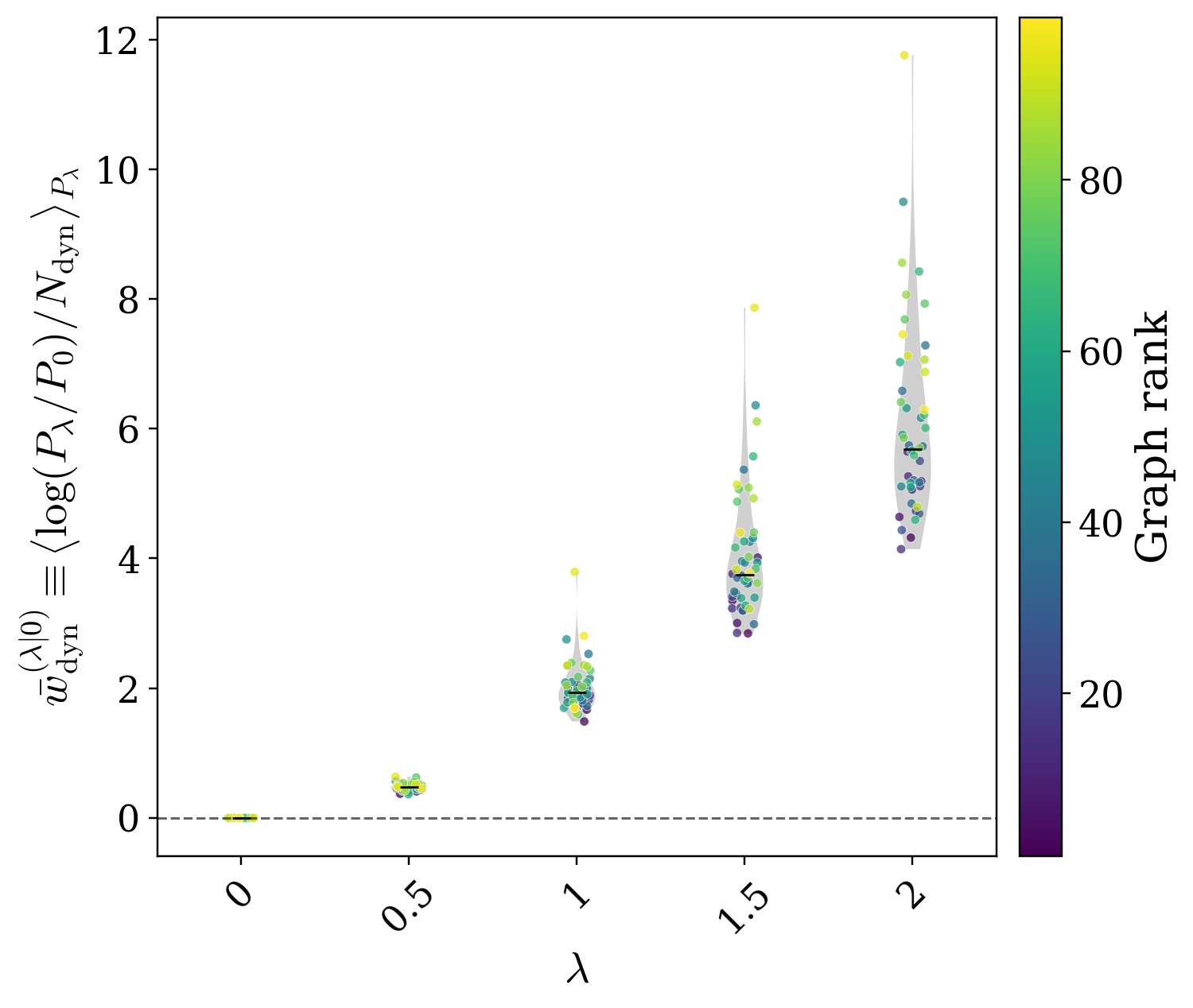}
\hfill
\includegraphics[height=0.45\textwidth]{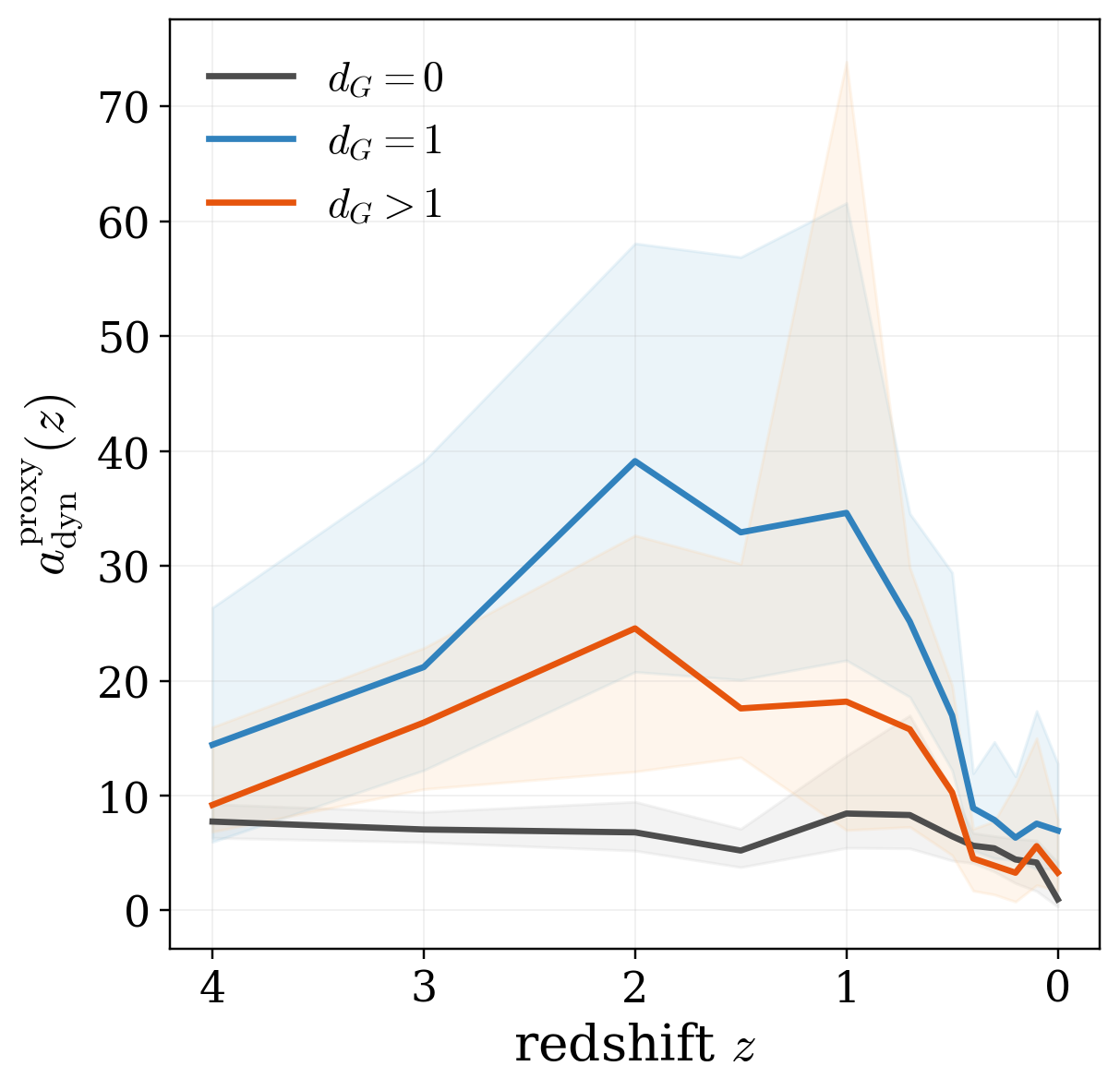}
\caption{\label{fig:noneq_counterfactual_proxy}
Left: graph-level path diagnostic $\bar w_{\rm dyn}^{(G)}(\lambda)$ defined relative to the environment-off reference $P_0$. It behaves like a path-level KL-like distance from the $\lambda=0$ ensemble. Right: redshift dependence of the fiducial local asymmetry proxy $a_{{\rm dyn},E}^{\rm proxy}(z)$ at $\lambda=1$ (gas channel), shown for the three environment bins $d_G=0$, $d_G=1$, and $d_G>1$. The solid lines show the median, while the shaded bands indicate the $16^{\rm th}-84^{\rm th}$ percentile range across graphs at fixed redshift. Larger values indicate stronger local forward-reverse asymmetry.}
\end{figure*}

In the left panel of Fig.~\ref{fig:noneq_counterfactual_proxy}, the distribution of $\bar w_{\rm dyn}^{(G)}$ shifts upward as $\lambda$ increases, and its scatter broadens at the same time. This is the expected behavior for a controlled deformation: as the host-edge gas response is strengthened, the deformed trajectory ensemble moves progressively farther from the $\lambda=0$ reference. The growing scatter indicates that the magnitude of this departure depends significantly on the graph.

The right panel probes a different aspect of the same framework. Instead of varying $\lambda$, it shows the redshift dependence of the fiducial local forward-reverse asymmetry proxy at $\lambda=1$. The strongest signal is observed in satellite populations, with a broad maximum around cosmic noon. A plausible interpretation is that this epoch corresponds to an overlap of rapid halo growth and intense baryonic activity. Around $z\sim2$, halos assemble quickly while galaxies remain gas-rich and experience strong star formation, feedback, and environmental effects. This makes graph-conditioned updates more history dependent and temporally biased, increasing the local forward-reverse asymmetry. The stronger signal in satellites also points to an important role for environmental effects. 

These two diagnostics therefore answer complementary questions. The statistic $\bar w_{\rm dyn}$ tells us how strongly the full trajectory ensemble departs from the environment-off reference under a controlled deformation, while the redshift-resolved $a_{\rm dyn}^{\rm proxy}$ identifies where in the fiducial graph-conditioned dynamics the local forward-reverse imbalance is strongest. Aside from these two, GPLM also enables access to additional path-level observables on fixed graphs, such as correlations among satellite tracks within the same system, which could be interesting to explore in future work.

\section{Discussion and Conclusions}
\label{sec:discussion}

This work formulates galaxy formation as an effective stochastic dynamics on layered halo graphs, represented as a graph-conditioned trajectory ensemble with an explicit path measure. On a fixed layered halo graph $G$, GPLM assigns the baryonic history $\mathbf{x}$ the conditional weight
\begin{equation}
P(\mathbf{x}\mid G) \propto
p_{\rm attach}(\mathbf{x}\mid G)\exp[-S_{\rm OM}(\mathbf{x}\mid G)],
\end{equation}
which extends to the joint measure $P(\mathbf{x},G)=P(\mathbf{x}\mid G)P(G)$ once a graph-sector model is specified. 

In the present implementation, a graph-based neural network model predicts residual drifts and diffusion on top of deterministic merger transport, yielding a discrete-time Gaussian OM sector on fixed simulation-extracted graphs. This path-space formulation enables graph-conditioned operator averages, controlled response deformations, and related diagnostics to be computed within one normalized probabilistic framework. 
Organizing inference at the trajectory level in this way makes GPLM conceptually distinct from summary-level forward models, field reconstructions, and graph-based emulators.

\subsection{Further Theoretical Directions}

The present construction suggests several natural theoretical extensions. In its current form the model is fixed-graph, two-field, and Gaussian, but the same path-action language can be developed further in a number of directions.

The proposed framework points to several theory-oriented directions that bring field-theoretic tools into astrophysical applications. In the MSRJD formalism, the operator content can be extended to include non-Gaussian terms, allowing a more flexible description of realistic scatter and higher cumulants in the modeled fields. The quadratic OM action also admits a natural geometric interpretation. Defining the inverse diffusion kernel as an effective metric, $g\equiv D^{-1}$, the action can be rewritten as $S=(1/2) \int dt (\dot{x}-b(x))^T g (\dot{x}-b(x))$. In this form, the stochastic dynamics can be viewed as a path optimization problem in a geometry shaped by the environment, where the most probable trajectories reduce to geodesics in the pure-noise, fixed-endpoint limit, and become forced, geodesic-like extremal paths in the presence of drift, with the drift acting as an effective force. At the level of evolving trajectory densities, the continuum stochastic dynamics can also connect to optimal transport theory in certain limits, where the evolution can be formulated as a gradient flow in the Wasserstein metric~\cite{Jordan1998OP,Otto31012001}.

A more controlled approach to constructing GPLM is to build it on a semi-analytic backbone~\cite{Alarcon:2022hzg,Alarcon:2025cij}. In that setting, the semi-analytic model would provide a structured baseline evolution, while the learned residual sector would capture the remaining unresolved dynamics and stochasticity. An operator basis with adjustable coefficients would then offer a natural way to organize controlled studies, allowing physically motivated deformations of transport, environmental response, or fluctuation terms to be explored within the same graph-conditioned framework.

It may also be useful to calibrate low-order operator expansions of graph-conditioned drift $b(x,G)$ and diffusion $D(x,G)$ against the trained model, with coefficients that depend on the local graph environment. Such expansions would provide a direct bridge between learned dynamics as well as more analytic stochastic descriptions, helping to identify which nonlinear, environmental and deviation structures are actually required by the data.

\subsection{Applications to Dark Matter Microphysics}

The likelihood formulation on layered halo graphs naturally connects to dark matter microphysics, such as self-interacting dark matter (SIDM). A direct continuation is an SIDM version of GPLM inspired by the parametric model developed in \cite{Yang:2023jwn,Yang:2024uqb,Yang:2024tba}. In that model, an integral approach applies a universal analytic kernel incrementally on a halo's $V_{\max}$ and $R_{\max}$ evolution history in cold dark matter, mapping the evolution trajectories into SIDM, incorporating the effect of gravothermal evolution along with the accretion history. These features naturally fit into the GPLM framework. The $V_{\max}$ and $R_{\max}$ are fields that evolve along merger histories, and a trained GNN model that works incrementally in a time-universal manner mirrors the analytic kernel.

For an increment at a specific layer, one can write schematically
\begin{equation}
\Delta\mathbf{y}_k = \Delta\mathbf{y}^{\rm tr}_k + \Delta\tau_k \mathbf{b}_{\rm SIDM}(s_k;\theta_{\rm DM})
  + \boldsymbol{\epsilon}_k,
\end{equation}
with $\mathbf{y}_k\equiv(V_{\max},R_{\max})_k$ and $\mathbf{y}^{\rm tr}_k$ a nonlinear transport map to propagate the two structural variables. Here $\mathbf{b}_{\rm SIDM}$ represents a residual drift in the spirit of the integral approach, while $\boldsymbol{\epsilon}_k$ captures residual scatter from limited conditioning and unresolved physics. 
The conditioning on local graph state, transported fields, redshift, etc., enables the proposed framework to capture the environmental stripping effect and improve upon human-calibrated analytic kernels. Depending on the structure of the unresolved variability, one may further extend the residual sector beyond the quadratic OM approximation through higher-order drift or noise operators. The path-action formulation also facilitates simultaneous exploration of the baryonic and dark matter contents. Models beyond elastic SIDM, such as dissipative SIDM and two-component SIDM with mass segregation, exhibit analogous developments and can thus be explored within the same formalism~\cite{Yang:2025xsp,Essig:2018pzq}.

\subsection{Toward a Full Multi-Field Galaxy-Formation Model}

A natural next step is to extend GPLM from the present two-field example to a multi-field model of galaxy formation on layered graphs. Relevant variables can include gas and star half-mass radii, star-formation activity, metallicity, black-hole mass, and halo structural properties. The graph construction and deterministic transport bookkeeping could remain the same, while the residual sector becomes multivariate, graph-local couplings are introduced where needed, and higher-order corrections are organized within a broader operator basis. Meanwhile, as mentioned for SIDM, non-linear maps could be included to transport fields other than masses.

A more comprehensive extension is to treat the full assembly history itself within a stochastic dynamical framework. In this view, not only the baryonic fields but also the graph sector and boundary sector are assigned effective stochastic dynamics. A schematic joint history action can then be written as
\begin{equation}
S_{\rm tot} = S_G[G_{0:K}] + S_x[\mathbf{x}_{0:K};G_{0:K}] + S_{\rm bnd}[\mathbf{x}_{0:K};G_{0:K}],
\end{equation}
with
\begin{equation}
P(G_{0:K}) \propto \exp\left[-\sum_k S_{G,k}\right],
\end{equation}
Here, $S_x$ is the conditional field action on a given graph history, $S_{\rm bnd}$ is a boundary action associated with node entry, and $S_G$ governs the stochastic evolution of the graph itself. This perspective is supported by classical excursion-set theory, which has already shown that major aspects of the halo accretion history can be modeled statistically as an effective stochastic process. In the present context, such a formulation would provide a route toward a unified stochastic treatment of graph histories and baryonic trajectories.

Such a multi-field path measure would also provide a principled basis for inverse problems based on incomplete observations~\cite{2021ApJS..254...22J,2018MNRAS.480.4379C}. The conditional path measure defines the latent distribution needed to ask which aspects of assembly history and environment are statistically recoverable from observed halo and galaxy properties, and which degeneracies remain intrinsic to the coarse-grained dynamics.

More broadly, one may ask whether a graph-conditioned path measure of this kind can eventually be refined not only by simulations but also by large observational samples across many redshifts~\cite{2025arXiv251215841T}. In schematic form, one would attempt to constrain a common latent dynamics through a likelihood of the form
\begin{equation}
\begin{split}
P(d_{\rm all}\mid \theta_{\rm GPLM}) &\sim \prod_z \int \mathcal D G \mathcal D \mathbf{x} \\
&\quad \times P(d_z\mid \mathbf{x}_z,G) P_{\theta_{\rm GPLM}}(\mathbf{x},G),
\end{split}
\end{equation}
Here $d_{\rm all}$ denotes the combined observational dataset across redshift, $d_z$ the observed sample in redshift slice $z$, $\mathbf{x}_z$ the latent graph-conditioned state projected to that slice, and $\theta_{\rm GPLM}$ the pretrained GPLM parameters, which could be partially updated with observational data within the same architecture. In this way, different redshift slices would probe different projections of the same underlying graph-conditioned evolution law. Whether this can yield substantial improvement beyond simulation pretraining, and which refinement strategies are statistically and physically effective, remains an open and challenging question. Even so, this possibility helps clarify one possible longer-term role of GPLM. Beyond serving as a generator of assembly histories, it may also offer a useful probabilistic setting in which merger structure, baryonic evolution, and dark-sector physics can be studied within a common graph-conditioned framework.

The training and testing codes used in this work are available at: \url{https://github.com/DanengYang/GraphPathLikelihood}

\acknowledgments
We acknowledge the IllustrisTNG collaboration for making the simulation data publicly available. 
This work is partially supported by the cosmology simulation database (CSD) in the National Basic Science Data Center (NBSDC-DB-10).
The authors used generative AI tools for code-assistance and language editing during software and manuscript development. All scientific content, methods, results, and conclusions were designed, verified, and approved by the authors.

\bibliography{refs}

\appendix

\onecolumngrid
\setcounter{section}{0}

\section{Layers of spatial halo graphs}
\label{sec:graph1}

\begin{figure*}[t]
  \centering
  \includegraphics[width=0.24\textwidth]{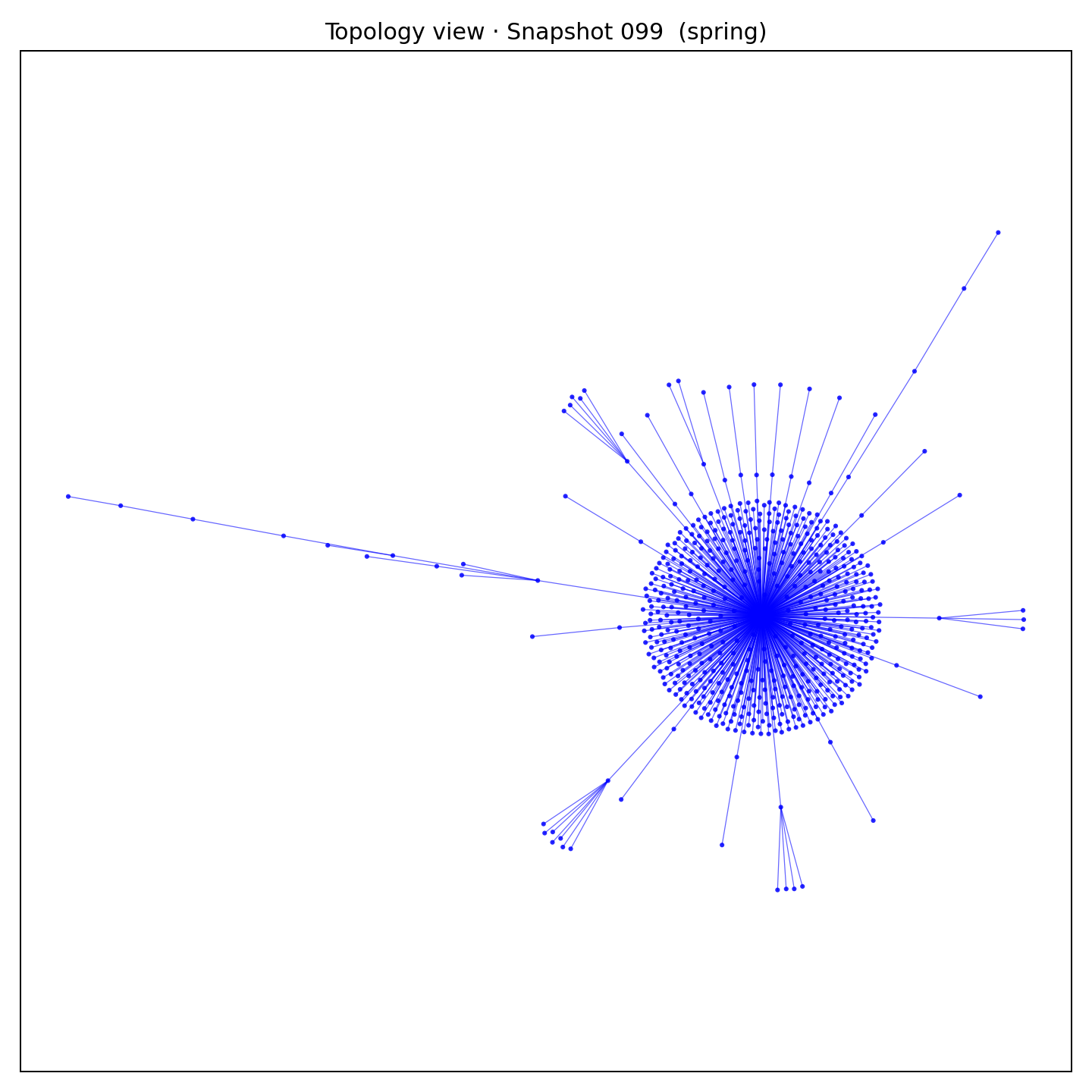}
  \includegraphics[width=0.24\textwidth]{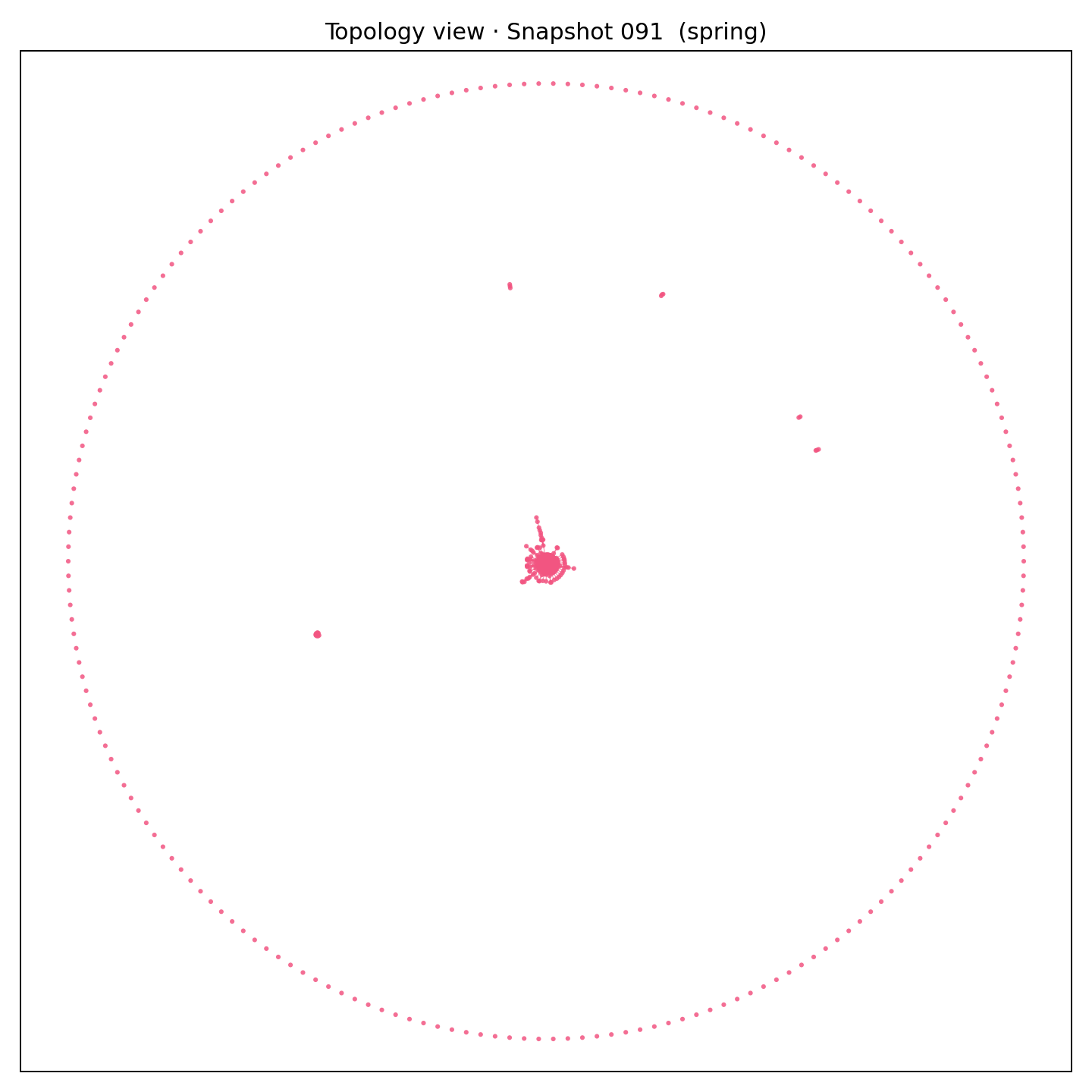}
  \includegraphics[width=0.24\textwidth]{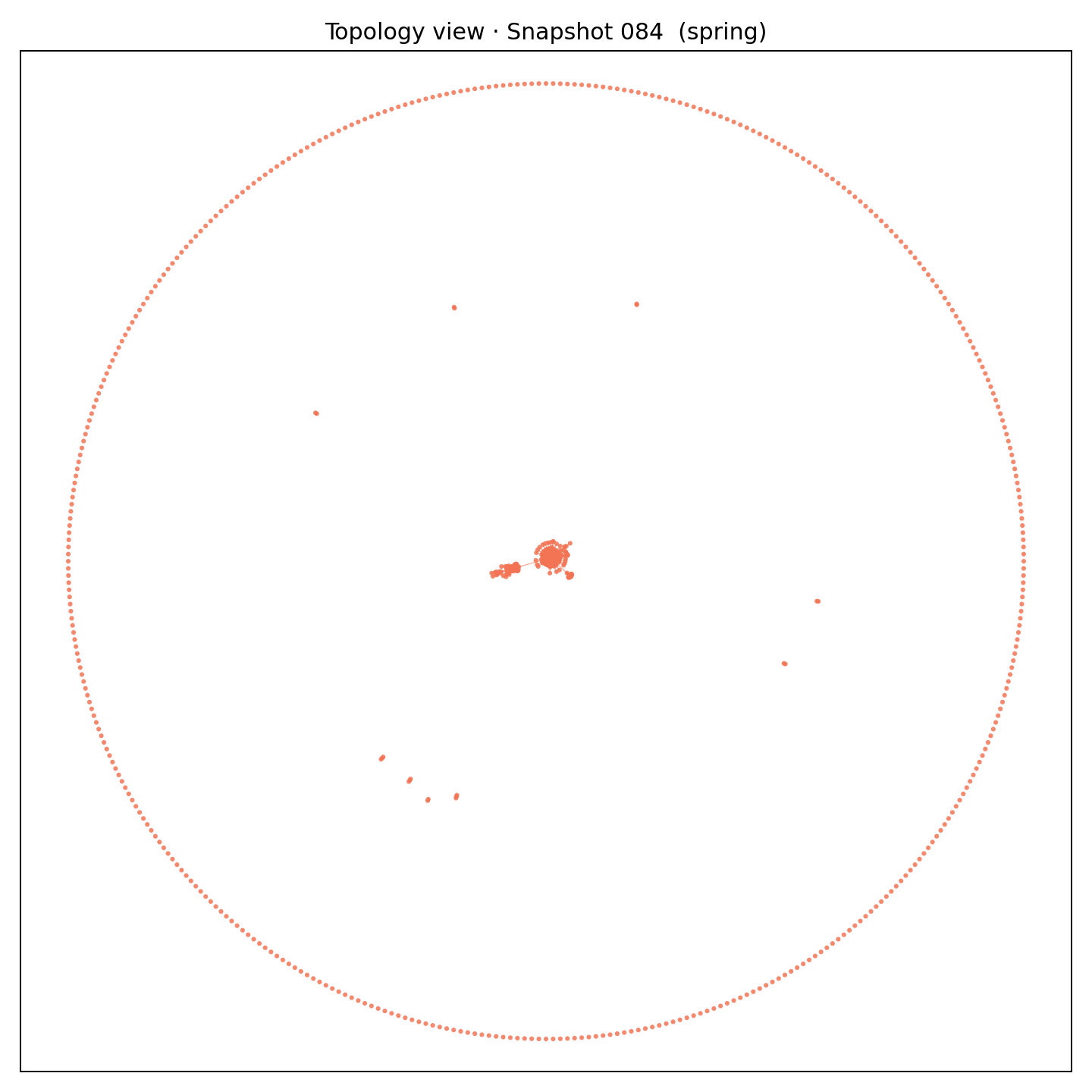}
  \includegraphics[width=0.24\textwidth]{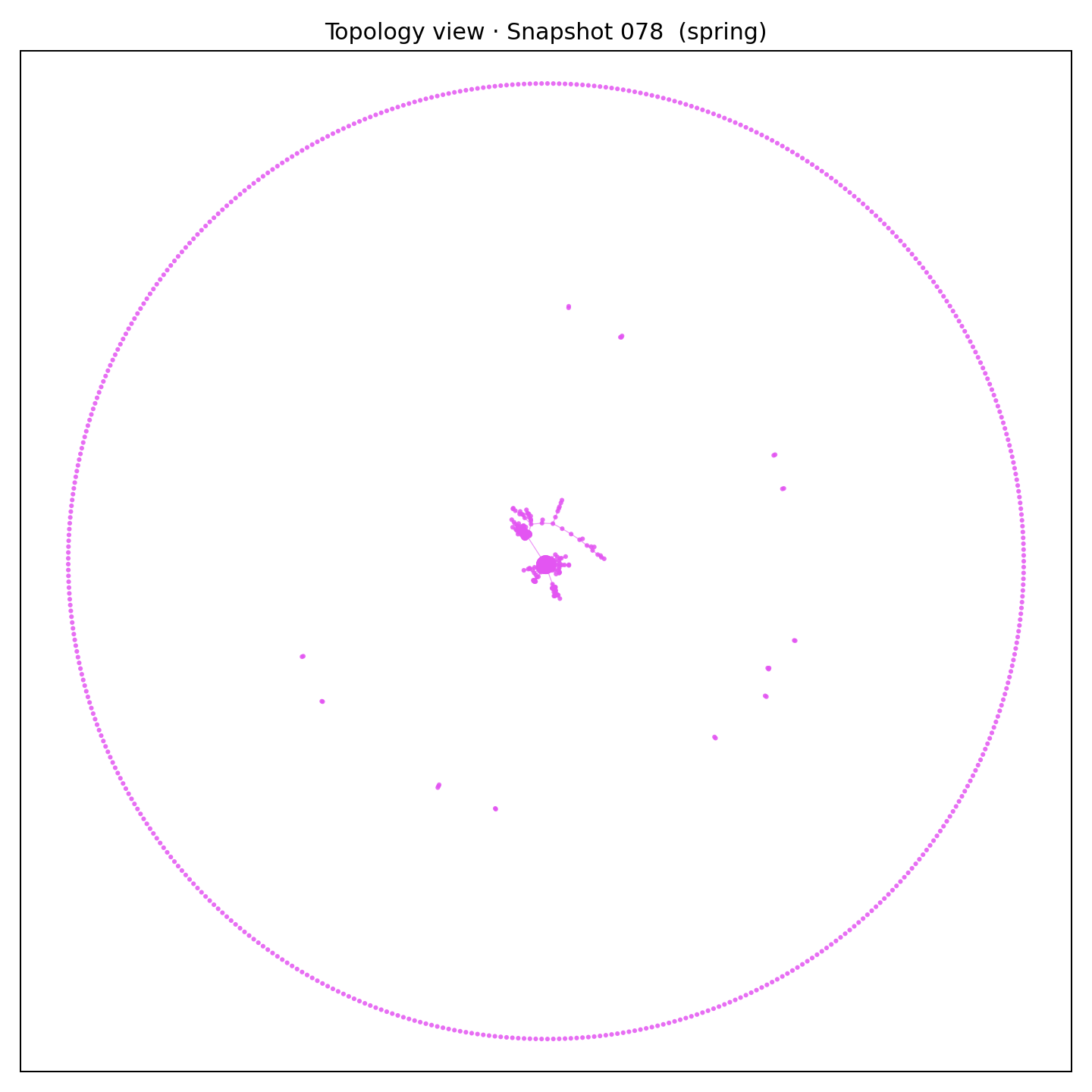} \\
  \includegraphics[width=0.24\textwidth]{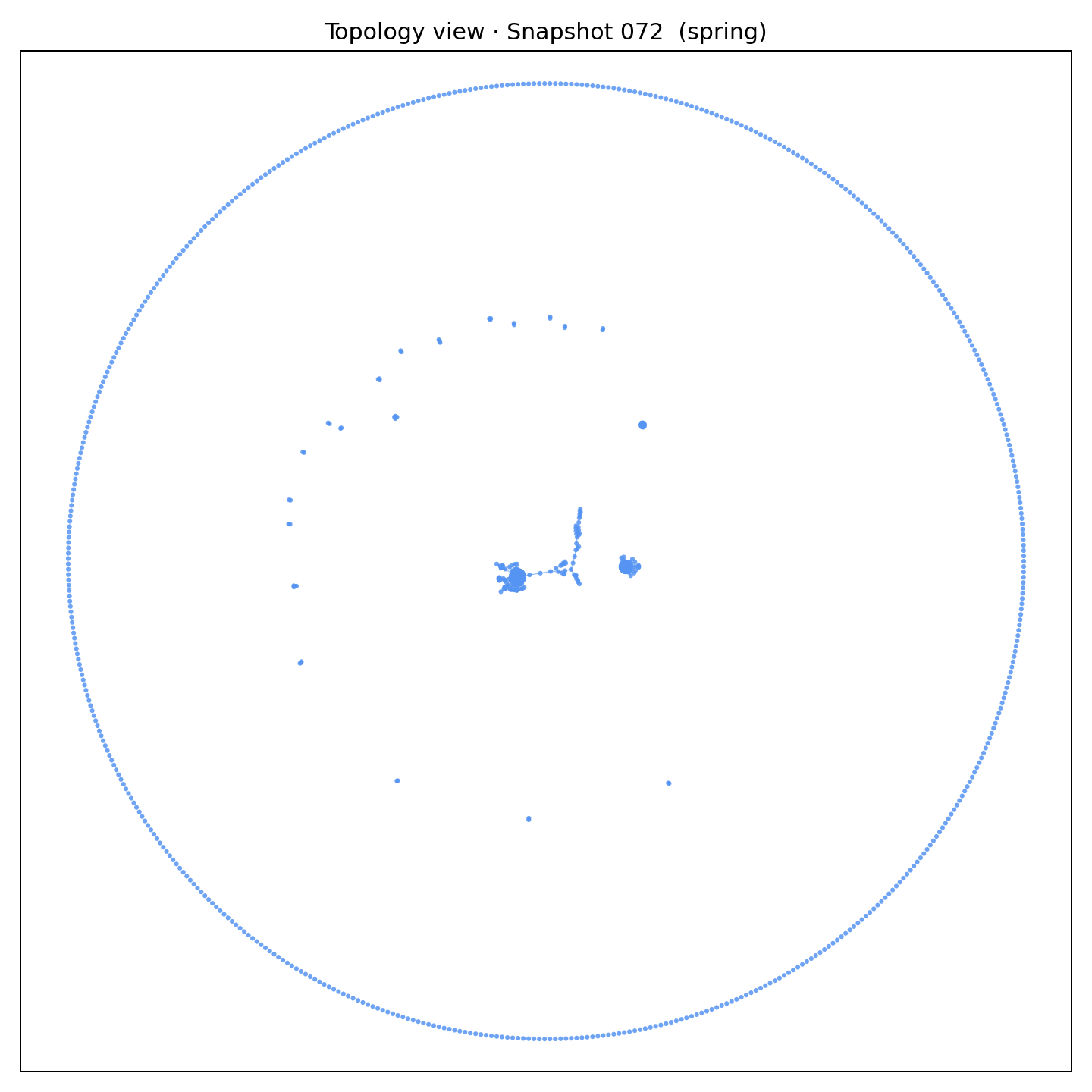}
  \includegraphics[width=0.24\textwidth]{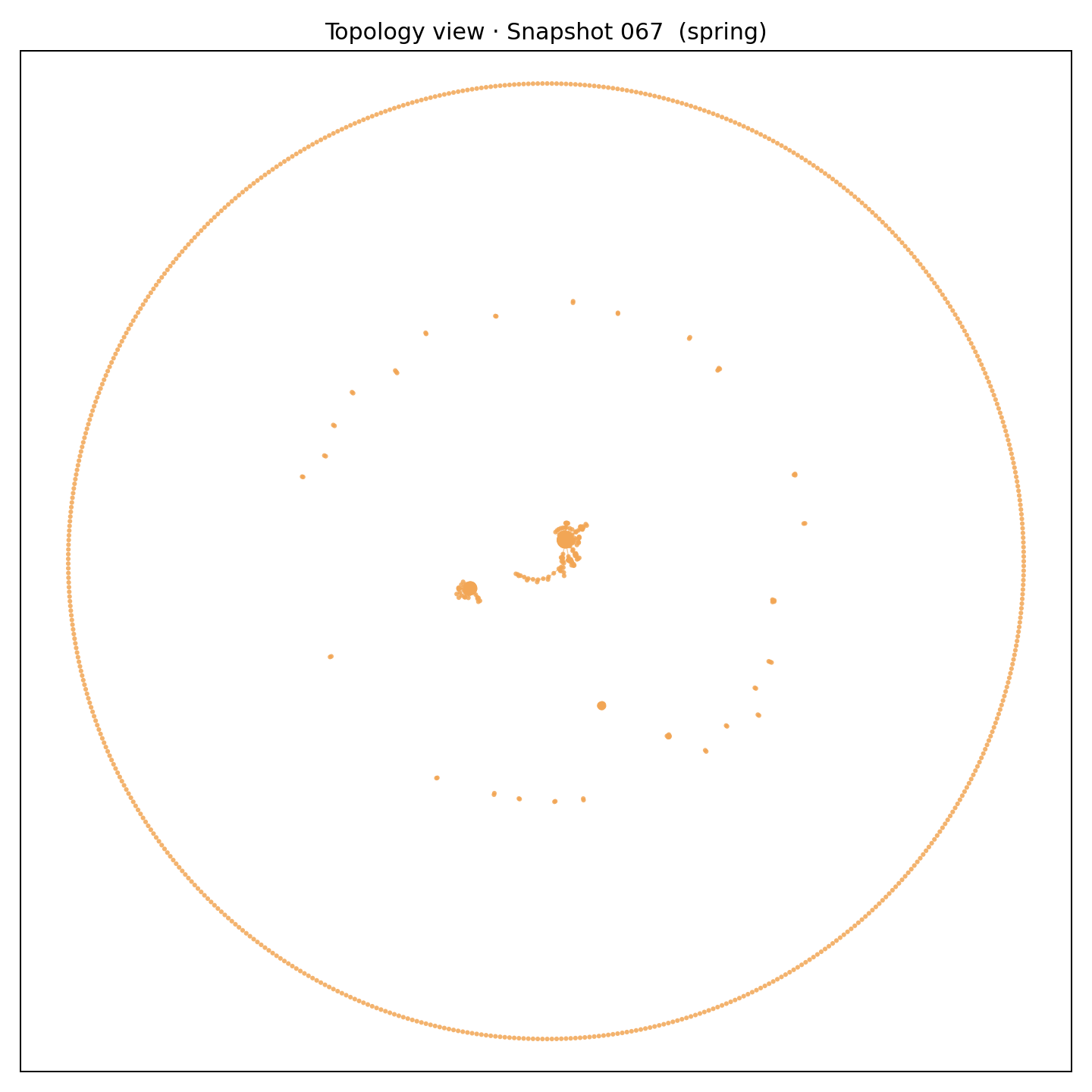}
  \includegraphics[width=0.24\textwidth]{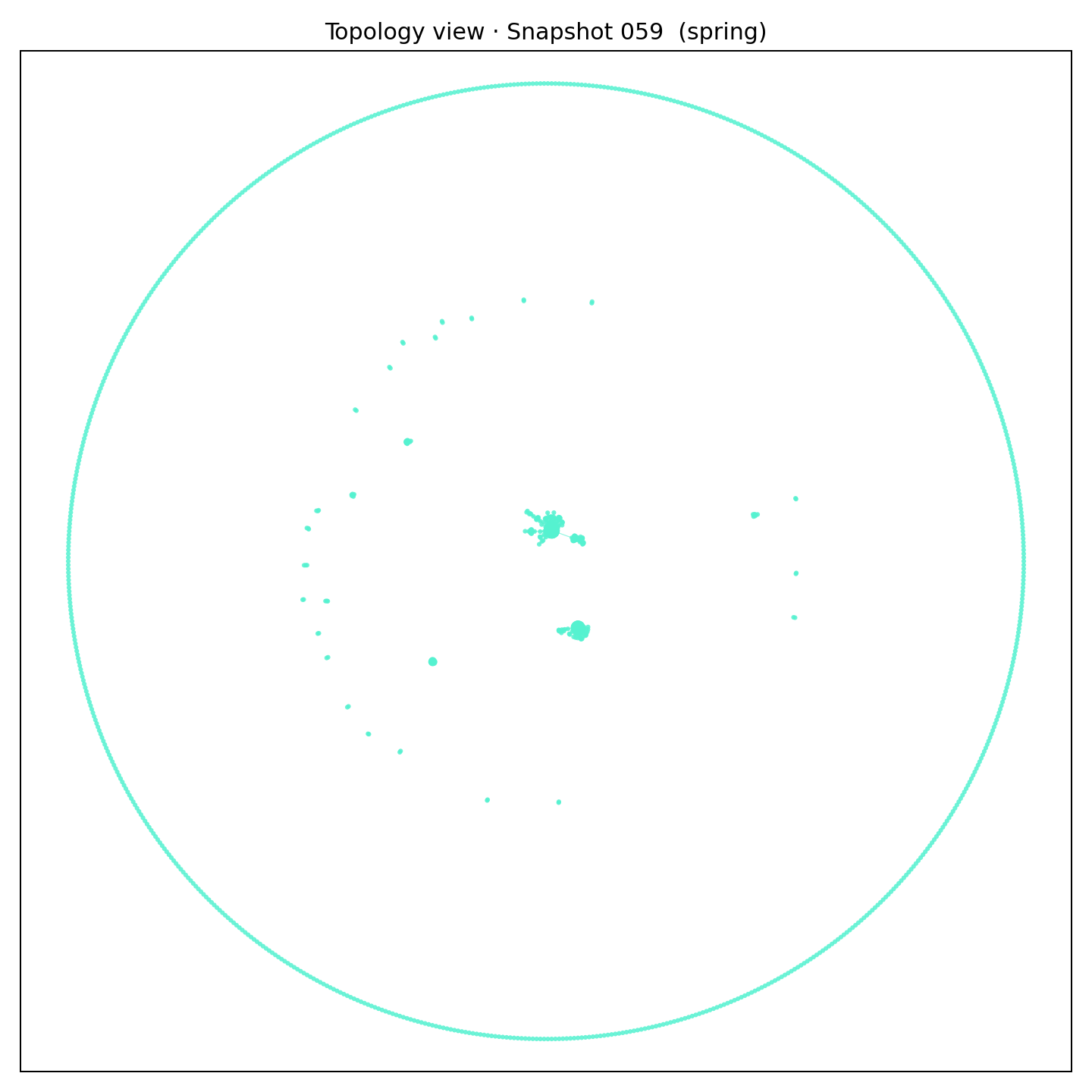}
  \includegraphics[width=0.24\textwidth]{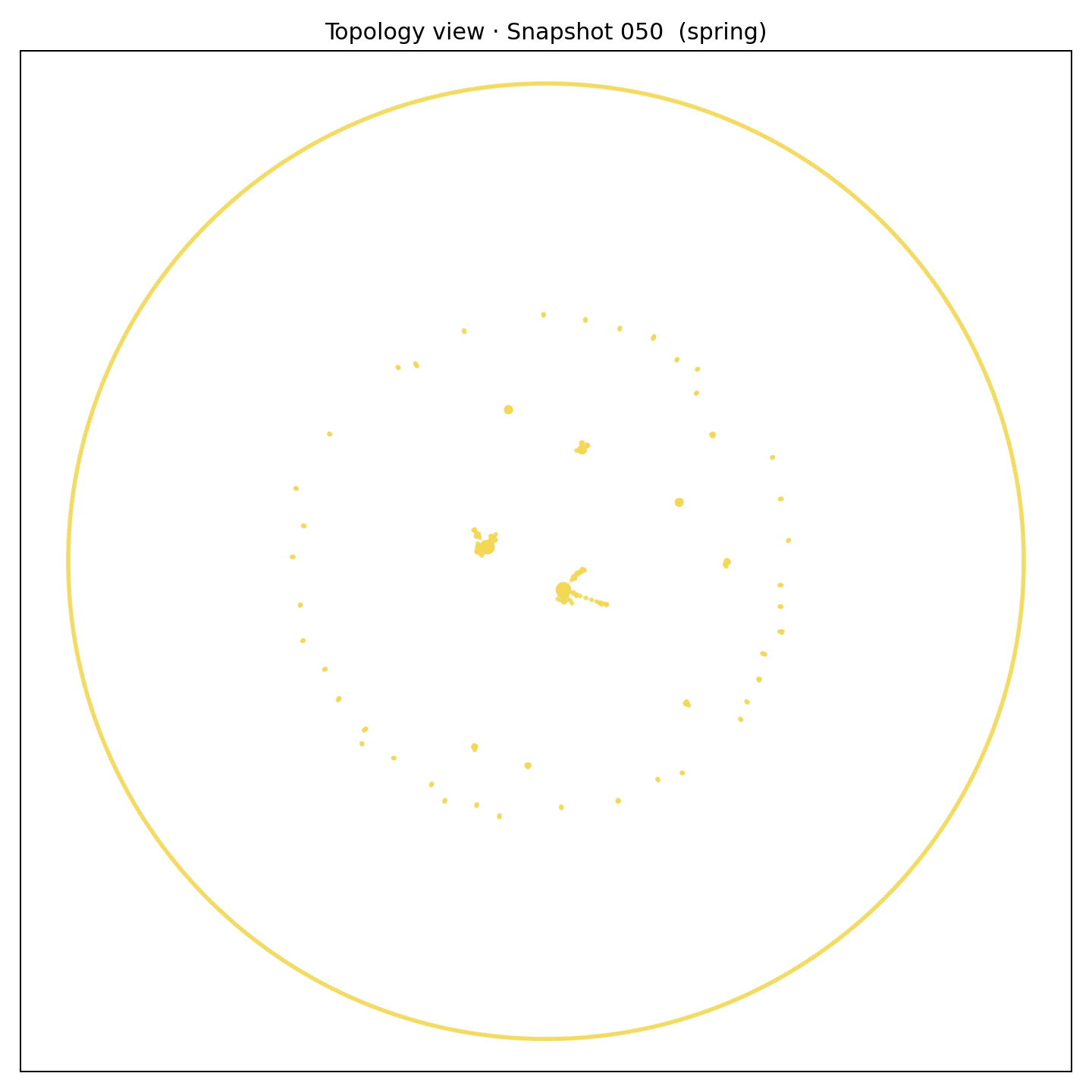} \\
  \includegraphics[width=0.24\textwidth]{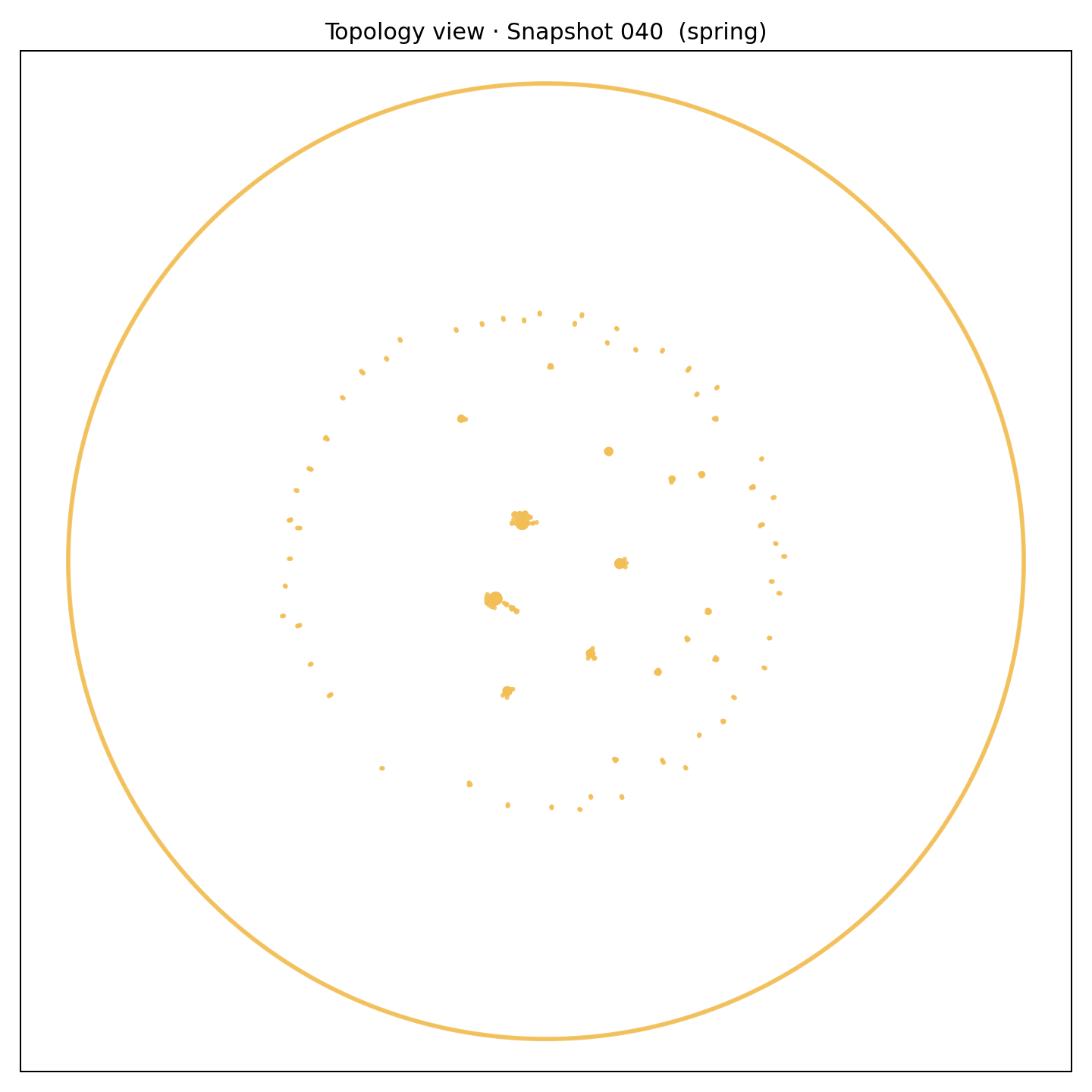}
  \includegraphics[width=0.24\textwidth]{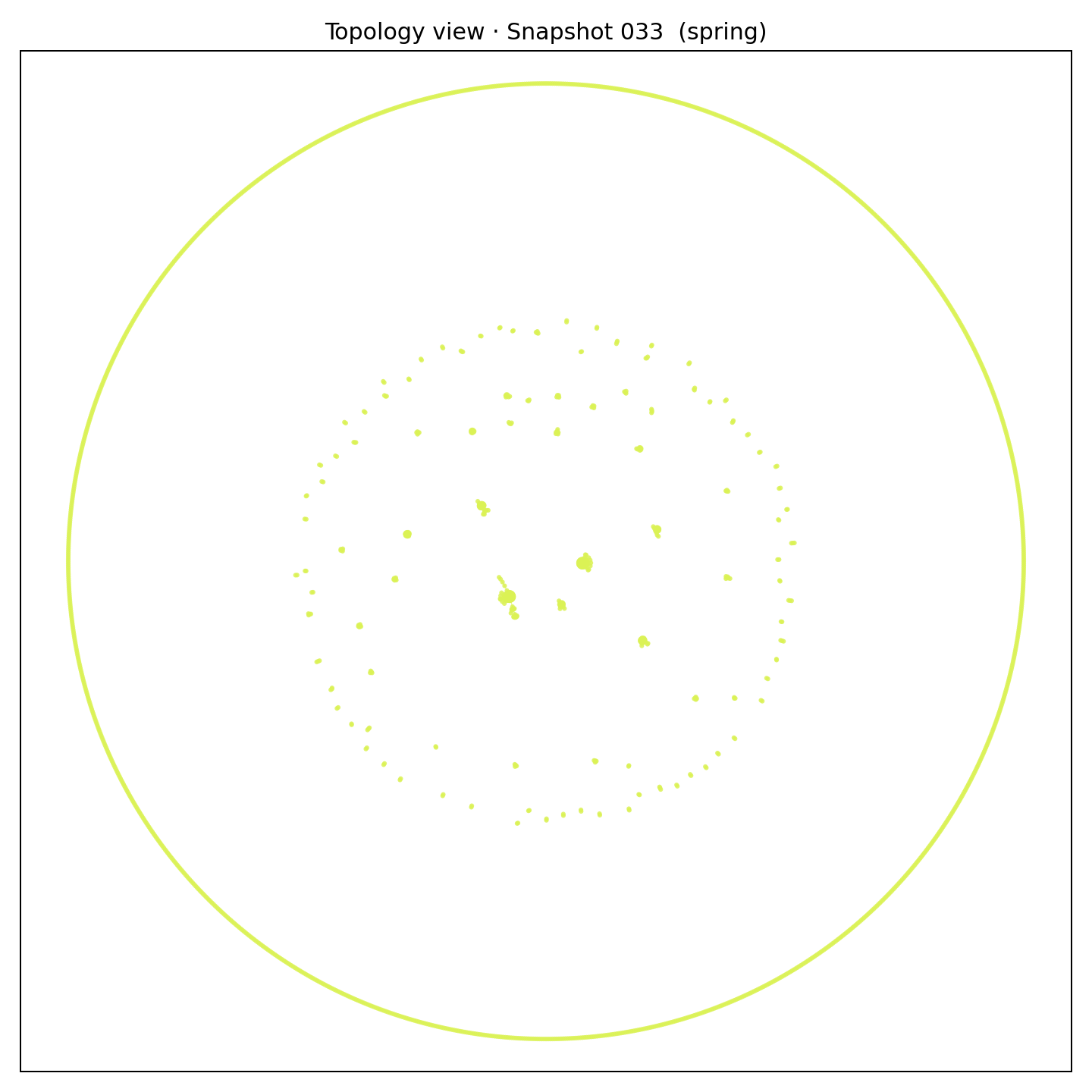}
  \includegraphics[width=0.24\textwidth]{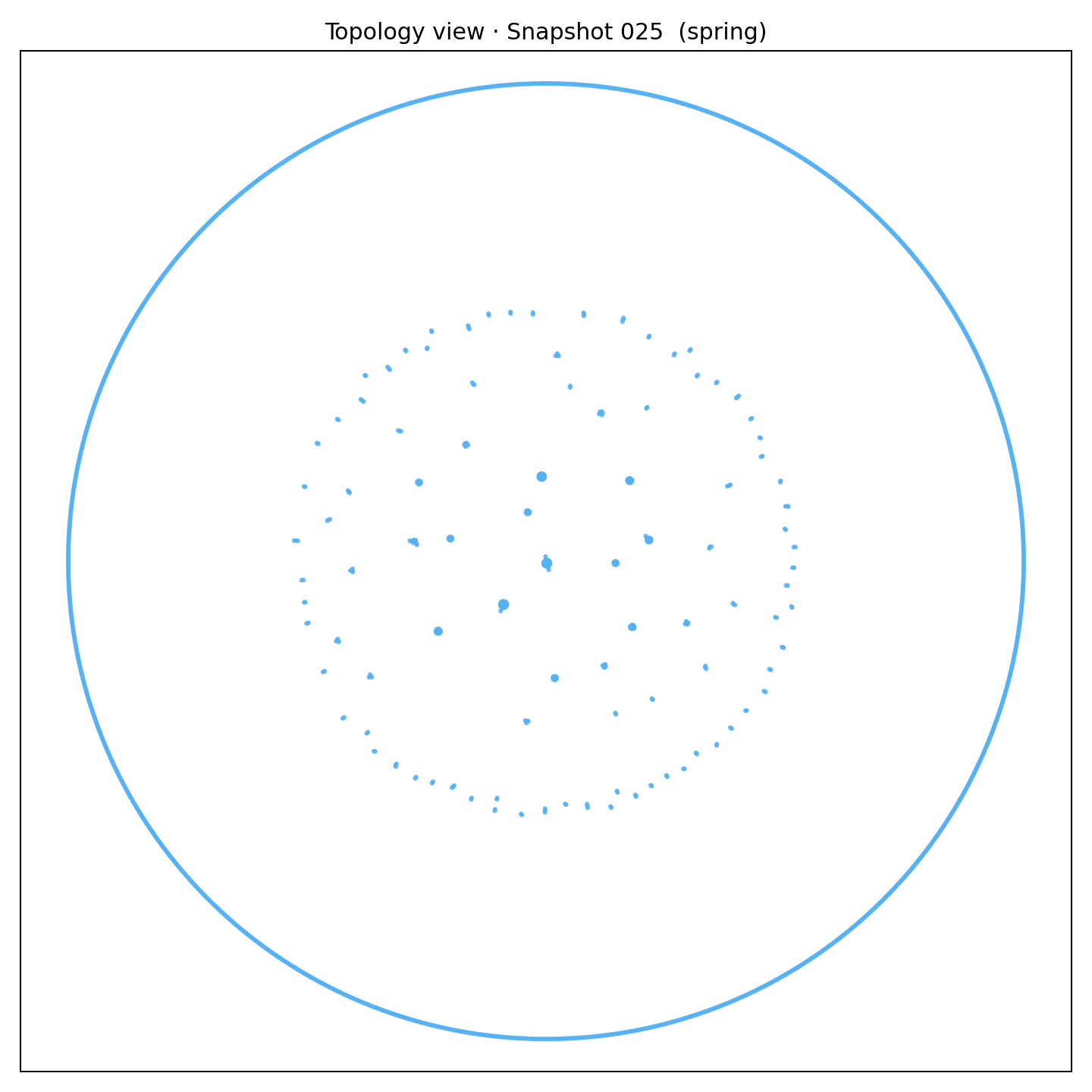}
  \includegraphics[width=0.24\textwidth]{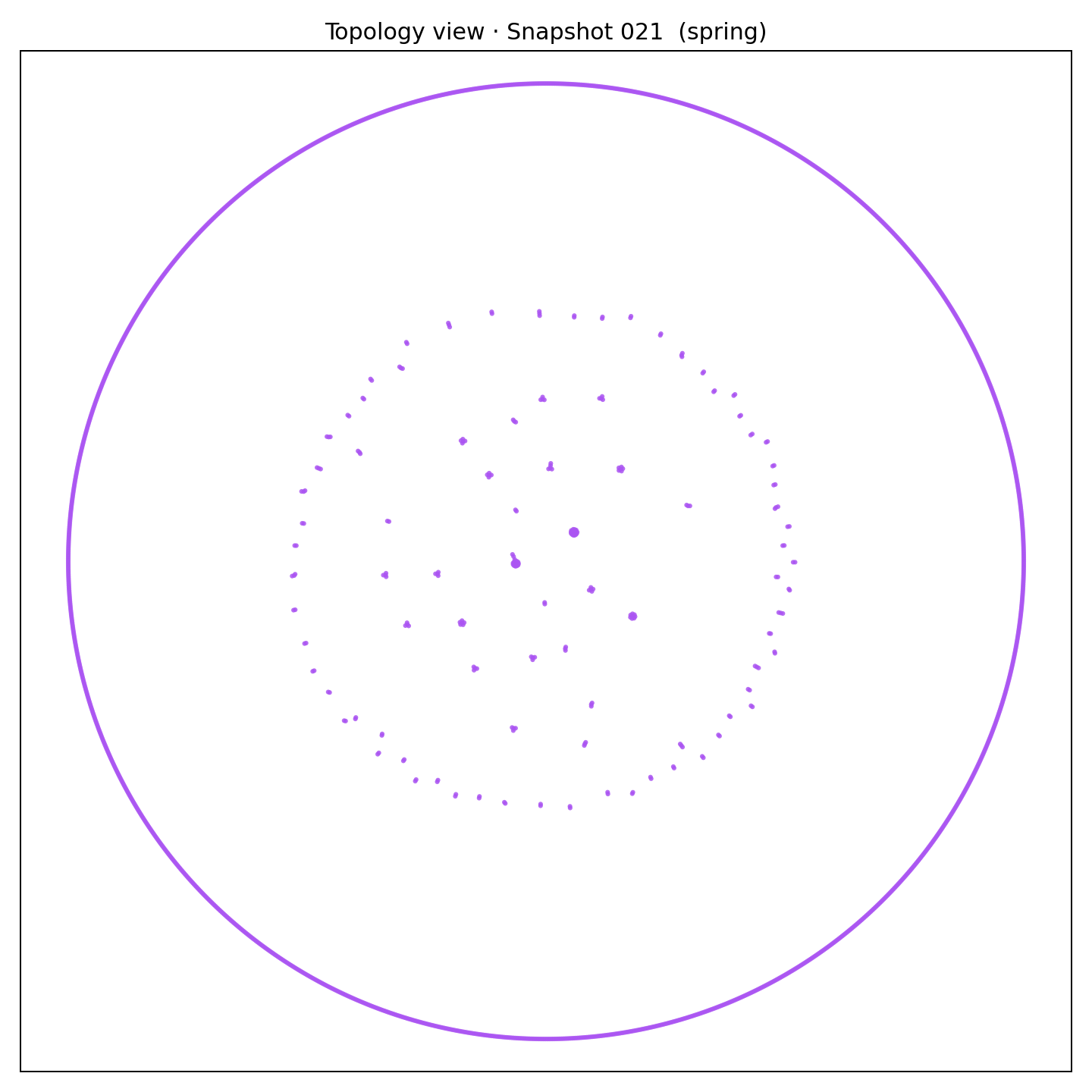}
  \caption{\label{fig:rank1_topology} Topological (spring-layout) view of the host-edge subgraphs for the rank-1 layered halo graph at multiple snapshots obtained by removing temporal edges. From top-left to bottom-right the panels correspond to increasing redshift: $z=0,0.1,0.2,0.3,0.4,0.5,0.7,1,1.5,2$. Only the $z=3$ snapshot is left over due to limited space. }
\end{figure*}

\begin{figure*}[t]
  \centering
  \includegraphics[width=0.24\textwidth]{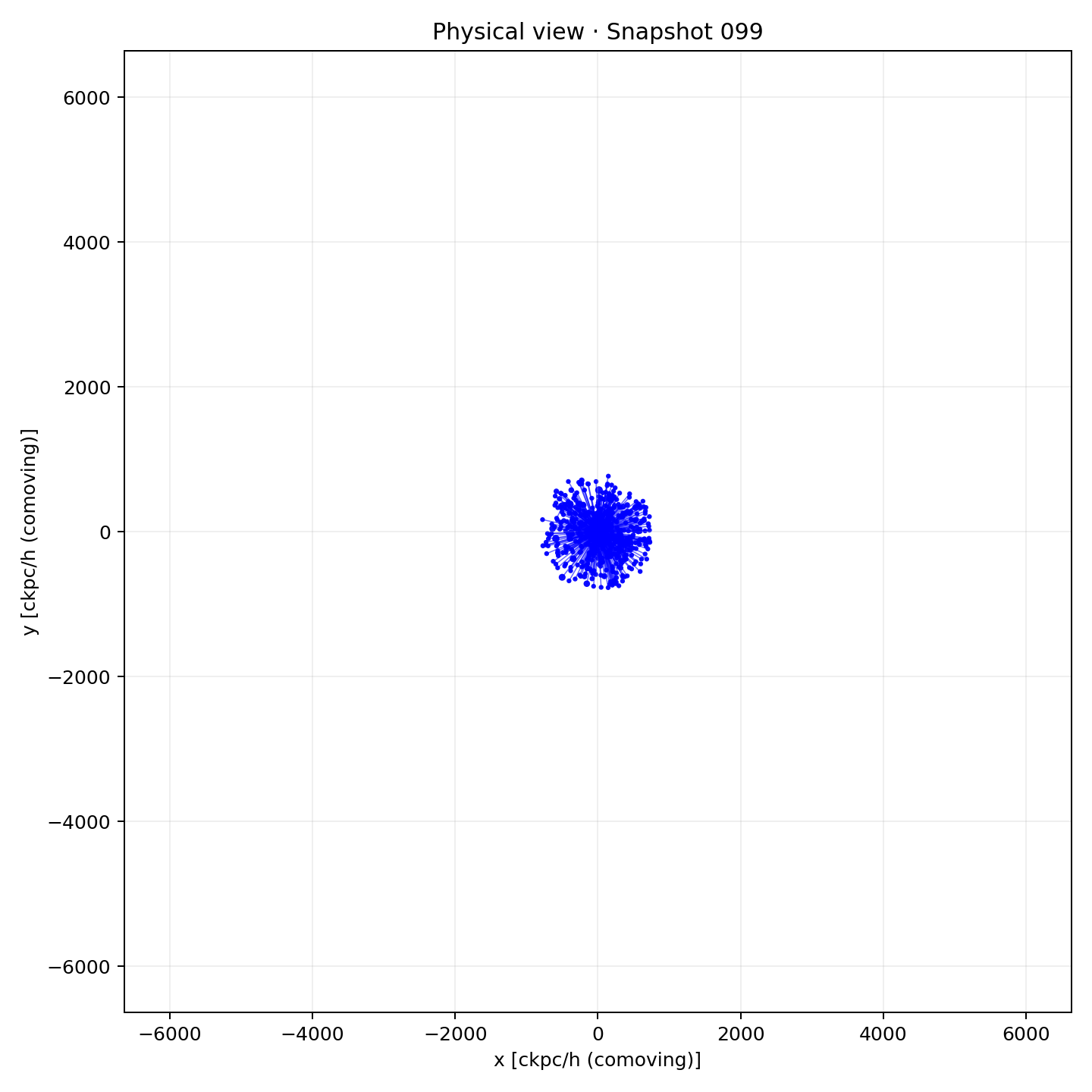}
  \includegraphics[width=0.24\textwidth]{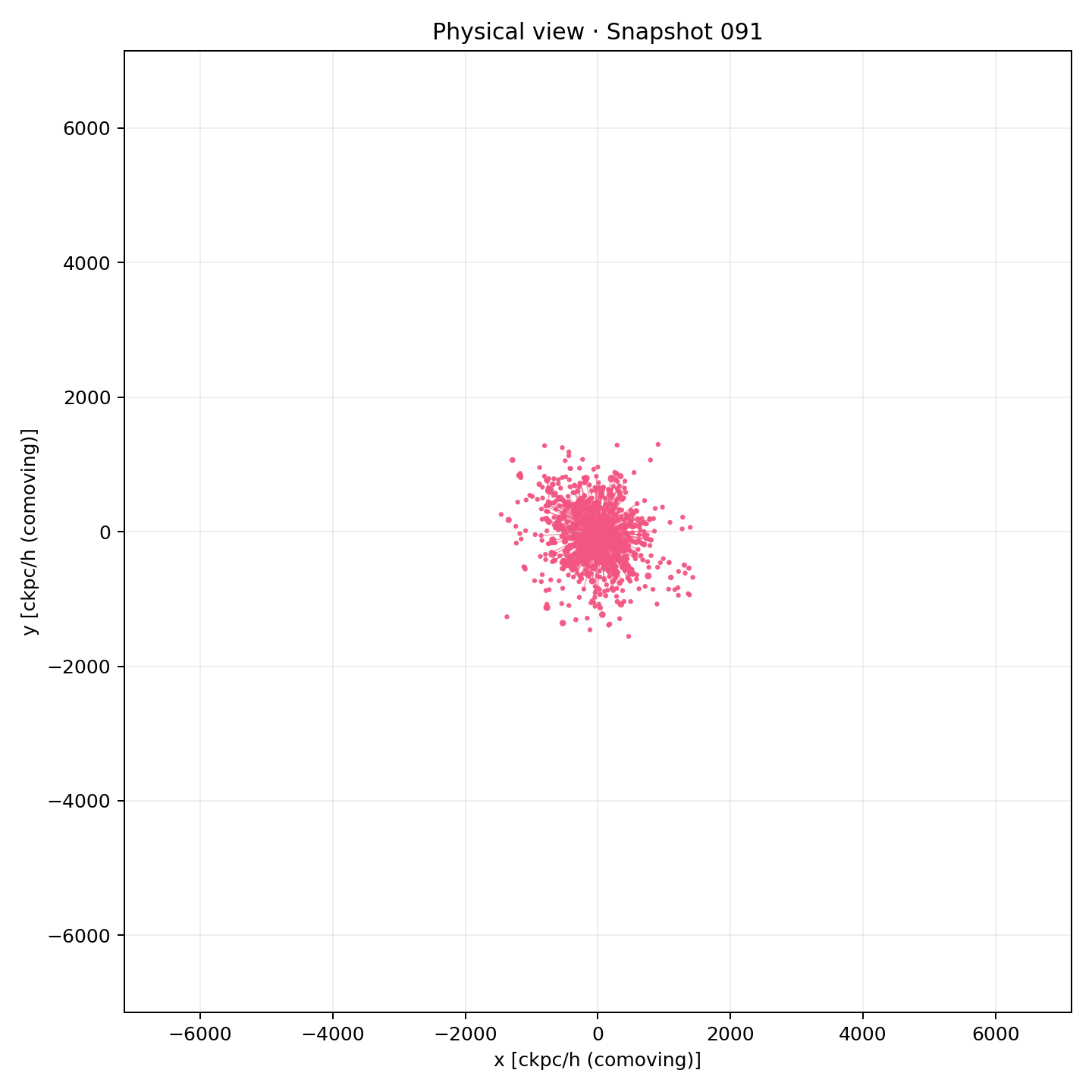}
  \includegraphics[width=0.24\textwidth]{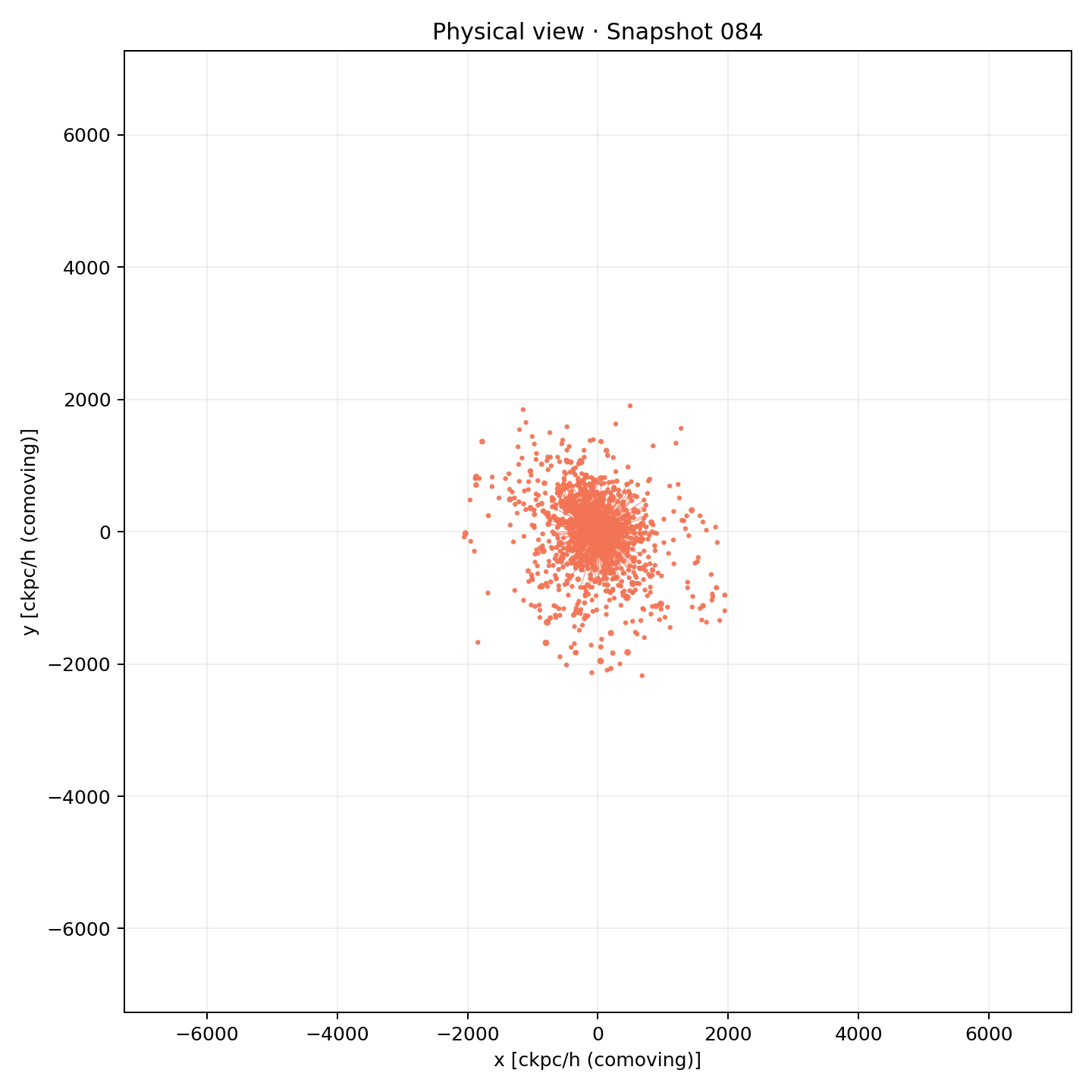}
  \includegraphics[width=0.24\textwidth]{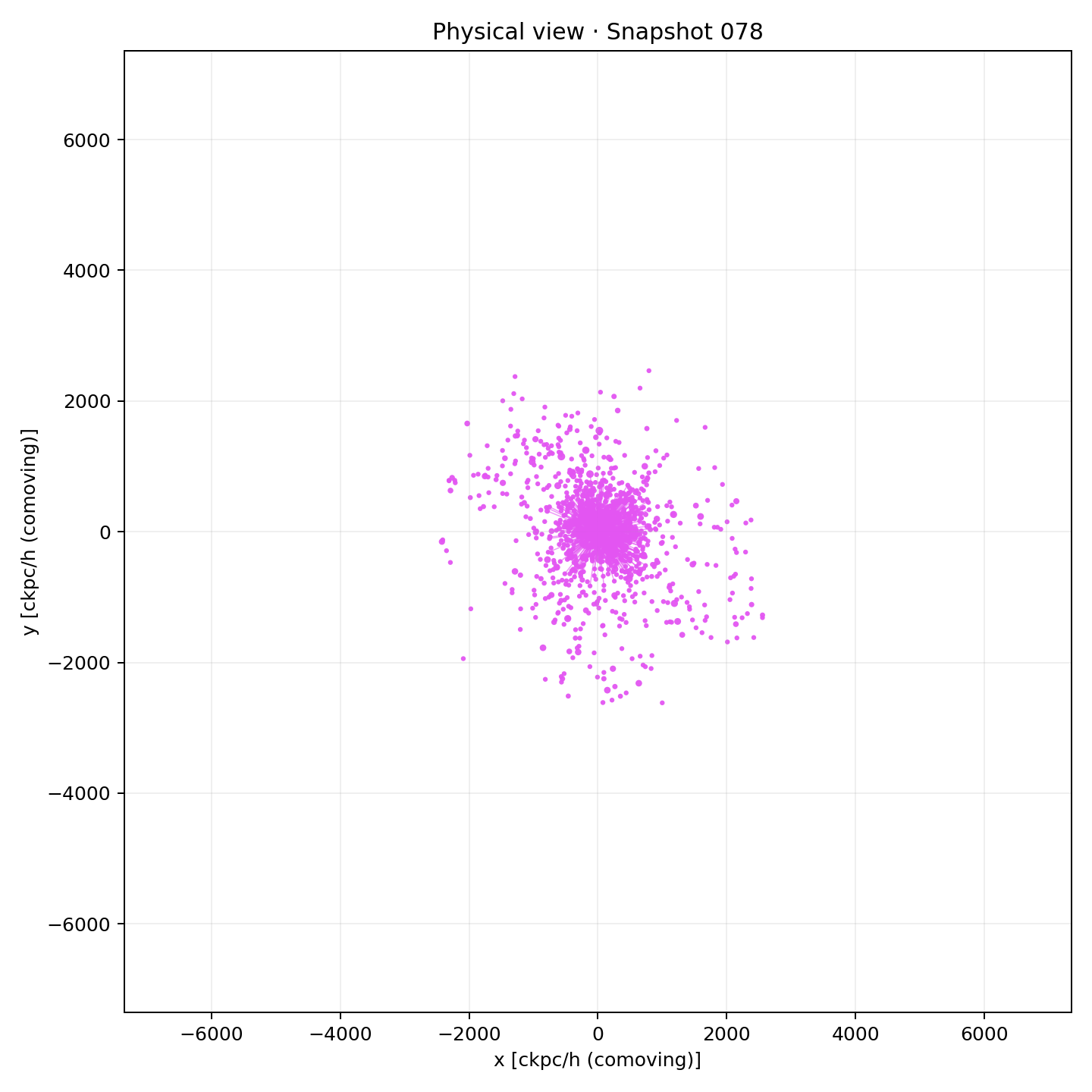} \\
  \includegraphics[width=0.24\textwidth]{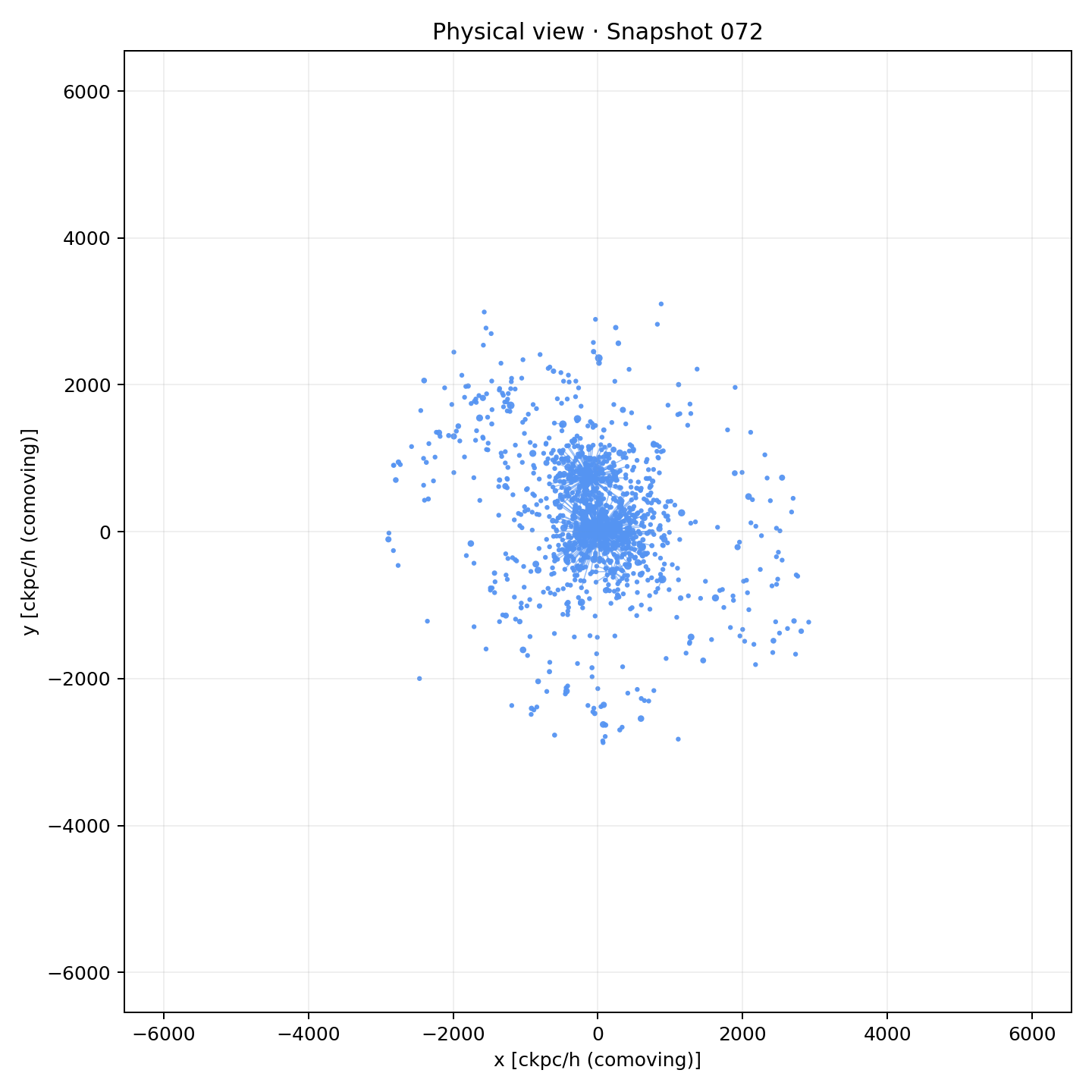}
  \includegraphics[width=0.24\textwidth]{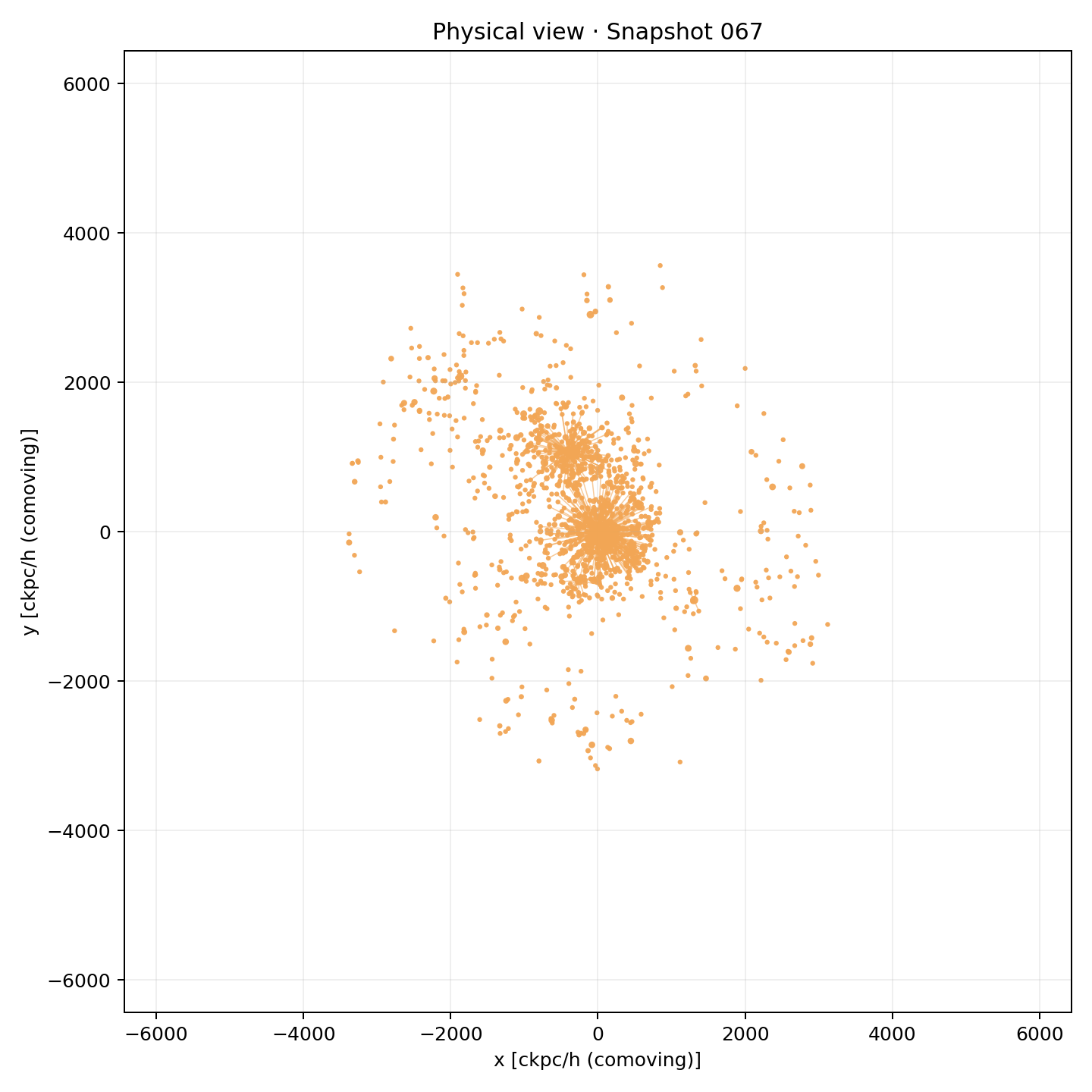}
  \includegraphics[width=0.24\textwidth]{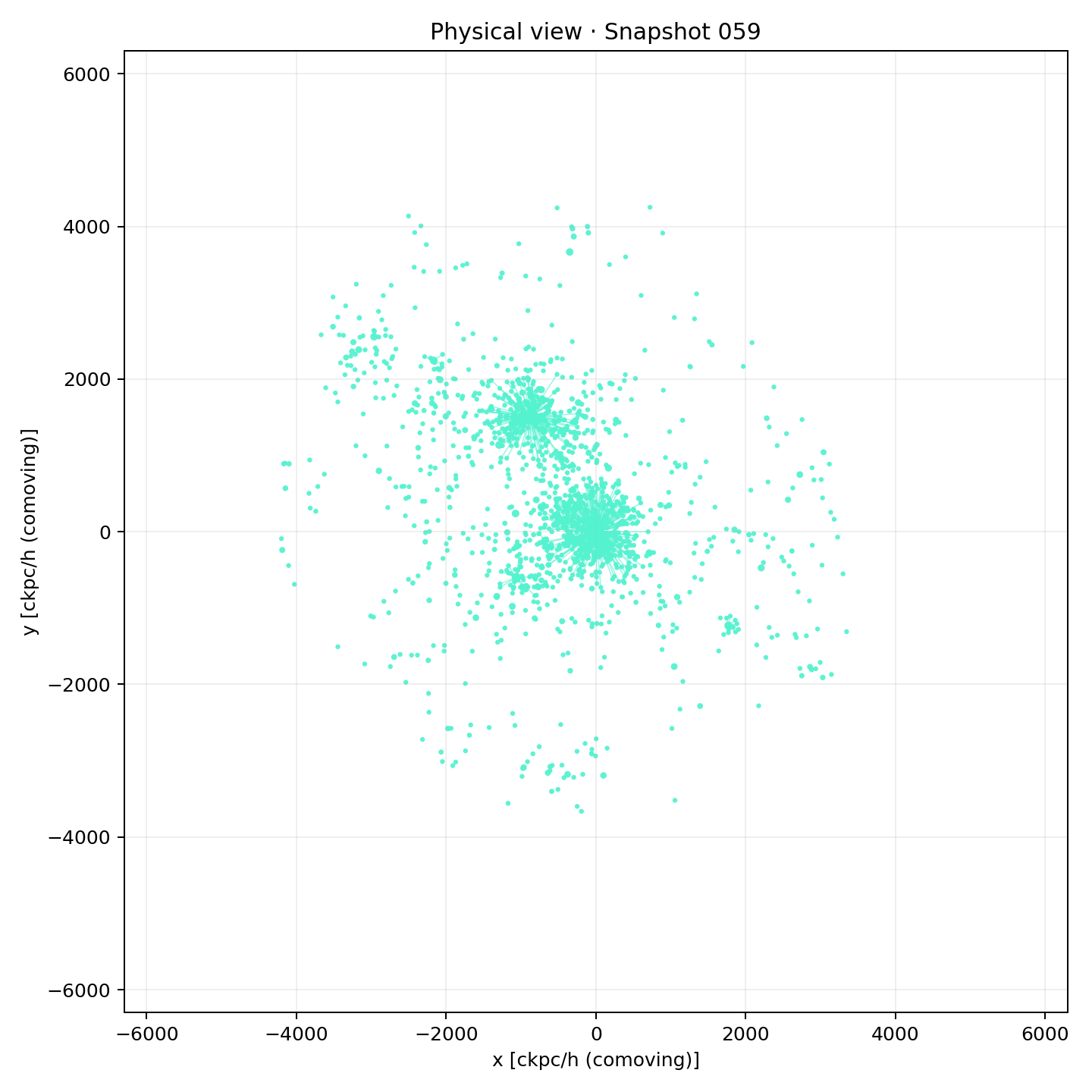}
  \includegraphics[width=0.24\textwidth]{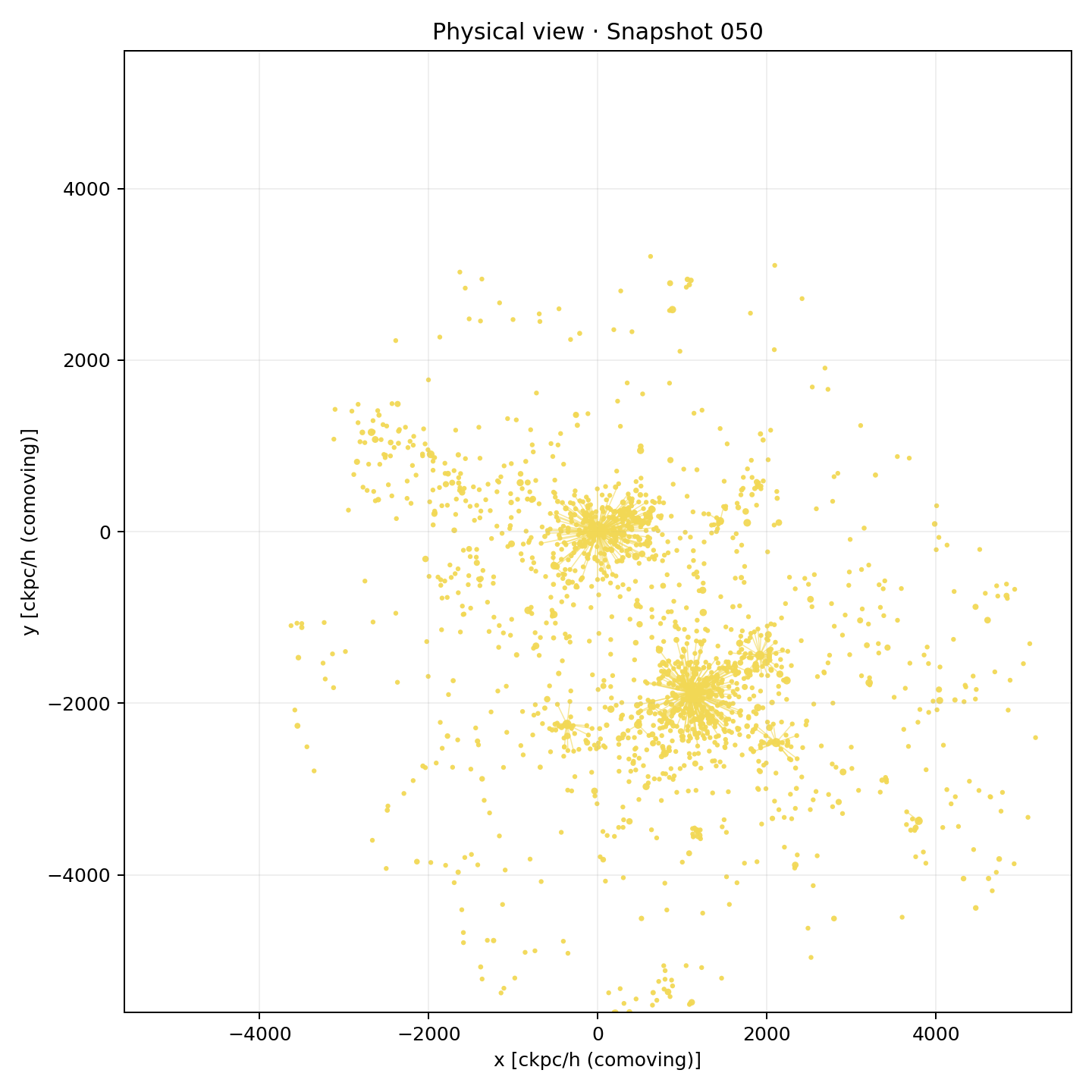} \\
  \includegraphics[width=0.24\textwidth]{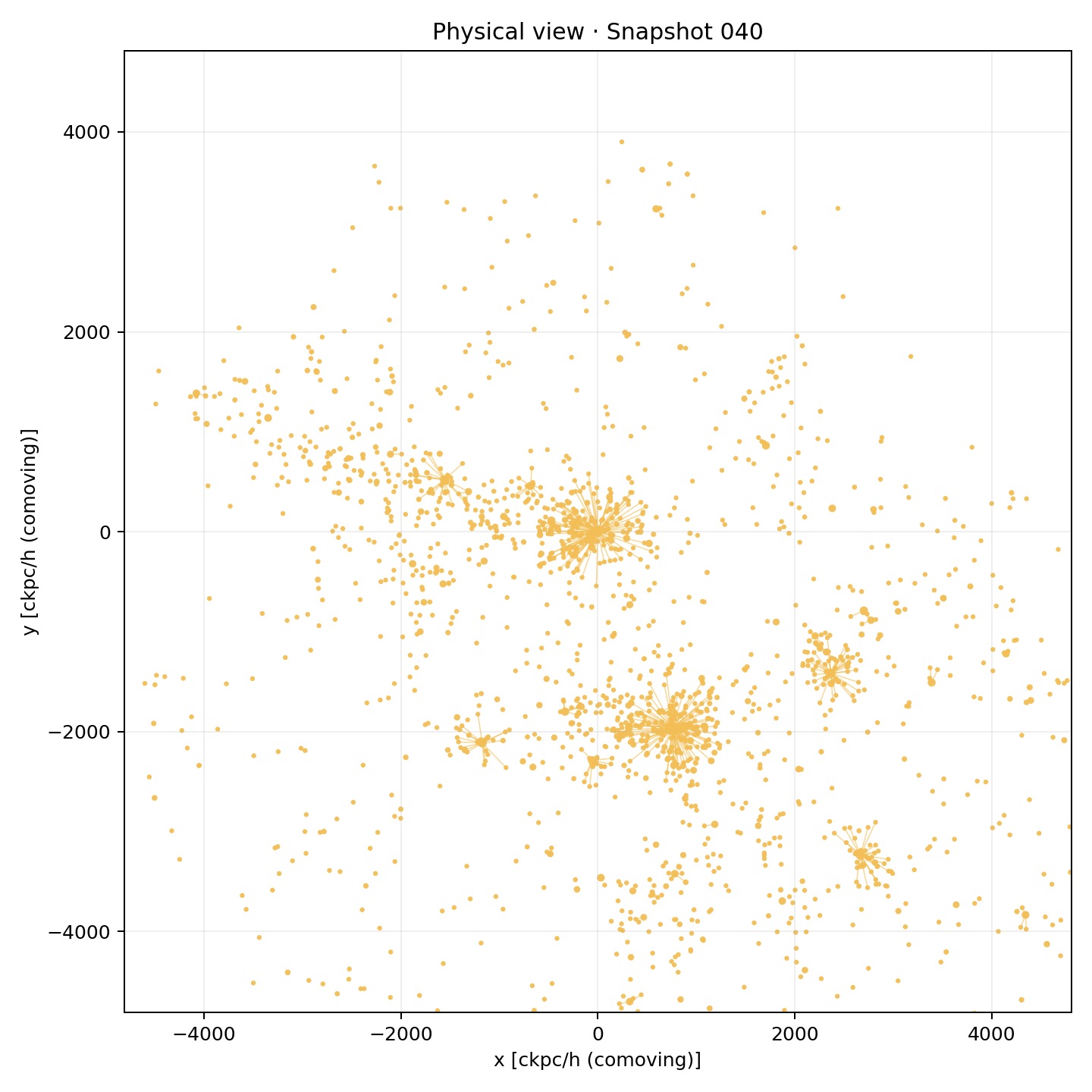}
  \includegraphics[width=0.24\textwidth]{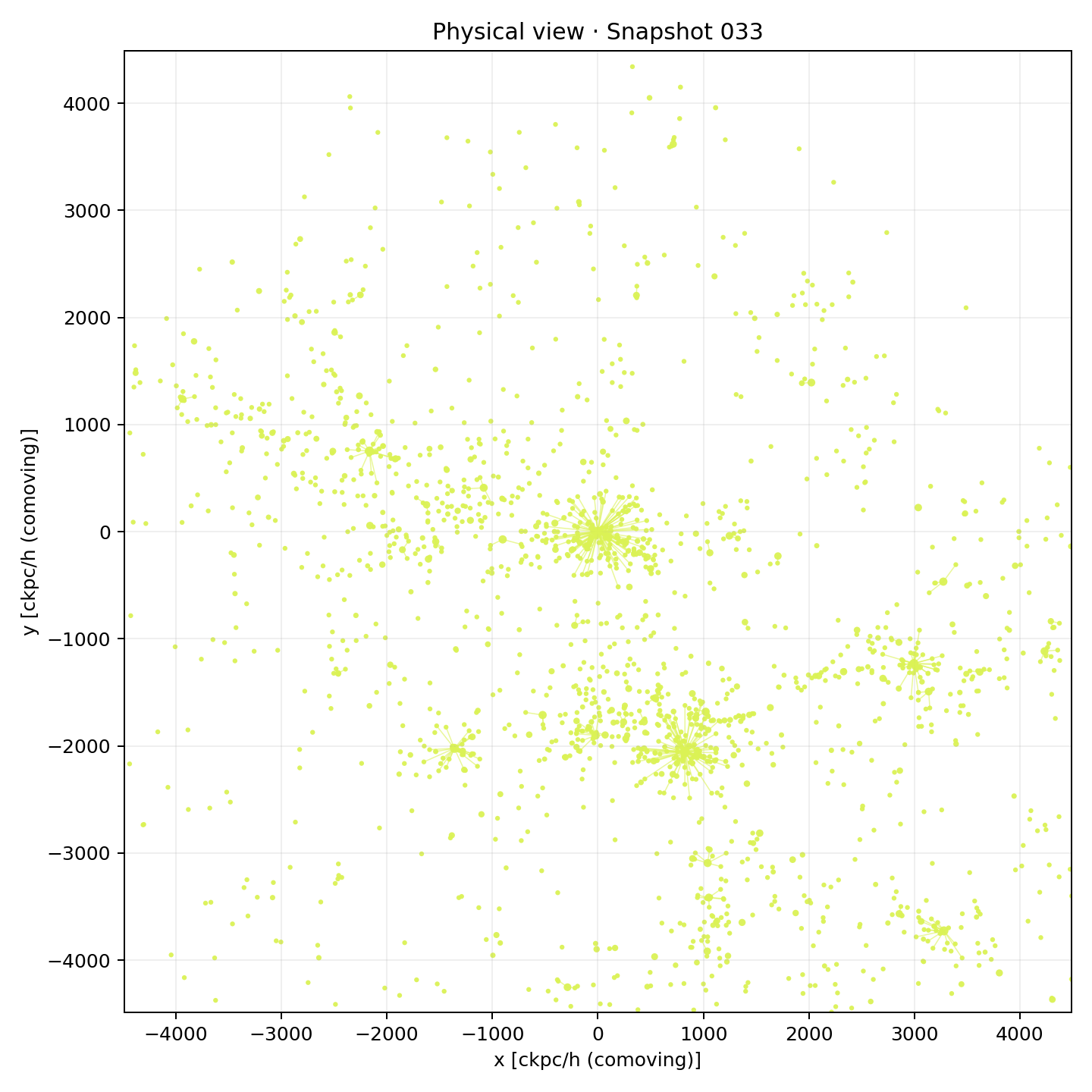}
  \includegraphics[width=0.24\textwidth]{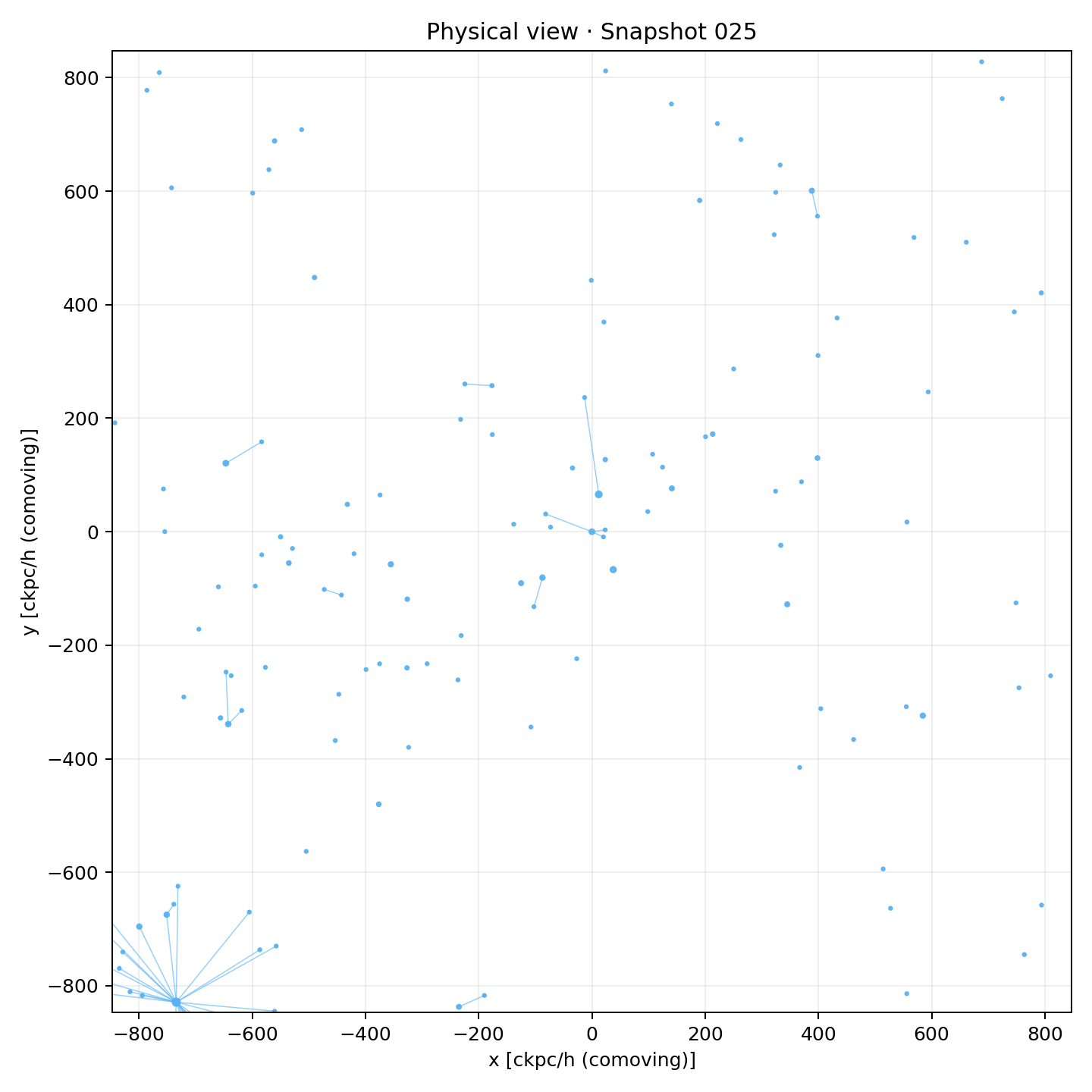}
  \includegraphics[width=0.24\textwidth]{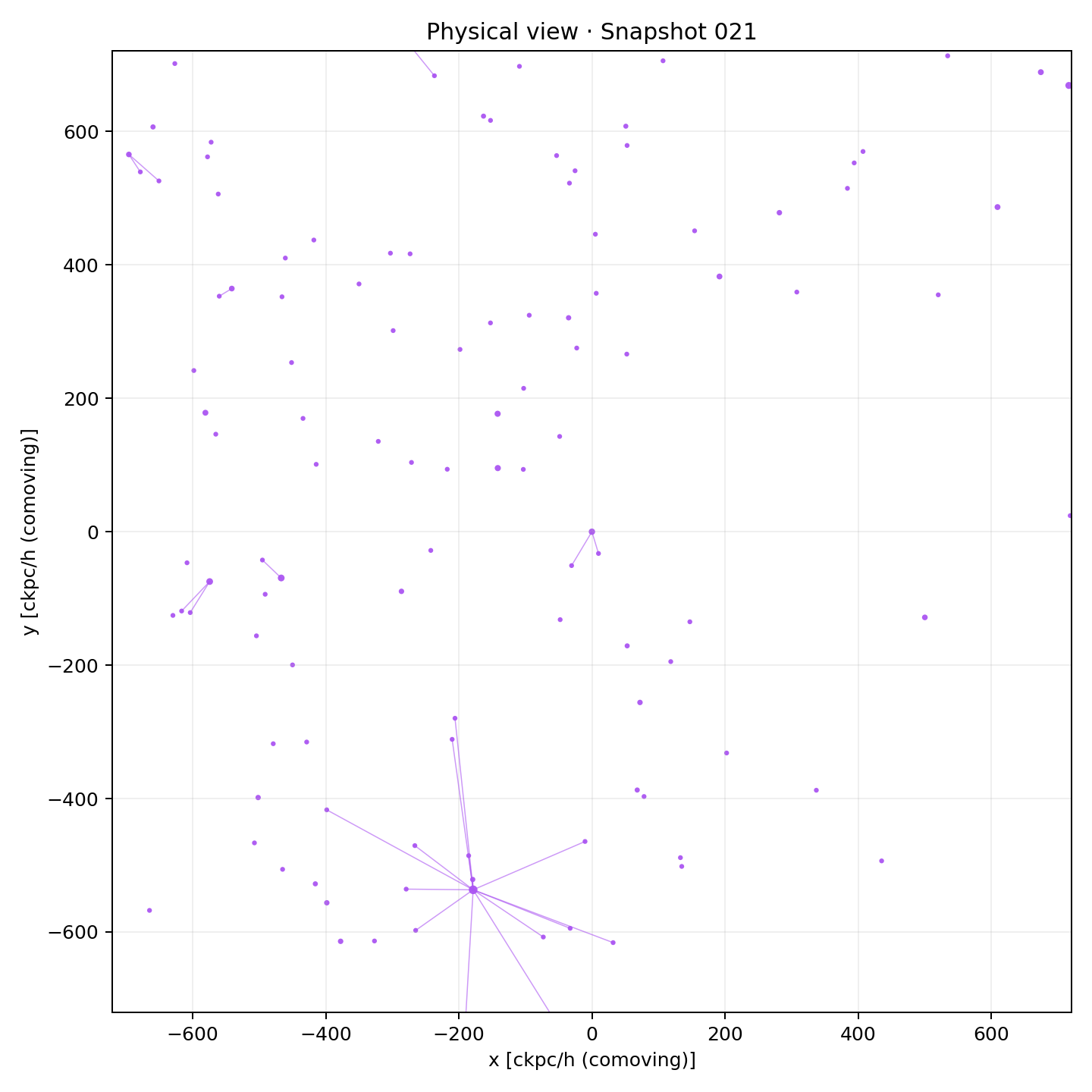}
  \caption{\label{fig:rank1_physical} Physical (comoving-coordinate) view of the same host–edge subgraphs shown in Fig.~\ref{fig:rank1_topology}. The snapshot ordering follows the topological view: $z=0, 0.1, 0.2, 0.3, 0.4, 0.5, 0.7, 1, 1.5,$ and $2$. The $z=3$ snapshot is omitted due to space limitations. This presentation enables a direct comparison between the graph connectivity and the corresponding spatial configuration at the same cosmic time.}
\end{figure*}

To visualize the layered construction, we show the host-edge subgraphs obtained by removing temporal edges from the rank-1 layered halo graph. 
Figures~\ref{fig:rank1_topology} and~\ref{fig:rank1_physical} illustrate these graphs at layers corresponding to redshifts from $3$ to $0$, in a topological view (spring layout) and a physical view (comoving coordinates), respectively.  
In the topological view, the system initially contains many isolated small graphs, most of them containing only one node. 
They merge into larger graphs and eventually form a single graph at $z=0$. 
In physical coordinates, their spatial clustering is more clearly displayed, with nodes gathering over time and moving into the virial radius of the endpoint ($z=0$) host halo. 
In all the panels, edges encode coeval host-satellite assignments that link satellites to their least-massive enclosing host halo~\cite{Yang:2022oqe}.

\section{Rollout parity and residual-trajectory diagnostics}
\label{sec:parity_appendix}

\begin{figure}[H]
  \centering
  \includegraphics[width=0.48\textwidth]{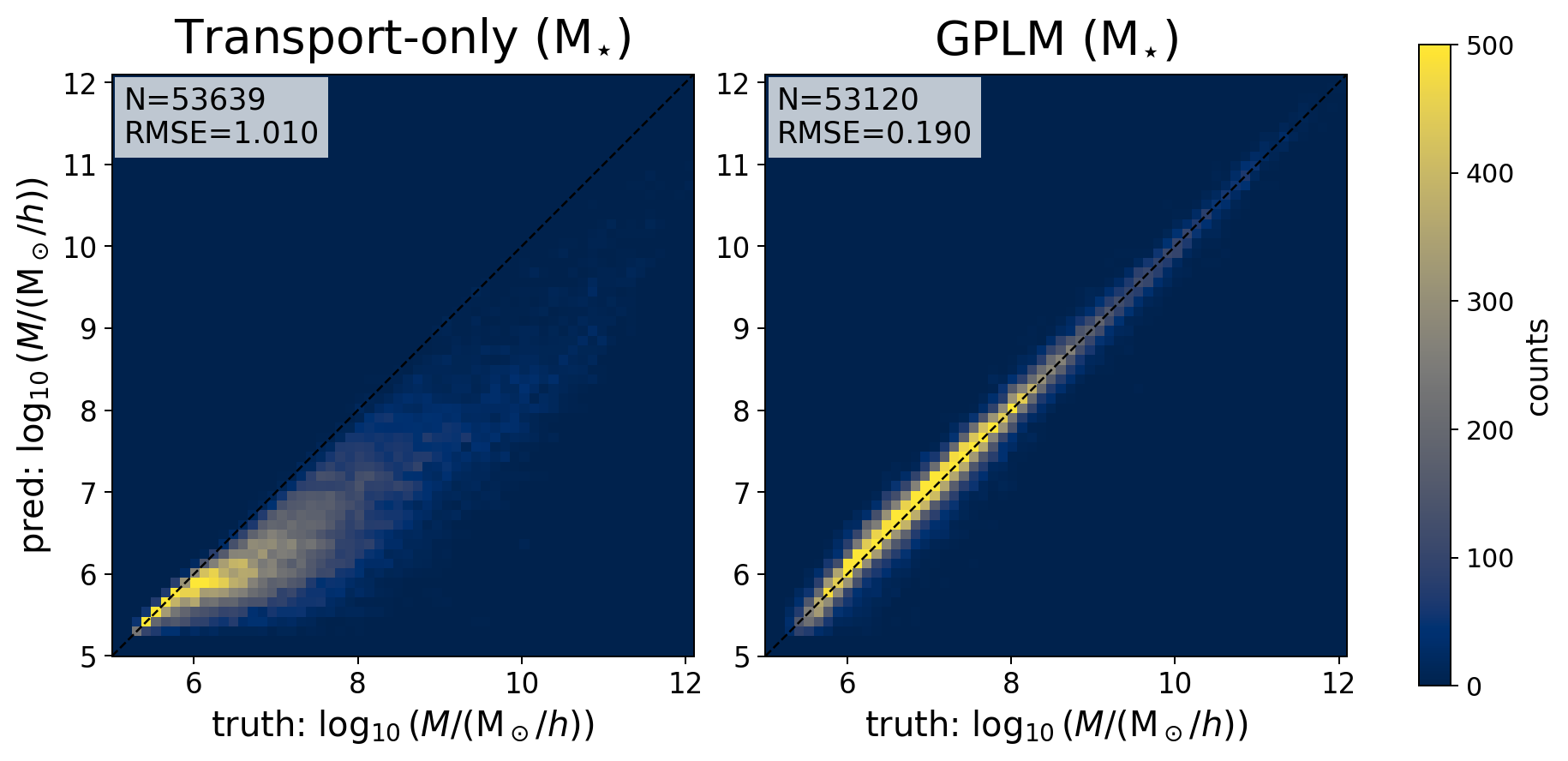}
  \includegraphics[width=0.48\textwidth]{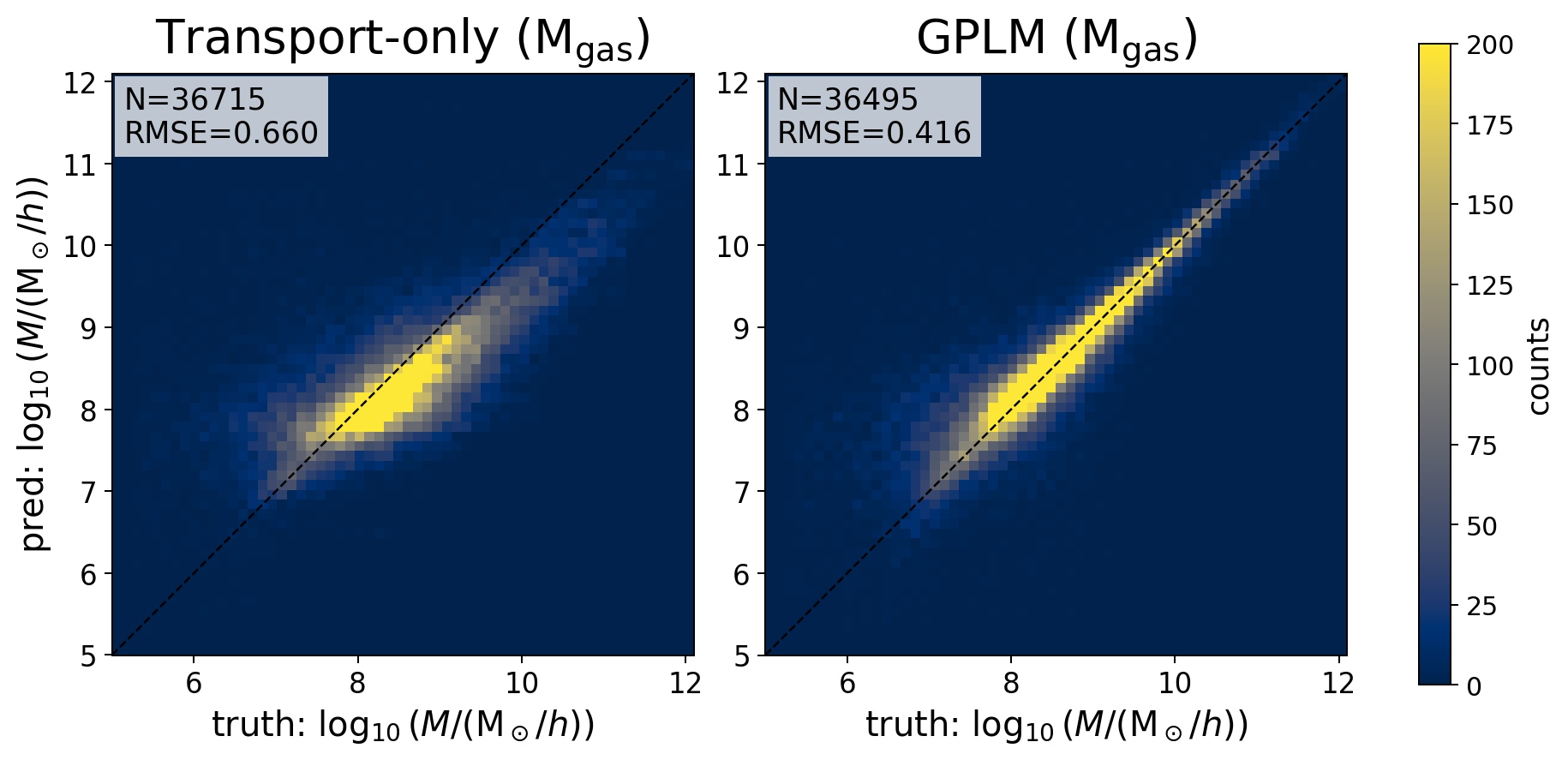}
  \caption{\label{fig:parity_stacked_main}
Stacked truth versus prediction parity across all snapshots for the stellar mass (left) and gas mass (right). In each figure, the left panel shows a transport-only reference and the right panel shows the GPLM prediction. Newly added nodes, whose values are directly seeded from the simulation truth at entry, are excluded from this comparison. Inset legends report the sample size and RMSE in $\log_{10}(M/(\mathrm{M}_{\odot}/h))$. }
\end{figure}

\begin{figure*}[t]
  \centering
  \includegraphics[width=0.95\textwidth]{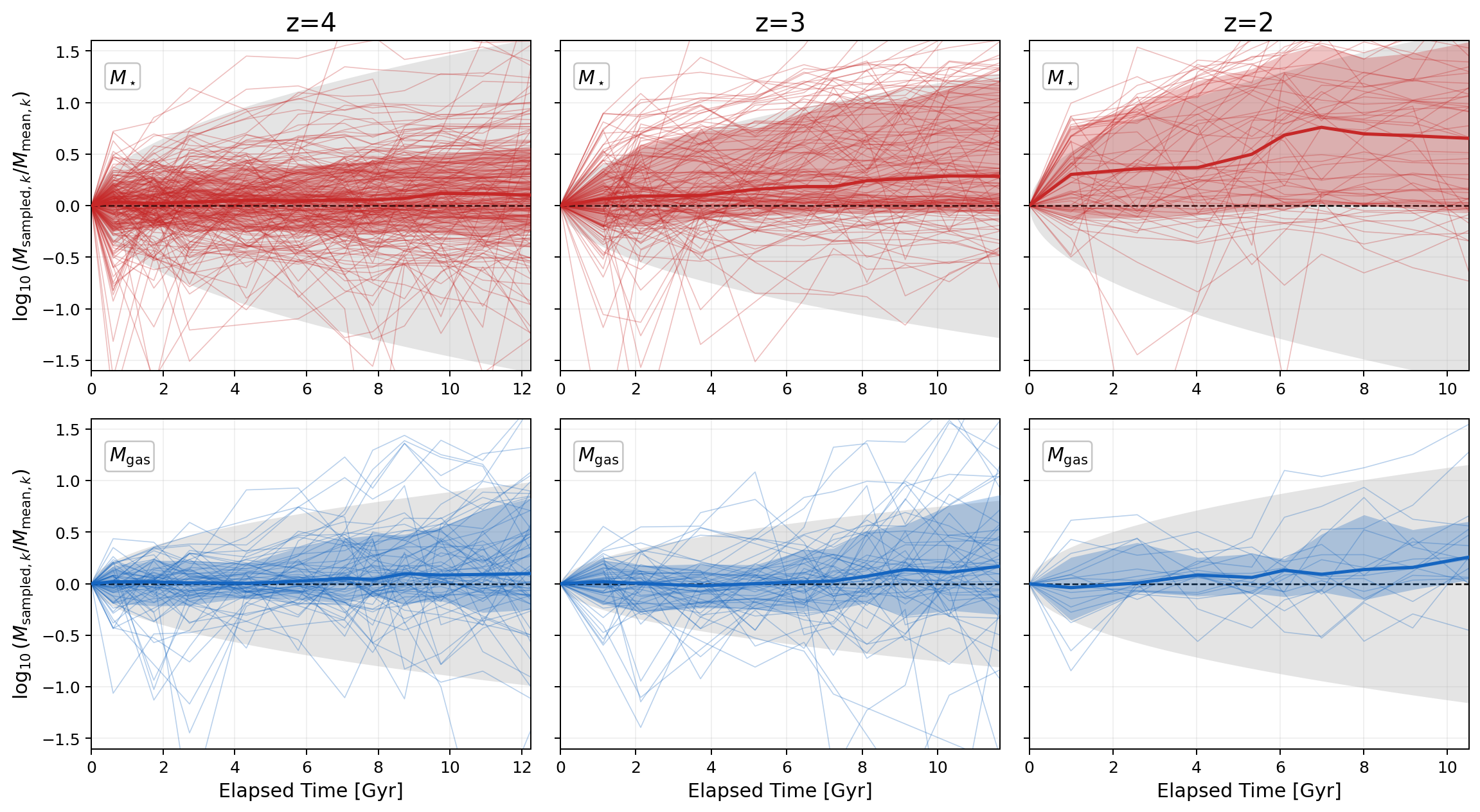}
  \caption{\label{fig:residual_trajectory_fan_appendix}
Residual rollout trajectories relative to the deterministic drift backbone, shown for entry groups beginning at $z=4$, $z=3$, and $z=2$, limited to tracks that persist to the final $z=0$ layer. The top row presents stellar mass residuals, while the bottom row displays gas mass residuals. The vertical axis shows $r_k=\log_{10}(M_{{\rm sampled},k}/M_{{\rm mean},k})$, and the horizontal axis indicates elapsed time since entry. Thin curves represent individual sampled trajectories, colored bands indicate the 16$^{\rm th}$-84$^{\rm th}$ percentile range, and thick curves show the median residual at each layer. The gray band gives the Brownian reference envelope, normalized using a $\sqrt{t}$ growth law based on the first propagated step of each group. This figure demonstrates that rollout scatter is driven by dynamic, state-dependent processes rather than by instantaneous broadening of the deterministic mean evolution.}
\end{figure*}

For completeness, this appendix collects two complementary rollout diagnostics. Figure~\ref{fig:parity_stacked_main} shows the stacked truth versus prediction parity comparison, while Fig.~\ref{fig:residual_trajectory_fan_appendix} visualizes sampled residual trajectories relative to the deterministic drift backbone. They are presented to illustrate point prediction performance and the time evolution of the learned stochastic sector. 

In Fig.~\ref{fig:parity_stacked_main}, the stacked Root-Mean-Square-Error (RMSE) over the displayed mass range decreases from $1.0$ dex to $0.19$ dex for $M_{\star}$ and from $0.66$ dex to $0.42$ dex for $M_{\rm gas}$. They are calculated based on nodes propagated from earlier retained layers, with newly entering truth-seeded nodes excluded from the comparison. 

To isolate the stochastic sector more directly, we also examine sampled rollout residuals relative to the deterministic mean-drift backbone. For each propagated node on a surviving trajectory, we define
\begin{equation}
  r_k \equiv \log_{10}\left(\frac{M_{{\rm sampled},k}}{M_{{\rm mean},k}}\right),
\end{equation}
where $M_{{\rm mean},k}$ is the autoregressive rollout obtained by evolving only the learned drift, and $M_{{\rm sampled},k}$ is the corresponding autoregressive rollout with sampled diffusion increments included at each retained layer pair. Figure~\ref{fig:residual_trajectory_fan_appendix} shows these residual trajectories for three entry-selected groups, defined by first appearance at $z=4$, $z=3$, and $z=2$, while retaining only tracks that survive to the final retained $z=0$ layer. The thin lines show individual trajectories, the colored bands show the 16$^{\rm th}$-84$^{\rm th}$ percentiles of the residual ensemble at each retained layer, and the thick curves show the corresponding medians. The gray band gives a Brownian reference envelope normalized to the first propagated step of each group.

Several qualitative features are visible in Fig.~\ref{fig:residual_trajectory_fan_appendix}. First, the residual ensemble broadens nontrivially with time rather than remaining a purely local one-step fluctuation, illustrating that the effective scatter is dynamical. Second, the growth of the colored bands is not a featureless $\sqrt{t}$ opening. In both channels, the residual spread evolves more irregularly than a stationary random walk would suggest, and in several panels the late-time growth appears visibly more regulated than the Brownian reference envelope. This behavior illustrates that the learned diffusion is an effective closure whose impact is modulated by the evolving graph-conditioned state, rather than a static, memoryless noise term with fixed amplitude. Because the model is trained on finite autoregressive windows, that local stochastic law is also constrained through its finite-horizon propagation.

\section{Attachment Fits for Newly Included Nodes}
\label{sec:attachment_fits}

The attachment factor $p_{\rm attach}$ supplies the boundary part of the graph-conditioned likelihood. Once a node is already linked to the layered graph by a temporal edge, its subsequent evolution is governed by the GPLM trajectory likelihood. 
By contrast, when a node first appears in the constructed graph without a progenitor in the adjacent earlier layer, its initial $(M_\star,M_{\rm gas})$ values must be specified separately. 
In the present implementation, that boundary sector is modeled through the approximate factorization
\begin{equation}
  p_{\rm attach}\approx P(M_\star\mid M_{\rm halo},z) P(M_{\rm gas}\mid M_{\rm halo},z),
\end{equation}
which provides a pragmatic closure for entry into the graph while leaving the in-graph environmental evolution to the learned trajectory measure.

The sample used for this fit should be interpreted in the same effective sense as the graph construction itself. 
Our layered halo graphs are designed to retain the halos most relevant to the $z=0$ host and to connect them through temporal and host edges.
However, due to technical and resolution limitations, they do not form a fully closed merger forest in every case. 
In particular, temporal links are only added between adjacent retained layers when a progenitor passing the mass threshold is identified. With the relatively sparse layer spacing used here, some halos could grow substantially between two retained layers and enter the constructed graph without incoming temporal edges from the previous layer. 
In the massive cases, their entries come together with the coeval host-edge structure, with the associated lower-mass nodes entering simultaneously. 
As a result, the attachment sample contains a spectrum of boundary nodes, from isolated low-mass entries to host-connected systems that appear after rapid growth between the retained layers.

Additional missing temporal edges can also arise from limitations of the merger construction itself. In particular, temporary irregularities in the SubLink histories can introduce spurious boundary nodes and blur the intended separation between the $p_{\rm attach}$ sector and the interior trajectory likelihood. In practice, this mainly widens the effective scatter in $p_{\rm attach}$. The GPLM dynamics is trained on simulation data embedded in the constructed graphs, and its accuracy is less affected.

The calibration is carried out on all constructed graphs using nodes with no incoming temporal edge from the previous retained layer, together with a halo-mass cut $M_{\rm halo}>10^8~\rm M_{\odot}/h$ and a mass-weighting scheme that reduces the dominance of the most numerous low-mass points. 
Table~\ref{tab:attach_counts} summarizes the number of entering nodes in each retained layer. The boundary sector is concentrated in the early retained snapshots. Out of 24,460 entering nodes in the full calibration sample, about 79\% appear by snapshot 50 ($z=1$) and about 82\% by snapshot 59 ($z=0.7$). This pattern indicates that $p_{\rm attach}$ mainly controls early-time entry into the constructed layered graphs, while its influence becomes progressively smaller at later times.
At the latest retained snapshot, the $M_{\rm gas}$ sample becomes too sparse to support a separate stable fit. In the present implementation, the snapshot-99 gas-channel entry relation is therefore taken from the snapshot-91 fit, while the $M_\star$ channel at snapshot 99 is fitted directly.

\begin{figure*}[t]
  \centering
  \includegraphics[width=0.24\textwidth]{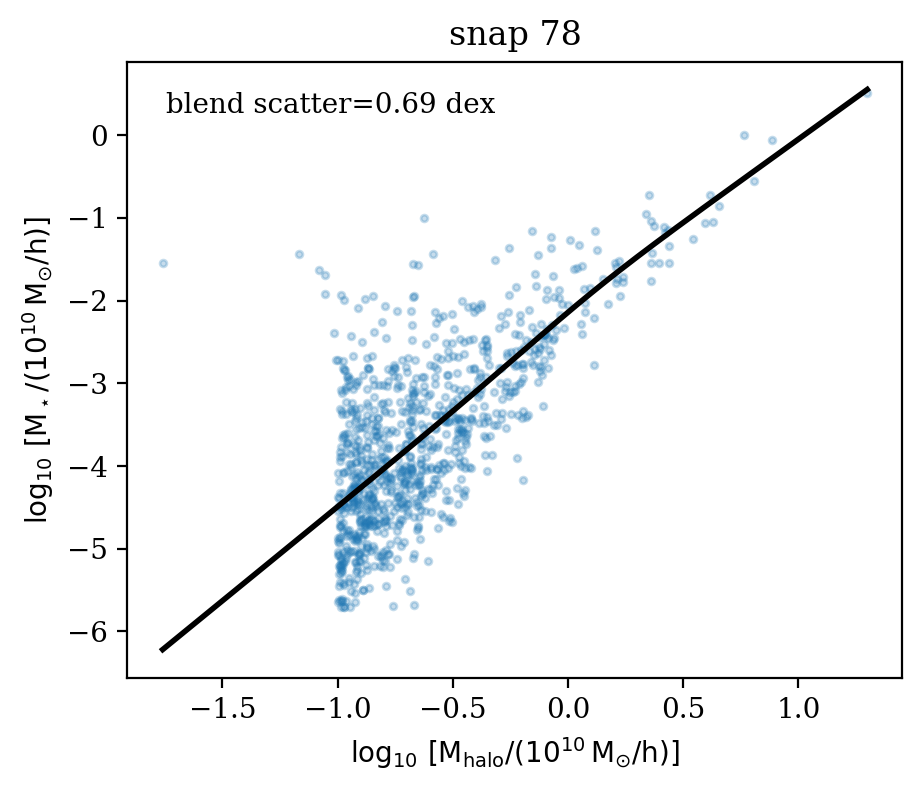}
  \includegraphics[width=0.24\textwidth]{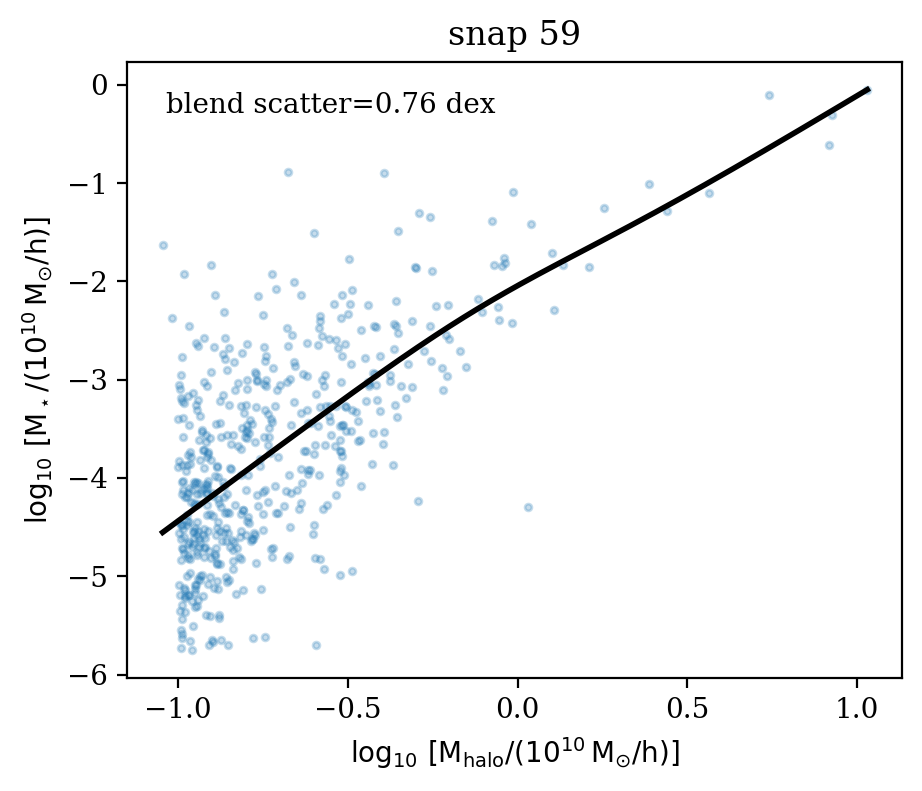}
  \includegraphics[width=0.24\textwidth]{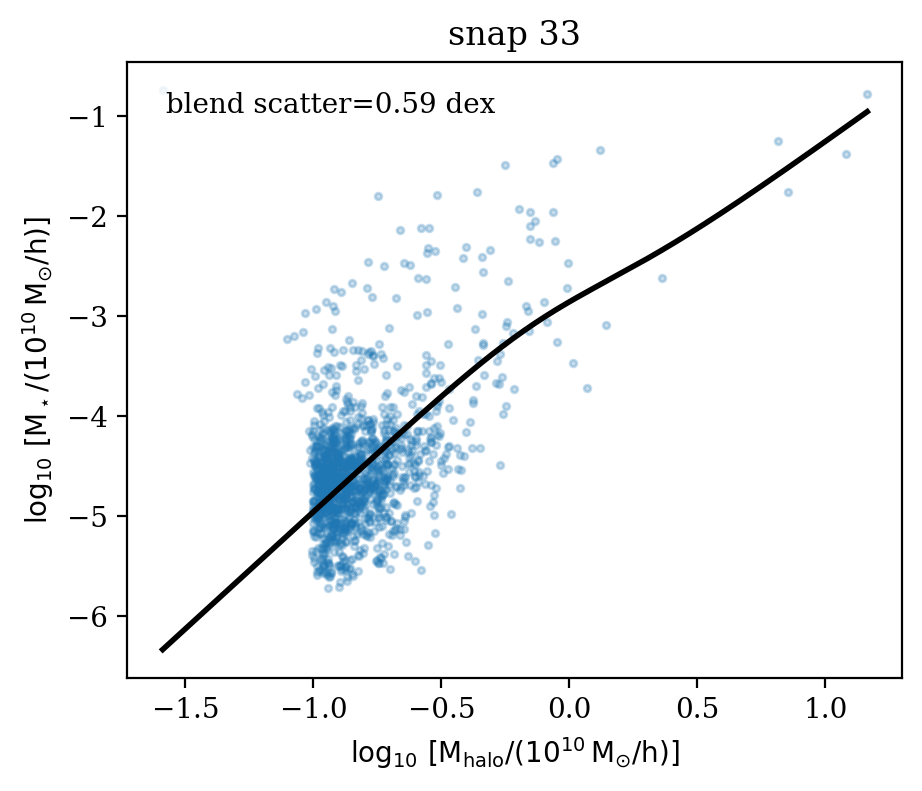} 
  \includegraphics[width=0.24\textwidth]{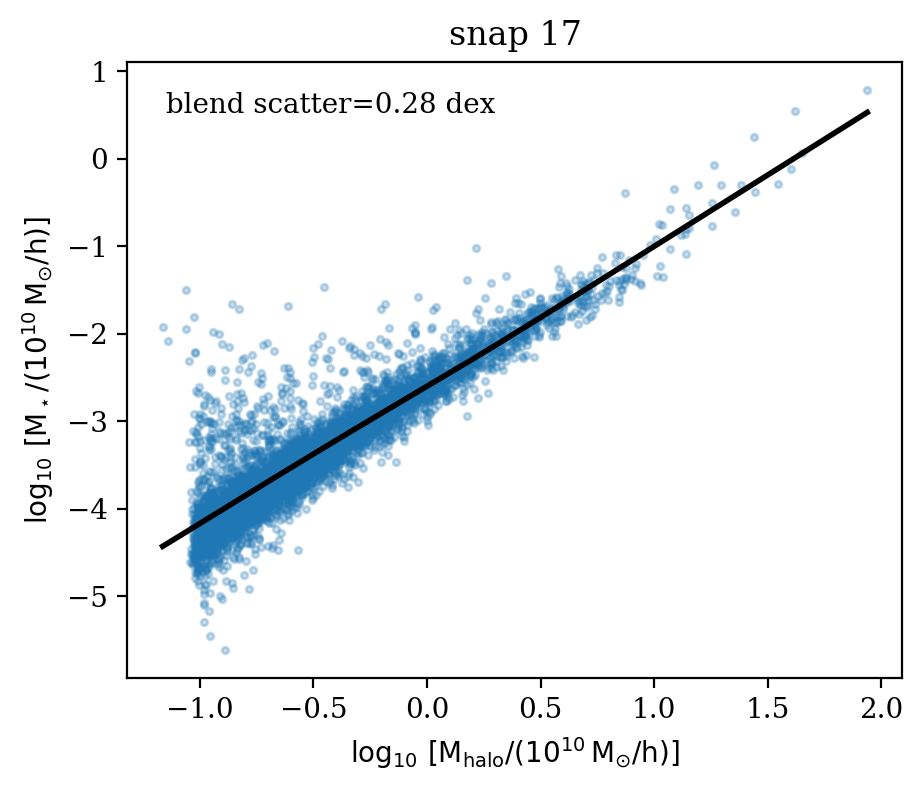} \\
  \includegraphics[width=0.24\textwidth]{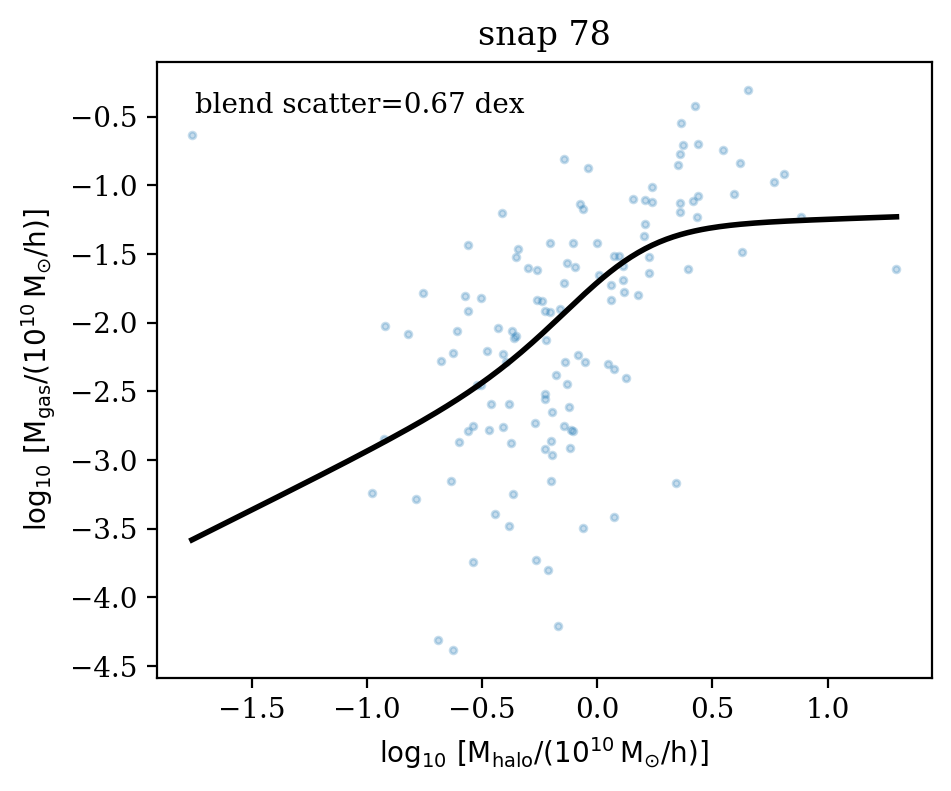}
  \includegraphics[width=0.24\textwidth]{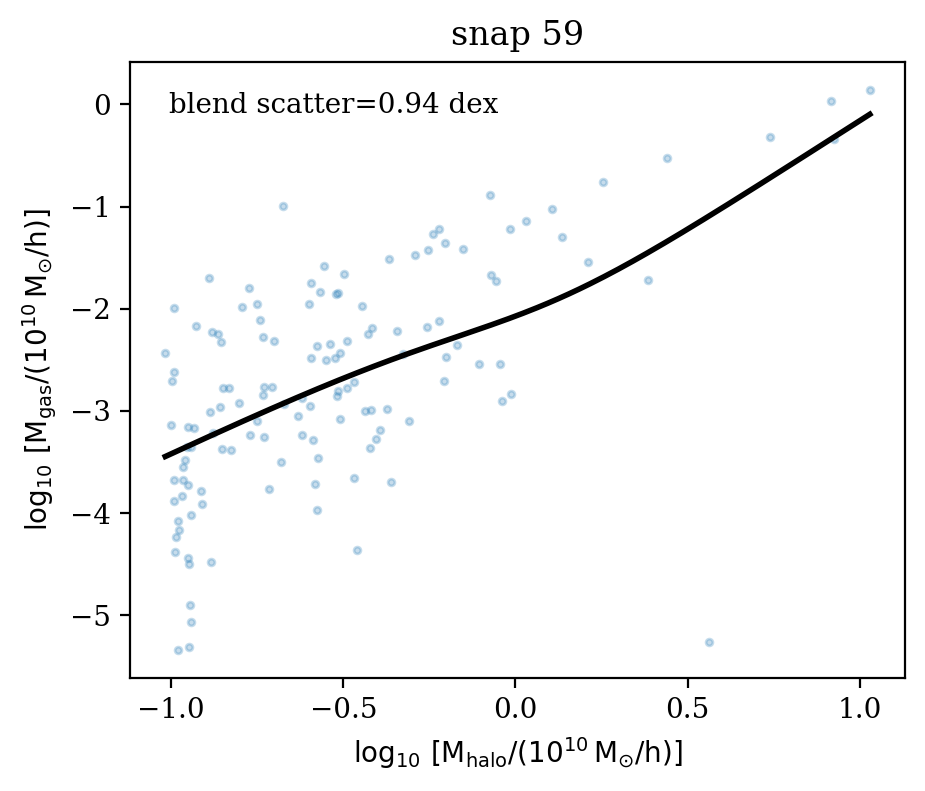}
  \includegraphics[width=0.24\textwidth]{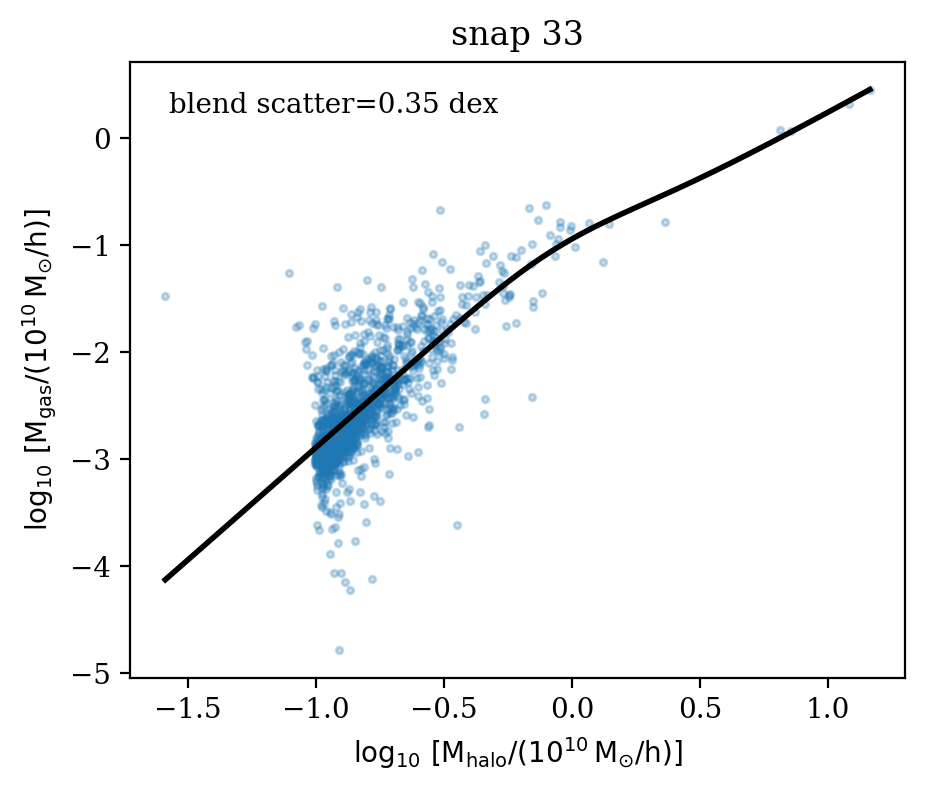}
  \includegraphics[width=0.24\textwidth]{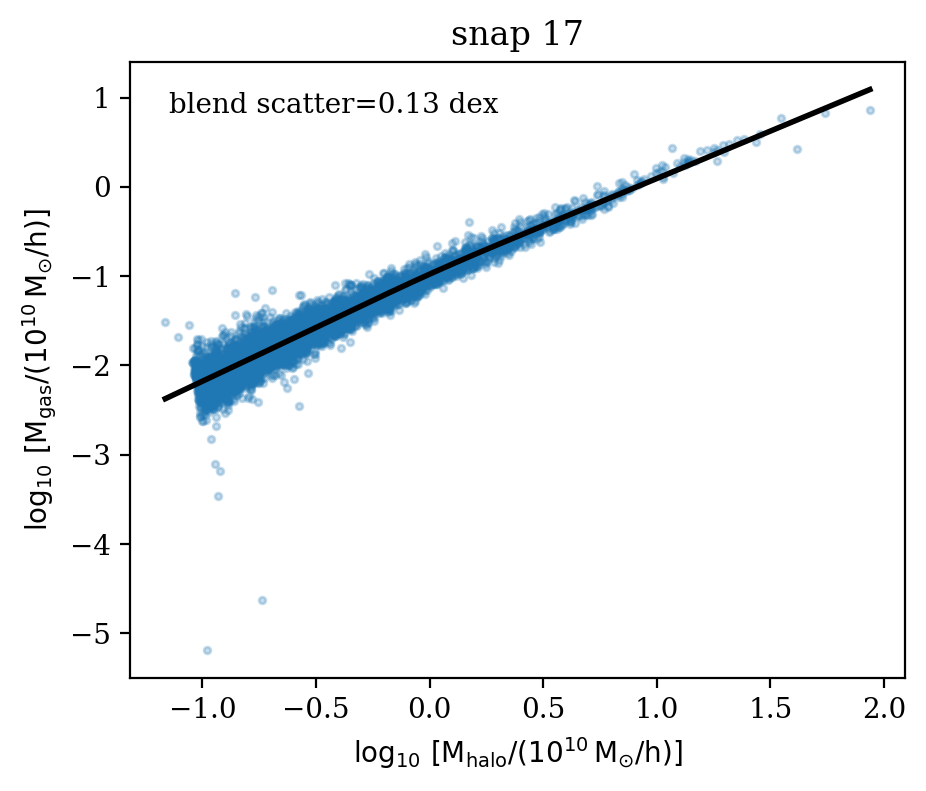}
  \caption{\label{fig:attachment_fit}
  Representative fits for newly included nodes in four snapshots. 
  Top (bottom) row illustrates the fits for $M_\star$ ($M_{\rm gas}$) versus $M_{\rm halo}$ for the newly added nodes.
  The solid curve in each panel shows the fit used to define the attachment likelihood.}
\end{figure*}

\begin{table}[t]
  \centering
  \caption{\label{tab:attach_counts} Number of entering nodes used in the $p_{\rm attach}$ calibration at each retained snapshot. Here $N_{\rm new}$ is the number of nodes with no incoming temporal edge from the previous retained layer, $N_{\rm layer}$ is the total number of nodes in that layer, and $f_{\rm new}\equiv N_{\rm new}/N_{\rm layer}$.}
  \begin{tabular}{cccc}
    \hline
    Snapshot & $N_{\rm new}$ & $N_{\rm layer}$ & $f_{\rm new}$ \\
    \hline
17 & 12007 & 12007 & 1.000 \\
21 & 2533 & 13962 & 0.181 \\
25 & 2078 & 14906 & 0.139 \\
33 & 1417 & 14147 & 0.100 \\
40 & 708 & 12915 & 0.055 \\
50 & 592 & 11269 & 0.053 \\
59 & 656 & 10031 & 0.065 \\
67 & 819 & 9091 & 0.090 \\
72 & 947 & 8531 & 0.111 \\
78 & 1095 & 8021 & 0.137 \\
84 & 1013 & 7224 & 0.140 \\
91 & 439 & 5841 & 0.075 \\
99 & 156 & 4017 & 0.039 \\
    \hline
  \end{tabular}
\end{table}

The attachment model is defined in log space as a two-branch linear relation,
\begin{equation}
  y_{\rm lo}(x) = \alpha_{\rm lo} + \beta_{\rm lo}x,
  \qquad
  y_{\rm hi}(x) = \alpha_{\rm hi} + \beta_{\rm hi}x,
\end{equation}
where $x\equiv\log_{10}[M_{\rm halo}/(10^{10}M_\odot/h)]$ and $y\equiv\log_{10}[M_\star/(10^{10}M_\odot/h)]$ or $y\equiv\log_{10}[M_{\rm gas}/(10^{10}M_\odot/h)]$. 
To avoid a discontinuity at the transition, we blend the two branches smoothly,
\begin{equation}
  w(x) = \frac{1}{1+\exp\left[-(x-x_{\rm split})/\Delta\right]},
  \qquad
  \mu(x) = [1-w(x)] y_{\rm lo}(x) + w(x) y_{\rm hi}(x),
\end{equation}
with $x_{\rm split}=0$ and $\Delta=0.2$.
We then model the conditional distribution as a lognormal (Gaussian in $y$) with mean $\mu(x)$ and scatter $\sigma(x)$. In the two-branch case, the scatter is blended in the same way,
\begin{equation}
  \sigma(x) = [1-w(x)] \sigma_{\rm low} + w(x) \sigma_{\rm high},
\end{equation}
where $\sigma_{\rm low}$ and $\sigma_{\rm high}$ are estimated from the weighted residuals in the low- and high-mass branches, respectively. 
Improved layered-graph construction with finer retained timesteps and more complete merger-history tracking should lead to a cleaner boundary sector and a more robust graph-path likelihood in future work.

Representative fits for four snapshots are shown in Fig.~\ref{fig:attachment_fit}. Top (bottom) panels show $M_\star$ ($M_{\rm gas}$) versus $M_{\rm halo}$. 
The solid curve shows the fit used to define the attachment likelihood for newly included nodes.

\end{document}